\documentclass[reprint, secnumarabic, amssymb, nobibnotes, aps, pra]{revtex4-1}
\usepackage[utf8]{inputenc}

 \usepackage{amsmath}
 \usepackage{bm}
 \usepackage{array}
 \usepackage{calc}
 \usepackage[dvipsnames]{xcolor}
 \usepackage{hyperref}
 \usepackage{ifthen}
 \usepackage{gensymb}
 \usepackage{graphicx}
 \usepackage{float}
 \usepackage{tikz}
 \usetikzlibrary{angles,calc,matrix,positioning,through,arrows.meta}
 \usepackage{tikz-layers}
 \usepackage{mathtools}
\DeclareMathOperator{\Arg}{Arg}

\DeclareMathOperator{\sgn}{sign}

\newcommand{\expected}[1]{\langle #1 \rangle}

\newcommand{\onlyif}{\quad \Longrightarrow \quad}

\newcommand{\bfhat}[1]{\boldsymbol{\hat{#1}}}

\newcommand{\odl}{(\Omega_{xy} \hat{L}_{xy} + \Omega_{zw} \hat{L}_{zw})}
\newcommand{\diff}[1]{\mathrm{d}#1}
\newcommand{\ta}{\theta}
\newcommand{\Ta}{\Theta}
\newcommand{\intTheta}{\ta}
\newcommand{\intTLim}{\ta_*}
\newcommand{\thetaFrac}{\chi}
\newcommand{\tiltR}{M}
\newcommand{\dint}{\int\displaylimits}

\newcommand{\cint}[1][0]{
    \ifthenelse{\equal{#1}{0}}{
        \dint_0^{2\pi}
    }{
        \dint_{#1}^{#1+2\pi}
    }
}
\newcommand{\cintm}[1][0]{
    \ifthenelse{\equal{#1}{0}}{
        \dint_{-\pi}^{\pi}
    }{
        \dint_{#1-\pi}^{#1+\pi}
    }
}

\newcommand{\cpt}{\cos\left(\bar{\ta}_1-\ta_2\right)}
\newcommand{\spt}{\sin\left(\bar{\ta}_1-\ta_2\right)}
\newcommand{\abs}[1]{\lvert#1\rvert}

\newcommand{\acr}{\acute{r}}
\newcommand{\act}{\acute{\ta}}
\newcommand{\grr}{\grave{r}}
\newcommand{\grt}{\grave{\ta}}

\newcommand{\cpolar}[2][]{r_{#2}#1e^{i\ta_{#2}#1}}
\newcommand{\ccpolar}[2][]{r_{#2}#1e^{-i\ta_{#2}#1}}

\newenvironment{stretchPmatrix}[1] {
    \renewcommand*{\arraystretch}{#1}
    \begin{pmatrix}
    } {
    \end{pmatrix}
}
\newcommand{\rot}[1]{
\begin{pmatrix}
    \cos#1 & -\sin#1 \\
    \sin#1 & \cos#1
\end{pmatrix}
}
\newcommand{\Rp}[1]{R(\phi_{#1})}

\newcommand{\en}{\text{--}}

\newcommand{\vDimens}{\frac{\hbar}{m}}

\begin{document}

\title{Curved vortex surfaces in four-dimensional superfluids: \\ I. Unequal-frequency double rotations}
\author{Ben McCanna}
	\email{benmccanna@outlook.com}
	\affiliation{School of Physics and Astronomy, University of Birmingham, Edgbaston Park Road, B15 2TT, West Midlands, United Kingdom}
	\author{Hannah M. Price}
			\email{H.Price.2@bham.ac.uk}
	\affiliation{School of Physics and Astronomy, University of Birmingham, Edgbaston Park Road, B15 2TT, West Midlands, United Kingdom}

\begin{abstract}
The study of superfluid quantum vortices has long been an important area of research, with previous work naturally focusing on two-dimensional and three-dimensional systems, where rotation stabilises point vortices and line vortices respectively. Interestingly, this physics generalises for a hypothetical four-dimensional (4D) superfluid to include vortex planes, which can have a much richer phenomenology. In this paper, we study the possibility of skewed and curved vortex planes, which have no direct analogue in lower dimensions. By analytically and numerically studying the 4D Gross-Pitaevskii equation, we show that such vortex surfaces can be stabilised and favoured by double rotation with unequal rotation frequencies. Our work raises open questions for further research into the physics of these vortex surfaces and suggests interesting future extensions to tilted vortex surfaces under equal-frequency double rotation and to more realistic 4D models. 
	\end{abstract}
\maketitle

\section{Introduction}

Quantum vortices are topological excitations that play an important role in the physics of superfluids~\cite{pitaevskii2003, pethick2002, cooper2008rapidly, fetter2009, madison2000, madison2001, matthews1999,abo2001observation,verhelst2017vortex}. Such vortices are characterised by the quantized circulation of the superfluid around a local density depletion, which is called the ``vortex core". As is well-studied, in a two-dimensional (2D) superfluid, the vortex core corresponds effectively to a 0-dimensional point, while for a three-dimensional (3D) superfluid, the core extends into a one-dimensional line or ring. As vortices are excitations, they are associated with an energy cost, but can be stabilised by either rotating the superfluid~\cite{fetter2009,cooper2008rapidly}, or equivalently by engineering artificial magnetic fields~\cite{dalibard2011colloquium,Cooper_2019,Ozawa2019Photonics,price2022roadmap}.

Recently, we began in Ref.~\cite{mccanna2021} to investigate the possible phenomenology of vortex structures in a four-dimensional (4D) superfluid, by studying a 4D generalisation of the Gross-Pitaevskii equation (GPE) including rotation~\cite{pitaevskii2003}. Interestingly, the extension to 4D considerably enriches the possible vortex structures as there are fundamental differences between rotations (or equivalently, magnetic fields) in different numbers of spatial dimensions. As we shall review further below, in 2D and 3D, all rotations are ``simple rotations" that can be characterised by a single rotation plane and rotation frequency, while in 4D, generic rotations are ``double rotations", meaning that two completely orthogonal planes of rotation, and hence two rotation frequencies, can be identified. In Ref.~\cite{mccanna2021}, we explored how the simplest case of a double rotation with equal frequencies can stabilise a new type of vortex structure in which the vortex core consists of two rigid orthogonal planes intersecting at a point, with no direct analogue in lower dimensions. 

In this paper, we shall go further to explore what happens as the two rotation frequencies in a 4D system are made unequal. As we shall show, this can lead to 4D vortex structures with cores composed of skewed non-orthogonal surfaces which curve to avoid the expected intersection point. We shall present both analytical and numerical calculations based on the 4D generalised GPE under rotation, and we shall develop and numerically test a theory to explain the skewed vortex planes in terms of a simplified competition between the rotational energy and the hydrodynamic vortex-vortex interaction terms. For unequal-frequency double rotations, we find skewed vortex surfaces that can be lower in energy than a pair of rigid orthogonal vortex planes~\cite{mccanna2021} for our system sizes and parameters. This lays the groundwork for a follow-on work in Ref.~\cite{mccannaequal}, which will apply a similar analysis to the case of equal-frequency double rotation.     

Looking further ahead, we note that we are studying a {\it minimal} 4D mathematical model, which is motivated as a natural extension of the standard GPE description of 2D and 3D superfluids. In the future, it will also be very interesting to explore if similar structures can be found in more experimentally-realistic models, connecting with recent theoretical and experimental advances in probing higher-dimensional physics~\cite{Kraus_2013,Price_2015,Ozawa2016,Lohse_2018,Zilberberg_2018,sugawa2018second,lu2018topological, kolodrubetz2016measuring, wang2020exceptional,wang2020circuit,Price2018, yu2019genuine,li2019emergence,ezawa2019electric,Weisbrich}, e.g. based on techniques such as ``synthetic dimensions"~\cite{Boada2012,Celi2014,Mancini2015,Stuhl2015,Gadway2015, An_2017,Price2017,salerno2019quantized, viebahn2019matter,barbiero2019bose, chalopin2020exploring,Ozawa2016,Yuan2016, ozawa2017synthetic, lustig2019photonic, Yuan2018rings,Yuan2018photonics,Yuan2019, yuan2020creating, Dutt2020,baum2018setting, price2019synthetic,crowley2019half,boyers2020exploring,ozawa2019topological,oliver2021bloch}, using which a 4D atomic quantum Hall system has recently been experimentally realized~\cite{bouhiron2022realization}. More generally, the rich phenomenology of curved vortex surfaces that we have begun to explore raises the possibility of finding other exotic topological excitations, such as closed vortex surfaces. Some of our results also suggest that vortices can lose some of their individual character in 4D, as the curved surfaces that we have found do not easily decompose into two separate but intersecting vortex states, unlike in Ref.~\cite{mccanna2021}. This opens interesting questions, for example, about what would happen at even higher rotation frequencies, where we may expect the number of vortices to become large. 

In this paper, we shall begin in Section~\ref{sec:vortexreview} by reviewing the basic physics of quantum vortices in 2D and 3D superfluids. Then in Section~\ref{sec:4DRot}, we shall discuss in more detail the different possible types of rotations in four spatial dimensions, including simple, double and isoclinic rotations. As part of this section, we shall discuss the rotation planes of isoclinic rotations in detail, as this will be useful for later sections of the paper. In Section~\ref{sec:4Dsuper}, we shall briefly review the 4D GPE and the physics of completely orthogonal 4D vortex planes, which were studied in 4D doubly-rotating superfluids with equal rotation frequencies in Ref.~\cite{mccanna2021}. We shall also introduce the numerical methods to be used throughout this work. In Section~\ref{sec:NonOrtho}, we shall then derive the hydrodynamic energy of a pair of non-orthogonal skewed 4D vortex planes that intersect at a point. As we then develop in Section~\ref{sec:Unequal}, our hydrodynamic theory predicts that, in a doubly-rotating 4D superfluid with two unequal rotation frequencies, a pair of rigid vortex planes will become skewed so as to start aligning with the higher frequency and with each other. Our numerical results verify this predicted skewness at large distances, but also show that the vortex surfaces curve near the origin to avoid the intersection point. Finally, in Section~\ref{sec:concl}, we shall summarize our results and discuss possible future extensions. 

\section{Review of superfluid vortices} \label{sec:vortexreview}

In this section, we shall briefly review the basic properties of quantized vortices in 2D and 3D superfluids. We shall begin by introducing the GPE and reviewing how this leads to the structure of a 2D superfluid vortex, before discussing the energy of such a state within a finite system. We shall then briefly discuss systems with multiple 2D vortices, before reviewing some aspects of vortex physics in 3D superfluids. This will lay the groundwork for the discussion of 4D vortices in Section~\ref{sec:4Dsuper} onwards. 

\subsection{Structure of a 2D Superfluid Vortex}
\label{sec:2DVortex}

We consider a system of weakly-interacting bosons in the absence of an external potential as can be described by the time independent Gross-Pitaevskii equation~\cite{pitaevskii2003}
\begin{equation}
	-\frac{\hbar^2}{2m}\nabla^2\psi + g|\psi|^2\psi = \mu\psi,
	\label{eq:GPE}
\end{equation}
where $\psi$ is the complex order parameter, \(m\) is the mass of the particle, $g$ is the interaction strength, and \(\mu\) is the chemical potential. Note that a harmonic trapping potential is 
also often included in the GPE, as this is present in many ultracold gas experiments~\cite{pitaevskii2003, pethick2002, cooper2008rapidly, fetter2009, madison2000, madison2001, matthews1999,abo-shaeer2001}; however, for simplicity we will neglect all such effects and restrict our discussion to infinite systems and finite systems with hard-wall boundary conditions, as specified below. 

From the order parameter, it is possible to directly define the superfluid density, phase, and velocity fields as $\rho = |\psi|^2$, $S = \Arg\psi$ and $\mathbf{v} = \frac{\hbar}{m}\nabla S$, respectively~\cite{pitaevskii2003}. Importantly, the form of the velocity field implies that the circulation of the superfluid around any closed loop $C$ has to be quantized with the circulation being given by
\begin{equation}
    \oint_C \mathbf{v}\cdot\diff\mathbf r = \frac{\hbar}{m} \oint_C \nabla S \cdot\diff\mathbf r,
\end{equation}
where the integral on the right hand side gives the change in the superfluid phase around the loop \(C\). This would be zero if \(S\) is single-valued, but since \(S\) is a phase and its value is only defined modulo \(2\pi\), so the superfluid circulation is quantized generally as~\cite{pitaevskii2003, cooper2008rapidly, fetter2009,  pethick2002, verhelst2017vortex}
\begin{equation}
    \oint_C \mathbf{v}\cdot\diff\mathbf r = 2\pi k\frac{\hbar}{m}, 
    \label{eq:winding}
\end{equation}
where \(k\) is an integer called the winding number.  Note that continuous deformations of the loop \(C\) within the superfluid will not change the integer winding number, since \(\mathbf{v}\) varies continuously so long as $\psi$ is non-zero. This means that $k \in \mathbb{Z}$ is a topological invariant, which will be equal to zero whenever \(C\) can be continously contracted to a point. Hence, a simply-connected superfluid (i.e. one in which all loops are contractible) cannot circulate. 

If a simply-connected superfluid is rotated, it therefore cannot behave as a rigid body but will instead form internal holes, called vortices, where the density goes to zero and around which the phase winds by a quantized amount~\cite{pitaevskii2003, cooper2008rapidly, fetter2009, pethick2002, madison2000, madison2001, matthews1999,abo-shaeer2001}. In 2D, the structure of a rotationally-symmetric vortex is described by 
\begin{equation}
    \psi (r,\ta) = |\psi(r)| e^{i k \ta}, 
    \label{eq:2D}
\end{equation}
where $(r, \ta)$ are 2D polar coordinates centered on the vortex core. The winding number $k$ corresponds to the topological charge of the vortex, and is conventionally taken to be positive for vortices and negative for anti-vortices.  From the above ansatz, the vortex velocity field is then~\cite{pitaevskii2003}
\begin{equation}
{\bf v} = \frac{\hbar}{m} \nabla  (k  \ta) =  \frac{\hbar}{m r} k  \bfhat{\ta},   
\label{eq:velocity}
\end{equation}
where $\bfhat{\ta}$ is the unit vector pointing along the $\ta$ direction. As required by Eq.~\eqref{eq:winding}, this velocity field falls off as $1/r$, and reverses direction when the sign of the winding number, $k$, is flipped. The angular momentum of the vortex in Eq.~\eqref{eq:2D} is also quantized as
\begin{equation}
{\bf L} = {\bf r } \times m {\bf  v} = \hbar k  \bfhat{z}
\end{equation}
with respect to the axis through the center of the vortex core, with $\bfhat{z}$ being the unit vector pointing out of the 2D plane, along the $z$ direction. More generally, in a system with axial symmetry, the angular momentum is quantized only for an on-axis vortex~\cite{pethick2002}. 

To complete this description of the vortex state [Eq.~\eqref{eq:2D}], the density-profile can be obtained numerically by solving the GPE [Eq.~\eqref{eq:GPE}]. When so doing, it is common to define the uniform background density as $n$ and then rescale $\psi \to \sqrt{n} \psi$ and $\mathbf{r} \to \xi \mathbf{r}$, where $\xi$ is the healing length, which satisfies $\hbar^2/m\xi^2 = gn = \mu$, and which physically is the distance over which $\rho$ typically varies. (Note that often a factor of $1/2$ is included in the definition of $\xi$~\cite{pitaevskii2003}.) Under these rescalings, Eq.~\eqref{eq:GPE} becomes dimensionless as
\begin{equation}
    -\frac{1}{2}\nabla^2\psi + |\psi|^2\psi = \psi
    \label{eq:DGPE},
\end{equation}
which is then solved numerically~\cite{pitaevskii2003,pethick2002} to obtain a dimensionless real-valued density profile function \(f_k(r)=|\psi|\). While the obtained $f_k (r)$ has no closed form, it can be shown that it asymptotically vanishes towards the vortex core as $f_k (r) = \mathcal{O} (r^{|k|})$ when $r \rightarrow 0$,  and asymptotically returns to the profile of a homogeneous superfluid as $f_k (r) = 1-  \mathcal{O} (r^{-2})$ when  $r \rightarrow \infty$~\cite{fetter2009}, with a crossover between these two behaviours around the healing length.   

% Detailed density profile less relevant, but the failure of a product approximation is important

\subsection{Energy of a 2D Superfluid Vortex}

Within a hydrodynamic description~\cite{pitaevskii2003, pethick2002, verhelst2017vortex}, the energy of a 2D vortex in the absence of an external potential is made up of a kinetic contribution 
\begin{equation}
    E_{\text{kin}} [\psi]= \frac{\hbar^2}{2m} \int (\nabla \sqrt{ \rho})^2 \diff^2 r + \frac{m}{2} \int \rho {\bf v}^2 \diff^2 r ,
    \label{eq:Kinetic}
\end{equation}
and an inter-particle interaction contribution
\begin{equation}
    E_{\text{int}} [\psi] = \frac{1}{2} g \int  \rho^2 \diff^2 r ,
    \label{eq:Interaction}
\end{equation}
where the integrals are carried out over the area of the 2D system, which we consider to be circular with radius $R$ and hard-wall boundary conditions. 

Both of the above contributions can be estimated analytically by assuming that the density is zero within the vortex core, which we take to be circular with a radius of our healing length $\xi$, and constant otherwise across the system. (Note that other definitions for the size of the vortex core can be used~\cite{pitaevskii2003, verhelst2017vortex}; however, the following argument is only approximate and such changes do not significantly affect the relative scaling and importance of the different energy terms.) For such a simple, so-called ``hollow-core", model for a vortex, the total energy difference, $\Delta E$, between the homogeneous and the vortex state is straightforward to calculate; firstly, in the hydrodynamic kinetic energy (Eq.~\eqref{eq:Kinetic}) introduced above, the first (``quantum pressure") term vanishes so that the extra kinetic energy due to a vortex is given simply by~\cite{pitaevskii2003} 
\begin{equation}
  \Delta  E_{\text{kin}} =  \frac{m}{2} \int \rho {\bf v}^2 \diff^2 r\simeq k^2 \pi n \frac{\hbar^2}{m} \ln \left( \frac{R}{\xi} \right),
  \label{eq:HollowKinetic}
\end{equation}
where $n$ is the constant density within the system outside of the vortex core. Similarly, the interaction energy (i.e. the energy needed to make a hole in the superfluid) can be calculated as
\begin{equation}
 \Delta   E_{\text{int}}  \simeq \frac{1}{2} gn^2\xi^2 \pi.
 \label{eq:inter}
\end{equation}
The latter can clearly be neglected for a large system with $R \gg \xi$, meaning that the total hydrodynamic energetic cost of a vortex can be approximated as~\cite{verhelst2017vortex}
\begin{equation}
 E_\text{h} \approx \Delta E_{\text{kin}} \simeq k^2 \pi n \frac{\hbar^2}{m} \ln \left( \frac{R}{\xi} \right)
\label{eq:hydrodynamic2D}
\end{equation}
A more accurate estimate of the energy cost relative to the uniform state can be found using the dimensionless numerical density-profile function, $f_k (r)$, in the vortex-state ansatz [Eq.~\eqref{eq:2D}]. Using the grand canonical energy at fixed chemical potential $\mu$ takes care of corrections to the background density \(n\) from the core depletion. Then, using that, \(\mu \!=\! \hbar^2/m\xi^2 \), and $n = N / \pi R^2$, in the uniform system, where $N$ is the total number of bosons, we may write this numerical vortex energy as
\begin{equation}
    E_k(R) = k^2 \mu N  \left(\frac{\xi}{R}\right)^2\ln\left(2.07\frac{R}{\xi}\right),
    \label{eq:VEnergy}
\end{equation}
which importantly is the same functional form as the simple hydrodynamic estimate [Eq.~\eqref{eq:hydrodynamic2D}], up to the numerical prefactor within the logarithm. (Note that if the healing length had instead been defined including a factor of $1/2$ as $\hbar^2/ 2 m\xi^2 = gn = \mu$, then this numerical prefactor becomes 1.46~\cite{pitaevskii2003}.) 

Vortices can be energetically stabilised by rotation (or equivalently an artificial magnetic field)~\cite{fetter2009,cooper2008rapidly}. In a rotating reference frame, the GPE [Eq.~\eqref{eq:GPE}] becomes
\begin{equation}
	\left[-\frac{\hbar^2}{2m}\nabla^2 + g|\psi|^2 - \mathbf{\Omega}\cdot\hat{\mathbf{L}}\right] \psi = \mu\psi, \label{eq:GPE2DR}
\end{equation}
 where $\hat{\mathbf{L}} = -i\hbar\mathbf{r}\times\nabla$ is the (3D) angular momentum operator, and $\mathbf{\Omega}$ is the rotation frequency vector~\cite{pitaevskii2003}. In 2D, we can assume that  $\mathbf{\Omega}= \Omega \bfhat{z}$, and hence the energy reduction from rotation is given by $\Delta E_{\text{rot}} = \Omega \langle \hat{L}_z \rangle$. As discussed above, vortices carry a finite amount of angular momentum and so are favoured by rotation. 

To leading order, we can assume the superfluid has a constant density and neglect the depletion of the core, so that the energy reduction from rotation can be calculated as~\cite{pitaevskii2003}
\begin{equation}
\Delta E_{\text{rot}} = \Omega \langle \hat{L}_z \rangle \simeq \Omega\hbar k \pi n R^2.
\label{eq:rot}
\end{equation}
(This approximation cannot be applied to the calculation of the hydrodynamic energy of a single vortex [Eq.~\eqref{eq:HollowKinetic}] as the \(1/r^2\) dependence of the integrand gives a singular contribution from the area around \(r=0\).)
As can be seen, this term reduces the energy of a state containing a vortex for which the circulation is aligned with the rotation, and raises the energy of a state (with opposite $k$) that is anti-aligned with the rotation. For a vortex to be energetically stabilised, the reduction in energy must be greater than (or equal to) the cost of making a vortex within the same approximations (e.g. Eq.~\eqref{eq:hydrodynamic2D}). This leads to an estimate of the critical frequency of~\cite{pitaevskii2003}
\begin{equation}
\Omega_c^{2D} \simeq k \frac{\hbar }{m R^2}   \ln \left(\frac{R}{\xi} \right) ,
\label{eq:crithollod}
\end{equation}
i.e. this is the minimal rotation frequency needed to stablise a vortex with winding number $k$. Using the energy for the numerical vortex profile (Eq.~\eqref{eq:VEnergy}) leads to a more accurate calculation for this frequency as
\begin{equation}
\Omega_c^{2D} = k \frac{\hbar }{m R^2}   \ln \left( 2.07 \frac{R}{\xi}\right).
\label{eq:2Dcrit}
\end{equation}
Note also that this critical frequency will depend on any external potentials that are present, and so will be different, e.g. with a harmonic trap~\cite{pethick2002}. However, in this paper, we will focus on untrapped systems with hard-wall boundary conditions, as mentioned above.

\subsection{Multiple Vortices in a 2D Superfluid}
\label{sec:Multiple2D}

As can be seen from Eq.~\eqref{eq:2Dcrit}, the critical rotation frequency is proportional to the winding number $k$, meaning that higher frequencies are required to stabilise vortices with higher winding numbers. However, by comparing the hydrodynamic energy with the rotation energy, it can be seen that, even at higher frequencies, it will always be energetically unfavourable (in the absence of additional external potentials) to produce a multiply charged vortex (i.e. with $|k|>1$) as compared to multiple singly-charged vortices (with $|k|=1$)~\cite{verhelst2017vortex}. 
%Consequently, we can focus our discussion on topological vortices with winding numbers, $|k|=1$. 

The above argument also suggests that a pair of similarly-charged 2D vortices will interact repulsively, as it is energetically unfavourable to bring them together and merge them into a single vortex with a higher winding number. Indeed, it can be shown that, in a sufficiently large system, the interaction energy between a pair of well-separated vortices, with charges $k_1$ and $k_2$ respectively, can be approximated as~\cite{pethick2002}
\begin{equation}
    \Delta E_{\text{pair}} \propto k_1 k_2 \ln \left(\frac{R}{\Delta r}\right)
    \label{eq:2DInt}
\end{equation}
where $\Delta r$ is the distance between the two vortex cores. As can be seen, this is attractive for oppositely-charged vortices (i.e. a vortex and anti-vortex pair) but repulsive for vortices with the same sign. In an infinite system, a pair of vortices can therefore continually lower their energy by moving apart, while a vortex and anti-vortex pair can lower their energy by coming together and annihilating. Note that Eq.~\eqref{eq:2DInt} is derived under the approximation that the density is constant everywhere in the system, i.e. ignoring the density depletion at the vortex core. Consequently, the calculated vortex-vortex interaction energy [Eq.~\eqref{eq:2DInt}] is only valid for separations \(\Delta r\gg\xi\), and attempting to take the limit \(\Delta r\to0\) gives a logarithmic divergence. In reality, when a pair of vortices with winding numbers \(k_{1,2}\) come together, they combine into a vortex with winding number \(k_1+k_2\). One can still obtain this correct result from Eq.~\eqref{eq:HollowKinetic} if we consider the vortices to be combined once their separation is similiar to the healing length \(\Delta r \sim \xi\). This is consistent with the constant density approximation, as the latter amounts to ignoring variations on the scale of \(\xi\) or below (except in the presence of a trap).

As the rotation frequency increases therefore above the critical frequency [Eq.~\eqref{eq:2Dcrit}], it will be energetically favourable to have more and more singly-charged vortices in the system. The effectively repulsive interactions between these vortices then mean that, at high enough rotation frequencies, the lowest energy state in the rotating frame exhibits a uniform array of vortices, known as an Abrikosov lattice~\cite{abrikosov1957magnetic, pethick2002}. 

\subsection{Vortices in 3D Superfluids}
\label{sec:3DVortices}

The above discussion can be straightforwardly generalised to describe vortices in a 3D superfluid~\cite{pitaevskii2003, pethick2002, verhelst2017vortex}. In 3D, a vortex core can be approximated as an extended 1D line, which must either begin and end on the surface of the system, or else form a closed loop within the superfluid. The former structures are often referred to as ``vortex lines" or ``vortex filaments", while the latter are typically called ``vortex rings"~\cite{verhelst2017vortex,Anderson2001,rosenbusch2002dynamics,komineas2007vortex,carretero2008nonlinear, bisset2015robust, wang2017single}. As our paper is concerned with the lowest-energy vortex structures to be stabilised by rotation, we shall hereafter focus on vortex lines, although it would also be very interesting to study the analogue of vortex rings in higher dimensions. 

In the simplest case, a cylindrically-symmetric vortex line in 3D can be described~\cite{verhelst2017vortex} e.g. by
\begin{equation}
        \psi (r, \ta, z) = |\psi(r, z)| e^{i k \ta}, 
\end{equation}
in cylindrical polar coordinates $(r, \ta, z)$, where we have assumed that the rotation axis lies along the $z$ direction and that the rotation is sufficiently strong so as to align and straighten the vortex core. In the absence of an additional potential, the vortex structure is then invariant along the $z$ direction and the dimensionless density profile is given by the radial function found numerically from the 2D GPE. Consequently, a 3D vortex line has the same velocity field as a 2D vortex [Eq.~\eqref{eq:velocity}], as well as the same critical frequency (in a cylindrical system)~\cite{verhelst2017vortex}. The latter point can be easily appreciated by noting that, in this case, the 2D calculation for the hydrodynamic energy follows through identically up to an overall multiplicative factor in both Eq.~\eqref{eq:hydrodynamic2D} and Eq.~\eqref{eq:rot}, to represent the height of the system~\cite{verhelst2017vortex}.

Similarly, when the rotation frequency becomes much higher than the critical frequency, many vortices enter the 3D system, and should eventually form a vortex lattice analogous to that in 2D, except with the vortex cores extended as straight lines along the rotation axis~\cite{abo2001observation,verhelst2017vortex}. It is also worth noting that, unlike in 2D, the shape and orientation of a 3D vortex line can depend, for example, on both the choice of rotation axis as well on the geometry and boundary conditions of the system~\cite{rosenbusch2002dynamics}. For example, in 3D there can be a competition between aligning the vortex core with the rotation axis in order to capitalise on energy reduction from rotation, and minimising the length of the vortex core in the superfluid so as to minimise the interparticle interaction energy. 

Another new phenomenon that emerges in 3D is the reconnection of vortex lines~\cite{schwarz1988three}; when two vortex lines are made to intersect in 3D, they will generically reconnect and move apart so as to remove the intersection point. Note that there are some special cases of metastable stationary states in 3D with intersecting vortex lines~\cite{meichle}. As we shall review later in Section~\ref{sec:4Dsuper}, a key difference between 3D and 4D superfluids is that, in the latter case there can be an intersection point between two vortex planes in a stationary state which is energetically stabilised by double rotation~\cite{mccanna2021}. However, as we shall go on to explore in Section~\ref{sec:NonOrtho} onwards, we can also find stationary states with curved vortex surfaces, in which the vortex core curves spatially in order to avoid the intersection point. Analogies between these surfaces and reconnections in 4D will be further explored in Ref.~\cite{mccannaequal}.

\section{Rotations in 4D}\label{sec:4DRot}

In order to further lay the groundwork for our discussion of 4D vortex structures in Section~\ref{sec:4Dsuper}, we shall now review the different types of rotations that are possible with four spatial dimensions, comparing these with 2D and 3D systems. We shall begin by introducing the concepts of simple, double and isoclinic rotations, before discussing the possible rotation planes of 4D isoclinic rotations in more mathematical detail. As we shall see, this will be relevant when considering the effects of rotation in a generalised 4D GPE in later parts of this paper.

\subsection{Simple, double, and isoclinic rotations}

In two dimensions, rotations are completely specified by their centre and rotation angle. The centre is the one fixed point of the rotation, while all other points are angularly displaced about the centre by the rotation angle. Represented as a matrix, any rotation of 2D space will be given as
\begin{align}
    \begin{pmatrix}
        \cos\alpha & -\sin\alpha \\
        \sin\alpha & \cos\alpha
    \end{pmatrix},
\end{align}
where \(\alpha\in (-\pi,\pi]\) is the angle of rotation and we are defining the origin as the centre of rotation, as we will throughout this paper.

Similarly, rotations in three dimensions are commonly described in terms of their axis and angle of rotation. The axis is both the line of points fixed by the rotation and the centre about which the rotation occurs. One can equally define rotations in 3D by their plane of rotation, which is orthogonal to the axis of rotation. All rotations in 3D are just 2D rotations of their plane of rotation, with the third direction left unchanged. This is obvious from the matrix representation  of a 3D rotation, which can always be brought into the following form
\begin{align}
    \begin{pmatrix}
        \cos\alpha & -\sin\alpha & 0 \\
        \sin\alpha & \cos\alpha & 0 \\
        0 & 0 & 1
    \end{pmatrix},
\end{align}
via a suitable choice of basis. The rotation plane is left invariant by the rotation but not pointwise invariant, unlike the axis. This means that points on the rotation plane remain on it after the rotation, but are rotated about the rotation axis.
% For example, a 3D rotation of angle \(\alpha\) in the \(x\en y\) plane is represented in the standard basis by the matrix
% \begin{align}
%     \begin{pmatrix}
%         \cos\alpha & -\sin\alpha & 0 \\
%         \sin\alpha & \cos\alpha & 0 \\
%         0 & 0 & 1
%     \end{pmatrix}
% \end{align}

Just as we can extend 2D rotations into a third direction to define 3D rotations, we may generate rotations of 4D space by extending 3D rotations into a fourth direction. In the 3D case this gave us every possible rotation, up to a change of basis. However, in 4D we can only generate a proper subset of rotations by extending our 3D definitions in this way. Members of this subset are commonly termed ``simple" rotations, since they reduce to the familiar three and two dimensional cases. Simple rotations have a single rotation plane just as in the 3D case, but are centred around a plane of fixed points as opposed to an axis. This fixed plane is completely orthogonal to the rotation plane, by which we mean that every vector in one plane is orthogonal to every vector in the other. In a matrix representation, any simple rotation of 4D space can take the following form in a suitable basis
\begin{align}
    \begin{pmatrix}
        \cos\alpha & -\sin\alpha & 0 & 0 \\
        \sin\alpha & \cos\alpha & 0 & 0 \\
        0 & 0 & 1 & 0 \\
        0 & 0 & 0 & 1
    \end{pmatrix}.
    \label{eq:4DSimple}
\end{align}
Note that in 4D, there are six Cartesian coordinate planes, meaning that the rotation group \(SO(4)\) of four-dimensional space has {\it six} generators, physically describing angular momentum. For this reason, the representation of these generators (and hence of angular momentum) as spatial vectors does not work in 4D, as it does in 3D. 

Moreover, generic elements of \(SO(4)\) are so-called ``double" rotations. These new types of rotations occur simultaneously through two completely orthogonal planes of rotation (e.g. the \(x\en y\) and \(z\en w\) planes), each with their own rotation angle. Represented as a matrix, any double rotation can be brought into the form
\begin{equation}
    M(\alpha,\beta) =
    \begin{pmatrix}
        \cos\alpha & -\sin\alpha & 0 & 0 \\
        \sin\alpha & \cos\alpha & 0 & 0 \\
        0 & 0 & \cos\beta & -\sin\beta \\
        0 & 0 & \sin\beta & \cos\beta
    \end{pmatrix},
    \label{eq:DoubleRot}
\end{equation}
by a suitable change of basis. This matrix form makes it clear that a double rotation can be thought of as two simultaneous simple rotations: in this case a rotation of angle \(\alpha \in (-\pi,\pi]\) in the \(x\en y\) plane, and one of angle \(\beta \in (-\pi,\pi]\) in the \((z\en w)\) plane. This means that any point on the \(x\en y\) or \(z\en w\) plane will remain on it but be rotated around the origin by an angle \(\alpha\) or \(\beta\), respectively. Points not on either rotation plane are rotated by an angle whose magnitude is strictly between \(|\alpha|\) and \(|\beta|\)~\cite{lounesto2001}, assuming that \(|\alpha|<|\beta|\). Consequently, the origin is the only fixed point, as long as neither rotation angle is zero. If either angle vanishes, we recover simple rotations as a special case of double rotations.

Besides simple rotations there is another very important special class of double rotations, called ``isoclinic" rotations, which will play an important role in the rest of this paper. These are the double rotations where both rotation angles are equal up to a sign, such as \(M(\alpha,\alpha)\) and \(M(\alpha,-\alpha)\). They come in two types known as right handed and left handed based on the relative senses of rotation in the two planes. For example, \(M(\alpha,\alpha)\) is a left isoclinic rotation of the \(x\en y\) and \(z\en w\) planes, while \(M(\alpha,-\alpha)\) is a right isoclinic rotation of these planes. All left isoclinic rotations commute with all right isoclinic ones, and any rotation of 4D space can be decomposed into a product of a left isoclinic rotation and a right isoclinic rotation \cite{lounesto2001}. However, this is not unique, as \(M=M_LM_R\) can also be written as \(M=(-M_L)(-M_R)\), where $M_L$ and $M_R$ denote left and right isoclinic rotations respectively. 

\subsection{Rotation Planes of an Isoclinic Rotation}
\label{sec:IsoPlanes}

In later sections of this paper, we will find it useful to take advantage of various mathematical properties of isoclinic rotations in our analysis of vortices in 4D superfluids. For that reason, we shall now discuss these special types of rotations in greater detail, focusing in particular on how to identify the rotation planes of left and right isoclinic rotations respectively. 

Recall that a general double rotation will rotate a vector through an angle with magnitude between \(|\alpha|\) and \(|\beta|\). However, for an isoclinic rotation \(\alpha=\pm\beta\), so every vector is displaced by the same given rotation angle, meaning that there is an infinite number of rotation planes. Each of these rotation planes can be described as the span of an arbitrary vector \(\mathbf{v}\) and its image under the rotation (i.e. either \(M_L\mathbf{v}\) or \(M_R\mathbf{v}\)), which means that every point in \(\mathbb{R}^4\) lies on one of these rotation planes~\cite{lounesto2001}. However, this does not imply that every possible 2D plane is a rotation plane (except for very special cases, as mentioned below), nor does it mean that these rotation planes are unique: any completely orthogonal pair of them can be used as a basis to define the particular isoclinic rotation. For example, from Eq.~\eqref{eq:DoubleRot} we can see that two of the rotation planes of $M(\alpha,\alpha)$ are given e.g. by the $x\en y$ and $z\en w$ rotation planes, although these are not the only rotation planes as we shall see below. This is in contrast to generic double rotations (i.e. $M (\alpha, \beta)$, with $\alpha \neq \beta \neq 0, \pi$), which have only  two unique rotation planes as discussed above.

Our aim is now to mathematically identify the rotation planes of a given left isoclinic rotation, which we shall denote as $M_L\equiv M (\alpha, \alpha)$, i.e. we chose our basis such that this particular rotation has the form given in Eq.~\eqref{eq:DoubleRot} with $\alpha\!=\!\beta$. As we shall see, an easy way to find the corresponding rotation planes is then to use the complex representation \(\mathbb{C}^2\) to represent \(\mathbb{R}^4\), such that the Cartesian position vector \((x,y,z,w)^T\) is represented by  \((x+iy,z+iw)^T\). 
% \begin{align}
%     r_1 &= \left(x^2+y^2\right)^{\frac{1}{2}}, \quad \ta_1 = \arctan(x,y), \\
%     r_2 &= \left(z^2+w^2\right)^{\frac{1}{2}}, \quad \ta_2 = \arctan(z,w),
% \end{align}
Note that the natural inner product in \(\mathbb{C}^2\), given in Cartesian coordinates by
\begin{align}
    \begin{pmatrix}
        x+iy \\
        z+iw
    \end{pmatrix}^\dagger
    \begin{pmatrix}
        x'+iy' \\
        z'+iw'
    \end{pmatrix}
    &= xx'+yy'+zz'+ww' \\
    + &i(xy'-yx' + zw'-wz'),
    \label{eq:}
    % \\ \begin{pmatrix}
    %     \cpolar{1} \\
    %     \cpolar{2}
    % \end{pmatrix}^\dagger
    % \begin{pmatrix}
    %     \cpolar[']{1} \\
    %     \cpolar[']{2}
    % \end{pmatrix}
    % &= r1r1'e^{i\left(\ta_1'-\ta_1\right)} + r_2r_2'e^{i\left(\ta_2'-\ta_2\right)}
\end{align}
contains the inner product in \(\mathbb{R}^4\) as its real part. This means that any unitary matrix acting on \(\mathbb{C}^2\) will be equivalent to some orthogonal matrix acting on \(\mathbb{R}^4\). However, the converse is not necessarily true as unitary matrices preserve both the real and imaginary parts of the complex inner product, while orthogonal transformations need only preserve the real inner product. Nevertheless, we can say that if an orthogonal transformation of \(\mathbb{R}^4\) (e.g. such as a 4D rotation) is represented by a matrix in the complex representation then that complex matrix is automatically unitary. To see this, note that the norm on \(\mathbb{C}^2\) agrees with the norm on \(\mathbb{R}^4\), that is
\begin{align}
    \begin{pmatrix}
        x+iy \\
        z+iw
    \end{pmatrix}^\dagger
    \begin{pmatrix}
        x+iy \\
        z+iw
    \end{pmatrix}
    =     \begin{pmatrix}
        x \\
        y \\
        z \\
        w
    \end{pmatrix}^T
    \begin{pmatrix}
        x \\
        y \\
        z \\
        w
    \end{pmatrix}.
\end{align}
As any orthogonal transformation will preserve the \(\mathbb{R}^4\) norm, so its \(\mathbb{C}^2\) representation will preserve the corresponding complex norm, meaning that if that representation is a complex matrix it must therefore be a unitary matrix.

Returning to the particular case of double rotations, we see that in the complex representation, Eq.~\eqref{eq:DoubleRot}  becomes:
\begin{equation}
    M(\alpha,\beta) =
    \begin{pmatrix}
        e^{i \alpha} & 0 \\
        0 &   e^{i \beta}  
    \end{pmatrix},\label{eq:c2rot}
\end{equation}
which is indeed unitary. It is also clear that the desired left isoclinic rotation \(M_L \) can simply be represented in \(\mathbb{C}^2\) as \(e^{i\alpha}\) times the identity. We will now use this to show how to construct and parametrise the rotation planes of this left isoclinic rotation using this complex representation, before also discussing the case of right isoclinic rotations. Note that, in the following, rather than Cartesian coordinates, we shall primarily use double polar coordinates  \((r_1,\ta_1,r_2,\ta_2)\), which are defined by \(\cpolar{1}\equiv x+iy\), and \(\cpolar{2}\equiv z+iw\), such that the complex position vector becomes \((\cpolar{1},\cpolar{2})^T\) in the complex representation. 

In general, a 2D plane in  \(\mathbb{R}^4\) can be defined as the set of solutions to a pair of simultaneous linear equations (e.g. $x=0$ together with $y=0$ defines the $z\en w$ plane passing through the origin). In contrast, in the complex representation we can define a plane using a single equation which is linear in \(\cpolar{1,2}\) and their complex conjugates. In other words, given four complex numbers \((a_1, a_2, b_1, b_2)\), the equation
\begin{equation}
    a_1\cpolar{1} + a_2\cpolar{2} + b_1\ccpolar{1}  + b_2\ccpolar{2} = 0
    \label{eq:planeEqnComplex}
\end{equation}
defines a plane passing through the origin, and any such plane can be defined (not uniquely) in this way. (For example, the above $z\en w$ plane can now be defined simply either as $\cpolar{1}=0$ or equivalently as $\ccpolar{1}=0$.) To get back to the real representation we then just take the real and imaginary parts of the complex equation. Note that we included the complex conjugates \(\ccpolar{1,2}\) in Eq.~\eqref{eq:planeEqnComplex} so that the complex equation can have the same number of parameters as the two real equations.  

As the rotation planes of \(M_L\) are invariant under \(M_L\), to find these rotation planes we must find the equations of the form~\eqref{eq:planeEqnComplex} that are also invariant in this way. The action of the left isoclinic rotation, \(M_L\), is given by \(\ta_{1,2}\to \ta_{1,2} + \alpha\) as introduced above, so that the image of Eq.~\eqref{eq:planeEqnComplex} under \(M_L\) is given by
\begin{align}
    &\begin{split}
        &e^{i\alpha}\left(a_1\cpolar{1} + a_2\cpolar{2}\right) \\ &\ \hphantom{e^{i\alpha}(a_1\cpolar{1}} + e^{-i\alpha}\left(b_1\ccpolar{1} + b_2\ccpolar{2}\right) = 0.
    \end{split}
    \label{eq:rotPlaneEqnComplex}
\end{align}
For this to reproduce Eq.~\eqref{eq:planeEqnComplex}, we require that, in general, either \(a_{1,2}=0\) or \(b_{1,2}=0\). Note that for the special angles of $\alpha=0, \pi$, we recover Eq.~\eqref{eq:planeEqnComplex} irrespective of the values of \((a_1, a_2, b_1, b_2)\), meaning that every single plane is a rotation plane of $M_L$ for these cases. However, these special cases are trivial as they physically correspond to, respectively, no rotation or to flipping the direction of all axes simultaneously. Focusing therefore on the general case, we identify two possibilities: either 
\begin{align}
    &a_1\cpolar{1} + a_2\cpolar{2} = 0,
    \label{eq:RotationPlane}
\end{align}
or 
\begin{align}
    &b_1\ccpolar{1} + b_2\ccpolar{2} = 0.
\end{align}
However, we can map the latter equation onto the former by taking the complex conjugate of both sides and identifying \(b_1^* = a_1\), and \(b_2^* = a_2\). Therefore, both cases are the same and so the rotation planes of \(M_L\) are given by the solutions to the equation \(a_1\cpolar{1}+a_2\cpolar{2}=0\), for arbitrary complex numbers \(a_{1,2}\).

In later sections of this paper, we will want to sometimes work in a coordinate system defined in relation to an arbitrary completely orthogonal pair of these rotation planes (which  we will denote as \(P_1\) and \(P_2\)), in the same way that the coordinates \((r_1,\ta_1,r_2,\ta_2)\) are defined in relation to the \(x\en y\) and \(z\en w\) planes. We shall therefore now go through how such a coordinate system can be defined. To begin, let \(P_j\) be given by the solutions to
\begin{align}
    a_{j1}\cpolar{1} + a_{j2}\cpolar{2} = 0,
    \label{eq:P12}
\end{align}
for \(j=1,2\), and let our coordinate system defined with respect to these planes be given by \((r_1',\ta_1',r_2',\ta_2')\). Note that the coefficients \(a_{jk}\) are not all independent: once the plane \(P_1\) is chosen, \(P_2\) is already fixed as the orthogonal complement of \(P_1\). For now we will not consider this constraint, but will effectively derive it later by comparing the primed and unprimed coordinate systems. 

We now proceed to define the primed coordinates. Recalling that the \(x\en y\) and \(z\en w\) planes are defined by \(\cpolar{2}=0\) and \(\cpolar{1}=0\) respectively, we see that in our new coordinate system the planes \(P_{1,2}\) should be given by \( r'_{1,2}e^{i \ta_{1,2}'}=0\) respectively. 
% \begin{align}
%     a_{j1}\cpolar{1} + a_{j2}\cpolar{2} = 0,
% \end{align}
Given Eq.~\eqref{eq:P12}, a simple way to achieve this is to define our coordinates as follows
\begin{align}
    \begin{pmatrix}
        \cpolar[']{1} \\
        \cpolar[']{2}
    \end{pmatrix}
    =
    \begin{pmatrix}
        a_{11} & a_{12} \\
        a_{21} & a_{22}
    \end{pmatrix}
    \begin{pmatrix}
        \cpolar{1} \\
        \cpolar{2}
    \end{pmatrix}.
    \label{eq:primed}
\end{align}
 Let us now determine the way in which the coefficients \(a_{jk}\) are constrained. This can be done by noting that the unprimed coordinates are an orthonormal system; for this to also be true of the primed coordinates, we must have that the total distance from the origin is preserved, that is 
\begin{align}
    r_1'^2+r_2'^2=r_1^2+r_2^2.    
\end{align}
This condition is equivalent to requiring that the coefficients \(a_{jk}\) furnish a unitary \(2\times 2\) (\(U(2)\)) matrix, such that \(a_{22} = e^{i\varphi}a_{11}^*\), \(a_{21} = -e^{i\varphi}a_{12}^*\), and \(|a_{11}|^2+|a_{12}|^2=1\). This can be satisfied with the following parametrisation
\begin{align}
    &\begin{pmatrix}
        a_{11} & a_{12} \\
        a_{21} & a_{22}
    \end{pmatrix} =
    \begin{pmatrix}
        e^{i\varphi_1}\cos\eta & -e^{-i\varphi_2}\sin\eta \\
        e^{i\varphi_2}\sin\eta & e^{-i\varphi_1}\cos\eta
    \end{pmatrix}
    e^{i\varphi_3}
    \label{eq:matrixParam}
\end{align}
with \(\varphi_{1,2,3}\in[0,2\pi)\), and \(\eta\in[0,\pi/2)\).
However, the \(U(1)\) factor \(e^{i\varphi_3}\) is redundant since this represents a left isoclinic rotation of the planes, just like the original rotation \(M_L\) but with a different angle. Since such a rotation leaves all the rotation planes invariant we may discard it, and we are left with the following
\begin{align}
    \begin{pmatrix}
        \cpolar[']{1} \\
        \cpolar[']{2}
    \end{pmatrix}
    =
    \begin{pmatrix}
        e^{i\varphi_1}\cos\eta & -e^{-i\varphi_2}\sin\eta \\
        e^{i\varphi_2}\sin\eta & e^{-i\varphi_1}\cos\eta
    \end{pmatrix}
    \begin{pmatrix}
        \cpolar{1} \\
        \cpolar{2}
    \end{pmatrix}. \
    \label{eq:SU(2)}
\end{align}
The above matrix is the general expression for a member of \(U(2)/U(1) = SU(2)\), i.e a special unitary \(2\times 2\) matrix. We can interpret this family of matrices as the group of right isoclinic rotations~\cite{Kim2016}. To see this note that Eq.~\eqref{eq:matrixParam} is an expression in the complex representation for rotations that commute with the given left isoclinic rotation, \(M_L\). (To see this, note that these are complex linear transformations and so they commute with \(i\), while \(M_L\) in the complex representation is simply multiplication by \(e^{i\alpha}\).) However, it is also well-known that all left isoclinic rotations commute with all right isoclinic rotations~\cite{lounesto2001}, whereas two isoclinic rotations of the same sense (or two generic double rotations) will only commute if they share the same rotation planes, as introduced above. In going from Eq.~\eqref{eq:matrixParam} to Eq.~\eqref{eq:SU(2)} we have factored out those left isoclinic rotations which commute with $M_L$, as they take the same form as $M_L$ in the chosen basis and so have the same rotation planes. The matrix in Eq.~\eqref{eq:SU(2)} therefore is a representation of the right isoclinic rotations, as these are the remaining rotation matrices that commute with $M_L$.  

We can also further simplify Eq.~\eqref{eq:SU(2)} by factoring out the subgroup of right isoclinic rotations which take the form \(M_R \equiv M(\alpha,-\alpha)\) in our chosen basis. Letting \(\varphi_2=\varphi_1+\varphi\), with \(\varphi\in[0,2\pi)\), we obtain
\begin{align}
    \begin{pmatrix}
        \cos\eta & -e^{-i\varphi}\sin\eta \\
        e^{i\varphi}\sin\eta & \cos\eta
    \end{pmatrix}
    \begin{pmatrix}
        e^{i\varphi_1} & 0 \\
        0 & e^{-i\varphi_1}
    \end{pmatrix},
    \label{eq:factorLeftIso}
\end{align}
where the second matrix can be recognised as $M_R$ [c.f. Eq.~\eqref{eq:c2rot}]. Since the diagonal factor $M_R$ just corresponds to initial rotations within the \(x\en y\), and \(z\en w\) planes, it is redundant in describing the coordinate transformation [Eq.~\eqref{eq:SU(2)}] from these planes to the arbitrary rotation planes, $P_j$, of \(M_L\). We can therefore discard it such that our final expression for the general transformation is
\begin{align}
    \begin{pmatrix}
        \cpolar[']{1} \\
        \cpolar[']{2}
    \end{pmatrix}
    =
    \begin{pmatrix}
        \cos\eta & -e^{-i\varphi}\sin\eta \\
        e^{i\varphi}\sin\eta & \cos\eta
    \end{pmatrix}
    \begin{pmatrix}
        \cpolar{1} \\
        \cpolar{2}
    \end{pmatrix},
    \label{eq:RotPlaneTransformation}
\end{align}
where \(\eta\in[0,\pi/2]\) and \(\varphi\in[0,2\pi)\). This is now the general form for a coordinate transformation from a fixed pair of rotation planes (e.g.  the \(x\en y\), and \(z\en w\) planes) to all other rotation planes of $M_L$, with all redundant parameters removed. 

Interestingly, it is clear from Eq.~\eqref{eq:RotPlaneTransformation} that \(\varphi\) is undefined when \(\eta=0\), because the off-diagonal elements vanish. Moreover, a careful analysis shows that this also occurs at the other endpoint, \(\eta=\pi/2\), as here the diagonal elements vanish and so we can eliminate \(\varphi\) as follows
\begin{align}
    \begin{pmatrix}
        0 & -e^{-i\varphi}\sin\eta \\
        e^{i\varphi}\sin\eta & 0
    \end{pmatrix} &=
    \begin{pmatrix}
        0 & -\sin\eta \\
        \sin\eta & 0
    \end{pmatrix}
    \begin{pmatrix}
        e^{i\varphi} & 0 \\
        0 & e^{-i\varphi} \nonumber
    \end{pmatrix}.
\end{align}
This means that when \(\eta=0\) or \(\pi/2\) every value of \(\varphi\in[0,2\pi)\) gives the same completely orthogonal pair of rotation planes. In other words, \(\varphi\) and \(\eta\) parameterise a 2-sphere (\(S^2\)), with the north and south pole given by \(\eta=0\) and \(\pi/2\), respectively.
Effectively, in going from Eq.~\eqref{eq:SU(2)} to Eq.~\eqref{eq:RotPlaneTransformation}, we have just taken the quotient \(SU(2)/U(1) = S^3/S^1 = S^2\), which is the celebrated Hopf fibration~\cite{Kim2016}. We can therefore say that the space of all rotation planes of any given left isoclinic rotation \(M_L\) is topologically equivalent to a 2-sphere \((S^2)\). 
For example, the orbits of the two points \(x=\pm1\), \(y=z=w=0\) under this set of transformations are
\begin{align}
    \begin{pmatrix}
        \cos\eta & -e^{-i\varphi}\sin\eta \\
        e^{i\varphi}\sin\eta & \cos\eta
    \end{pmatrix}
    \begin{pmatrix}
        \pm1 \\
        0
    \end{pmatrix}
    &= 
    \begin{pmatrix}
        \pm\cos\eta \\
        \pm e^{i\varphi}\sin\eta
    \end{pmatrix},
\end{align}
i.e. \(x=\pm\cos\eta\), \(y=0\), \(z=\pm\cos\varphi\sin\eta\), \(w=\pm\sin\varphi\sin\eta\), which are the north and south hemispheres of the 2-sphere given by \(x^2 + z^2 + w^2 = 1\), \(y=0\).

Now that we have finished this derivation, we will conclude this section with a few observations. Firstly, there is a much quicker way of deriving Eq.~\eqref{eq:matrixParam}, that the transformations from a fixed pair of rotation planes of \(M_L\) to all other rotation planes are represented in \(\mathbb{C}^2\) by the unitary group, based on the following argument. Let the rotation which takes one rotation plane of \(M_L\) into another be given by \(U\). The fact that \(M_L\) acts on all its rotation planes in the same way means that \(M_L\) should be invariant under a change of basis by the rotation \(U\). In other words, we have the equation \(U^{-1}M_LU = M_L\), which means that \(M_L\) and \(U\) commute. In the complex (\(\mathbb{C}^2\)) representation \(M_L\) is simply \(e^{i\alpha}\) multiplied by the \(2\times 2\) identity matrix [c.f. Eq~\eqref{eq:c2rot}], and so in this representation, \(U\) must commute with \(i\). The most general way for \(U\) to satisfy this is if \(U\) is simply a complex matrix. However, we know that \(U\) is also a rotation, and we derived earlier that if a 4D rotation (or more generally an orthogonal transformation) in the \(\mathbb{C}^2\) representation is given by a matrix, then that matrix is unitary.

Secondly, note that all of the arguments of this section can also be applied to a given right isoclinic rotation \({M_R}\!\equiv\! M(\alpha,-\alpha)\), provided we use a different complex representation of \(\mathbb{R}^4\) where the position vector is given by \((x+iy,z-iw)^T = (\cpolar{1},\ccpolar{2})^T\). Transformations between rotation planes of \(M_R\) --- which form the left isoclinic subgroup of \(SO(4)\) --- then take the same form as Eq.~\eqref{eq:SU(2)}, but with \(e^{i\ta_2}\) and \(e^{i\ta_2'}\) replaced by their complex conjugates. Then, the left isoclinic rotations of the form \(M_L\equiv M(\alpha,\alpha)\) can be factored out, just as rotations of the form of \(M_R\) could be factored out of the Eq~\eqref{eq:SU(2)}. Thus we can obtain an equivalent expression to Eq.~\eqref{eq:RotPlaneTransformation} for the general coordinate transformation from a fixed pair of rotation planes of \(M_R\) to all other rotation planes, with all redundant parameters removed. It is explicitly given by
\begin{align}
    \begin{pmatrix}
        \cpolar[']{1} \\
        \ccpolar[']{2}
    \end{pmatrix}
    =
    \begin{pmatrix}
        \cos\eta & -e^{-i\varphi}\sin\eta \\
        e^{i\varphi}\sin\eta & \cos\eta
    \end{pmatrix}
    \begin{pmatrix}
        \cpolar{1} \\
        \ccpolar{2}
    \end{pmatrix}.
    \label{eq:RotPlaneTransformationR}
\end{align}
Again this equation is identical to Eq.~\eqref{eq:RotPlaneTransformation} except that the second element of each of the position vectors is replaced by its complex conjugate.

Finally, consider a fluid undergoing constant rigid rotation associated with a left isoclinic rotation in time, i.e. taking \(M_L(t) \equiv M (\alpha(t), \alpha (t)) \) with $\alpha (t) = \omega t$, where $\omega$ is a constant frequency. Such a system can be described using any completely orthogonal pair of the rotation planes of \(M_L(t)\); to see this, we define one such pair by \(r_je^{i\ta_j}=0\) and another by \(r_j'e^{i\ta_j'}=0\), with \(j=1, 2\), where the primed and unprimed coordinates are related by Eq.~\eqref{eq:RotPlaneTransformation}. Then we take the \(\mathbb{R}^4\) gradient of Eq.~\eqref{eq:RotPlaneTransformation} as follows
\begin{align}
    &\begin{pmatrix}
        \nabla\left[\cpolar[']{1}\right] \vspace{0.5em} \\
        \nabla\left[\cpolar[']{2}\right]
    \end{pmatrix} =
    \begin{pmatrix}
        \cos\eta & -e^{-i\varphi}\sin\eta \\
        e^{i\varphi}\sin\eta & \cos\eta
    \end{pmatrix}
    \begin{pmatrix}
        \nabla\left[\cpolar{1}\right] \vspace{0.5em} \\
        \nabla\left[\cpolar{2}\right]
    \end{pmatrix},
\end{align}
and then take the complex inner product of this equation with Eq.~\eqref{eq:RotPlaneTransformation} to obtain
\begin{align}
    &\begin{pmatrix}
        \cpolar[']{1} \\
        \cpolar[']{2}
    \end{pmatrix}^\dagger
    \begin{pmatrix}
        \nabla\left[\cpolar[']{1}\right] \vspace{0.5em} \\
        \nabla\left[\cpolar[']{2}\right]
    \end{pmatrix} =
    \begin{pmatrix}
        \cpolar{1} \\
        \cpolar{2}
    \end{pmatrix}^\dagger
    \begin{pmatrix}
        \nabla\left[\cpolar{1}\right] \vspace{0.5em} \\
        \nabla\left[\cpolar{2}\right]
    \end{pmatrix}.
\end{align}
Expanding the inner product and evaluating the gradient we end up with a complex equation with a real part given by
\begin{align}
    r_1'\bfhat{r}_1' + r_2'\bfhat{r}_2' = r_1\bfhat{r}_1 + r_2\bfhat{r}_2
\end{align}
and an imaginary part given by
\begin{align}
    r_1'\bfhat{\ta}_1' + r_2'\bfhat{\ta}_2' = r_1\bfhat{\ta}_1 + r_2\bfhat{\ta}_2,
    \label{eq:RigidRot}
\end{align}
where the hats above symbols indicates the unit vectors in those directions. The first equation shows that the \(\mathbb{R}^4\) position vector takes the same form in both bases, as expected. The second equation is less trivial, and can be physically interpreted as equating two velocity fields; once Eq.~\eqref{eq:RigidRot} is multiplied by the frequency \(\omega\), the RHS is a velocity field describing rigid left isoclinic rotation through the unprimed planes, while the LHS describes the same thing through the primed planes. That these two are equal shows that either pair can be used to describe such a rigidly rotating fluid, and therefore the fluid exhibits symmetry with respect to all right isoclinic rotations. 

However, as mentioned above, a superfluid does not behave as a rigid body under rotation, but instead forms quantized vortices~\cite{pitaevskii2003}. Indeed, for all the 4D superfluid vortex states that we shall study in the remainder of this paper, the 4D velocity field is significantly different to that of rigid rotation (Eq.~\eqref{eq:RigidRot}). Instead, for these states, the \(SU(2)\) symmetry, associated with the set of equivalent rotation planes, is naturally broken. This leads to degeneracies between states that are oriented with respect to the different rotation planes of an isoclinic double rotation. We shall see this, first of all, in the next section where we review the case of orthogonal 4D vortex planes, which we previously studied in Ref.~\cite{mccanna2021}.

\section{Orthogonal Vortex Planes in 4D Superfluids}
\label{sec:4Dsuper}

So far, we have reviewed the well-known physics of vortices in 2D and 3D, and introduced the different types of rotation that become possible in 4D systems. We shall now combine these ideas in order to discuss some of the vortex structures that can emerge in a 4D superfluid under double rotation. In this section, we shall focus, in particular, on the case of orthogonal vortex planes, which we earlier described in Ref.~\cite{mccanna2021}. After briefly reviewing the main findings of this previous work, we shall proceed to re-derive the hydrodynamic energy of two completely orthogonal vortex planes within a hyperspherical system, and to introduce our numerical methods, illustrating these by presenting numerical results that complement those already published. The intention of this section is to establish a basis of comparison for when we extend our discussion to non-orthogonal 4D vortex planes in Section~\ref{sec:NonOrtho}. 

\subsection{Structure of 4D Orthogonal Vortex Planes}

As in Section~\ref{sec:vortexreview}, we want to consider a superfluid described by the GPE without external potentials, but now with atoms free to move in four spatial dimensions. In the absence of rotation, a generalised 4D GPE can be written in the same form as in lower dimensions [Eq.~\eqref{eq:GPE}], namely~\cite{mccanna2021}:
\begin{equation}
	-\frac{\hbar^2}{2m}\nabla^2\psi + g|\psi|^2\psi = \mu{\psi},
\end{equation}
except now with $\nabla^2$ corresponding to the 4D Laplacian. This serves as a minimal model in which to explore 4D vortex physics, and is a plausible mathematical description of low-temperature interacting bosons in a hypothetical universe with four spatial dimensions~\cite{Wodkiewicz,Stampfer,Trang,mccanna2021}. In the future, it will also be interesting to consider a more tailored model moving towards a realistic 4D experiment, based, for example, on adding one or more synthetic dimensions to an ultracold bosonic gas~\cite{Boada2012,Celi2014,Mancini2015,Stuhl2015,Gadway2015, Price_2015,An_2017,Price2017,salerno2019quantized, viebahn2019matter,barbiero2019bose, chalopin2020exploring,oliver2021bloch, bouhiron2022realization}. However, the form of such a model will depend strongly on the details of the particular experimental implementation chosen, and will likely include other effects, such as lattices, unusual interaction terms and asymmetries between real and synthetic dimensions. These more experimental models therefore go beyond our current work, but raise interesting opportunities for future research as discussed briefly in our conclusions in Section~\ref{sec:concl}. 

While the above 4D GPE is identical to that in lower dimensions, in the rotating frame [c.f. Eq.~\eqref{eq:GPE2DR}], we have to be more careful as the angular momentum operator can no longer be treated as a vector as discussed in Section~\ref{sec:4DRot}. Instead, in 4D, the angular momentum operator is a 4x4 antisymmetric tensor, with components $\hat{L
}_{\gamma \delta}$ that correspond to the angular momentum in the $\gamma\en \delta$ plane (with ${\gamma, \delta} \in \{x, y, z, w\}$). The general form of the rotating-frame GPE then takes the form
\begin{equation}
	\left[-\frac{\hbar^2}{2m}\nabla^2 + g|\psi|^2 - \sum_{ \gamma \delta} \Omega_{\gamma \delta}\hat{L
}_{\gamma \delta}  \right] \psi = \mu\psi, \label{eq:GPE4DR}
\end{equation}
where $\Omega_{\gamma \delta}$ is the rotation frequency associated with the $\gamma\en \delta$ plane, and the sum runs over the six different Cartesian planes in 4D. 

As can be seen from Eq.~\eqref{eq:GPE4DR}, the simplest situation is when there is only one plane with a non-zero rotation frequency, e.g. $\Omega_{xy} \equiv \Omega \neq 0$. This corresponds to the case of simple rotation, which can be understood as a usual three-dimensional rotation extended into 4D [c.f. Eq.~\eqref{eq:4DSimple}]. As we previously showed in Ref.~\cite{mccanna2021}, this sort of rotation can stabilise a single ``vortex plane", where the dimensionless order parameter can be described by:
\begin{equation}
    \psi = f_k(r_1)e^{ik\ta_1}, 
\end{equation}
where $(r_1,\ta_1)$ are plane polar coordinates in the plane of rotation (e.g. $x\en y$), and $f_k(r)$ is independent of the coordinates not involved in rotation (e.g. $z$, $w$) such that the radial function is the same as that found numerically from the 2D GPE [c.f. Section~\ref{sec:vortexreview}]. In this case, the vortex core is a single plane (defined e.g. by $x=0$, $y=0$), as was also verified numerically in Ref.~\cite{mccanna2021}. Physically, this can be understood as the natural extension of point vortices from 2D and line vortices from 3D into 4D, as the extra dimension plays no role. 

In contrast, double rotations are an intrinsically 4D (or higher) phenomenon and so can lead to much richer vortex physics, as will be our focus in the remainder of this article. Specifically, we will focus on the 4D GPE in a doubly rotating frame
\begin{equation}
	\left[-\frac{\hbar^2}{2m}\nabla^2 + g|\psi|^2 - \Omega_1 \hat{L}_1 - \Omega_2 \hat{L}_2\right] \psi = \mu\psi,
	\label{eq:GPER}
\end{equation}
where $\Omega_j$ and $\hat{L}_j$ are respectively the rotation frequency and angular momentum operator in plane $j$. For example, we could choose plane 1 as the $x\en y$ plane (i.e. $\Omega_1 \equiv \Omega_{x y}$, $\hat{L}_1 \equiv \hat{L}_{xy} = -i\hbar(x\partial_y - y\partial_x$)), and plane 2 as the $z\en w$ plane (i.e. $\Omega_2 \equiv \Omega_{z w}$, $\hat{L}_2 \equiv \hat{L}_{zw} = -i\hbar(z\partial_w - w\partial_z)$). Note that such a set-up is related to certain 4D quantum Hall models in which a nontrivial second Chern number is generated by applying magnetic fields in two completely orthogonal planes \cite{Price_2015,Ozawa2016,Lohse_2018,Zilberberg_2018,Mochol_Grzelak_2018}.

To proceed, for simplicity we will henceforward adopt double polar coordinates $(r_1,\ta_1,r_2,\ta_2),$ defined by
\begin{equation}
    (x,y,z,w) = (r_1\cos\ta_1,r_1\sin\ta_1,r_2\cos\ta_2,r_2\sin\ta_2), \nonumber
\end{equation}
such that $\hat{L}_j = -i\hbar\partial_{\ta_j}$. As $\hat{L}_1$ and $\hat{L}_2$ describe a double rotation, they commute with each other, meaning that we are able to look for simultaneous eigenstates of both angular momentum operators. As we showed in Ref.~\cite{mccanna2021}, for suitable equal-frequency ($\Omega_1 = \Omega_2$) rotations, a reasonable ansatz for the (dimensionless) ground-state is
\begin{equation}
    \psi = f_{k_1, k_2} (r_1, r_1)e^{i k_1 \ta_1 + ik_2 \ta_2},
    \label{eq:OrthogonalAnsatz}
\end{equation}
where $k_1$ and $k_2$ are the integer winding numbers in the respective rotation planes and $f_{k_1, k_2} (r_1, r_1)$ describes the 4D superfluid density profile, which is assumed to just be a function of the radii of the two planes. This ansatz describes a pair of completely orthogonal vortex planes which intersect at the origin, with the superfluid circulating simultaneously and independently in the two rotation planes with a velocity field given by
\begin{equation}
{\bf v} = {\bf v}_1 + {\bf v}_2 =  \frac{\hbar}{m} \left( \frac{k_1}{r_1}  \bfhat{\ta}_1 + \frac{k_2}{r_2}  \bfhat{\ta}_2 \right),   
\label{eq:velocity4D}
\end{equation}
corresponding to a superposition of 2D vortex-velocity fields in each rotation plane [c.f. Eq.~\eqref{eq:velocity}].
Such a vortex structure is therefore topologically characterised by the \(\mathbb{Z}\times \mathbb{Z}\) topological winding numbers~\cite{mccanna2021}.
Note that this ansatz preferentially picks out the \(x\en y\) and \(z\en w\) planes; however, equal frequency rotations are isoclinic and hence have an infinite number of rotation planes, as we discussed in Sec.~\ref{sec:IsoPlanes}. This means that suitable ansatzes could be defined with respect to any of these planes, and our choice is arbitrary.

The function $f_{k_1, k_2} (r_1, r_1)$ in Eq.~\eqref{eq:OrthogonalAnsatz} can be found numerically from solving the 4D GPE [Eq.~\eqref{eq:GPER}]. As we previously showed~\cite{mccanna2021}, this function appears to be close to a product ansatz $f_{k_1, k_2} (r_1, r_1) \approx f_{k_1} (r_1) f_{k_2} (r_2)$ where $ f_{k_i} (r_i)$ is the 2D density profile of a vortex with winding number $k_i$ in plane $i$; however, this separable approximation fails significantly near the intersection of the vortex planes near the origin due to the intrinsic nonlinearity of the GPE equation. Before presenting a numerical example of such a vortex structure, we shall first discuss the associated hydrodynamic energy and critical frequency, in an extension of the standard textbook discussion for 2D vortices that was presented in Section~\ref{sec:vortexreview}.

\subsection{Hydrodynamic Energy of Completely Orthogonal Vortex Planes}
\label{sec:OrthoHydro}

As it will be helpful in the following sections, we shall here derive the hydrodynamic energy for a pair of completely orthogonal vortex planes in a 4D superfluid. As discussed in Ref.~\cite{mccanna2021}, we have previously studied the energy of the orthogonal-vortex structure [Eq.~\eqref{eq:OrthogonalAnsatz}] in a ``duocylinder" geometry defined by hard-wall boundaries at \(r_1=R_1\) and \(r_2=R_2\), where $R_{j}$ are the radii in the $j=1,2$  planes. In this geometry, the energy could be approximated by a decomposition as
\begin{equation}
    E_{k_1,k_2}(R_1,R_2) = E_{k_1}(R_1) + E_{k_2}(R_2),
    \label{eq:EnergySum}
\end{equation}
where \(E_{k_j}(R_j)\) is the 2D energy [Eq.~\eqref{eq:VEnergy}] associated with having a vortex with winding number $k_j$ in a 2D disc of radius $R_j$ and hard-wall boundary conditions. 

In this paper, we shall focus on a 4D  hypersphere (or ``4D ball") geometry, which is defined by having hard-wall boundaries at $r_1^2 + r_2^2 = R^2$, where $R$ is the hyperspherical radius. This geometry is theoretically interesting in the following sections as it preserves the symmetry of isoclinic rotations [c.f. Section~\ref{sec:4DRot}] unlike the duocylinder geometry, which has boundary conditions that preferentially pick out two planes as being special.

Nevertheless, as we shall now show, we can also approximate the energy of completely orthogonal vortex planes in such a hypersphere as a decomposition into a sum of the energies of each individual plane, in an analogous manner to Eq.~\eqref{eq:EnergySum}. To see this, we consider the hydrodynamics of a simplified ``hollow core" vortex model, similar to that used in 2D as reviewed in Section~\ref{sec:vortexreview}. Specifically, we will consider a pair of completely orthogonal vortex planes, which intersect at the origin, and we will approximate the density profile as zero within one healing length of each vortex core and at the system boundary, and equal to a constant \(N/V\) everywhere else, where \(N\) is the particle number and \(V=\pi^2R^4/2\) is the volume of the 4D ball. Following Section~\ref{sec:vortexreview}, we can again neglect contributions to the hydrodynamic energy from density variations [c.f. Eq.~\eqref{eq:HollowKinetic}] and from interparticle interactions [c.f. Eq.~\eqref{eq:inter}], leaving only  
\begin{equation}
    E_{\text{h}} \approx  \frac{m}{2} \int \rho {\bf v}^2 \diff^4 r = \frac{m}{2} \int \rho ( {\bf v}_1 + {\bf v}_2 ) ^2 \diff^4 r ,
\end{equation}
where in the second equality, we have used that the velocity field is well-described by a sum of the individual velocities for each vortex plane [c.f. Eq.~\eqref{eq:velocity4D}]. Furthermore, as \(\mathbf{v}_j\) lies in plane \(j\), it follows that \(\mathbf{v}_1\cdot\mathbf{v}_2 = 0\), i.e. that the hydrodynamic vortex-vortex interaction term vanishes, leaving us with
\begin{align}
    E_{\text{h}} = \frac{N\hbar^2}{2Vm} \dint_\xi^Rr_1\diff{r_1} \cintm\diff{\ta_1} \smashoperator{ \dint_\xi^{\sqrt{R^2-r_1^2}} } r_2\diff{r_2} \cintm\diff{\ta_2} \left(\frac{k_1^2}{r_1^2} + \frac{k_2^2}{r_2^2}\right), \label{eq:eho}
\end{align}
where we have used that \(v^2 = (\hbar^2/m^2)(k_1^2/r_1^2 + k_2^2/r_2^2)\) from Eq.~\eqref{eq:velocity4D}.

The integration region is symmetric with respect to swapping \(r_1\) and \(r_2\), so each term in the integrand gives the same result, just with a different coefficient \(k_j^2\). This means we really only need to consider one of these terms, and with the order of integration we have above, it is easiest to compute the \(1/r_1^2\) term. We will also rescale \(r_j\to Rr_j\) and evaluate the \(\ta\) integrals to give
\begin{align}
    E_{\text{h}} &= \frac{4N\hbar^2}{mR^2} \left(k_1^2 + k_2^2\right) \dint_{\frac{\xi}{R}}^1\frac{\diff{r_1}}{r_1} 
    \smashoperator{ \dint_{\frac{\xi}{R}}^{\sqrt{1-r_1^2}} } r_2\diff{r_2} \\
    &= \frac{2N\hbar^2}{mR^2} \left(k_1^2 + k_2^2\right)\dint_{\frac{\xi}{R}}^1\frac{\diff{r_1}}{r_1}\left(1-r_1^2 - \frac{\xi^2}{R^2}\right) \\
    &= \frac{2N\hbar^2}{mR^2} \left(k_1^2 + k_2^2\right) \left(1-\frac{\xi^2}{R^2}\right)\left[\ln\left(\frac{R}{\xi}\right) - \frac{1}{2}\right],
\end{align}
which to leading order in \(\xi/R\) gives us
\begin{align}
    E_{\text{h}} = 2N\frac{\hbar^2}{mR^2} \left(k_1^2 + k_2^2\right)\ln\left(\frac{R}{\xi}\right),
    \label{eq:HollowOrthoHydro}
\end{align}
 This corresponds to a sum like that in Eq.~\eqref{eq:EnergySum}, but for the hydrodynamic energy in this simplified constant-density model in a 4D hypersphere. Note that this equation is very similar to the point vortex energy in 2D [Eq.~\eqref{eq:hydrodynamic2D}], except we have \(k_1^2+k_2^2\) instead of \(k^2\), and there is a geometric factor of \(2\) coming from the difference between the area of a disk and the 4D volume of a hypersphere. Again, just as in the 2D case, we can obtain a more accurate energy for these orthogonal intersecting vortices by using the dimensionless numerical density-profile function \(f_{k_1,k_2}(r_1,r_2)\) from our ansatz [Eq.~\eqref{eq:OrthogonalAnsatz}]. Using the grand canonical energy relative to the uniform state with a chemical potential given by \(\mu=\hbar^2/m\xi^2\), we obtain
\begin{align}
    E_h \approx 2\mu N\frac{\xi^2}{R^2} \left(k_1^2 + k_2^2\right)\ln\left(2.07\frac{R}{\xi}\right),
    \label{eq:OrthoHydro}
\end{align}
numerically via a fitting procedure to the form of Eq.~\eqref{eq:HollowOrthoHydro}, with the logarithmic prefactor as the fit parameter.

This result, without the initial factor of 2, was obtained the exact same way in our previous paper~\cite{mccanna2021} for a duocylinder geometry given by \(r_1 \leq R_1\), and \(r_2\leq R_2\). In that instance, however, the analytical calculation yielded precisely this form since the boundary conditions in the two planes were decoupled, and the energy integral therefore decomposed into a sum in the two planes. In the hyperspherical geomtery, there are less than leading order terms that we have ignored that do not have the same form. Therefore we believe that the result for a spherical geometry [Eq.~\eqref{eq:OrthoHydro}] is more approximate.
The energy reduction from equal-frequency double rotation (\(\Omega_1=\Omega_2=\Omega\)) can be calculated as 
\begin{align}
    \Delta E_\text{rot} = \Omega \left( \langle \hat{L}_1 \rangle + \langle \hat{L}_2 \rangle \right) = \Omega N  \hbar  (k_1 + k_2) , 
    \label{eq:orthrot}
\end{align}
where we have used that each vortex plane independently contributes angular momentum equal to $N \hbar k_j$ [c.f.~Eq.\eqref{eq:rot}]. 
This gives a critical frequency of
\begin{align}
    \Omega_c \approx 2\frac{\hbar}{mR^2}\ln\left(2.07\frac{R}{\xi}\right)
    \label{eq:4Dcrit}
\end{align}
to stabilise an orthogonal pair of \(k=1\) vortex planes. Naturally, we will present most of our results with frequencies in units of \(\Omega_c\). Note that Ref.~\cite{mccanna2021} worked in units of \(\Omega_c^{2D}\), the 2D critical frequency [Eq.~\eqref{eq:2Dcrit}] for a \(k=1\) point vortex in a disk of radius \(R\). To convert between these unit conventions, we may use the fact  \(\Omega_c = 2\Omega_c^{2D}\).

\begin{figure}[!]
    \centering
    \includegraphics[width=0.4\textwidth]{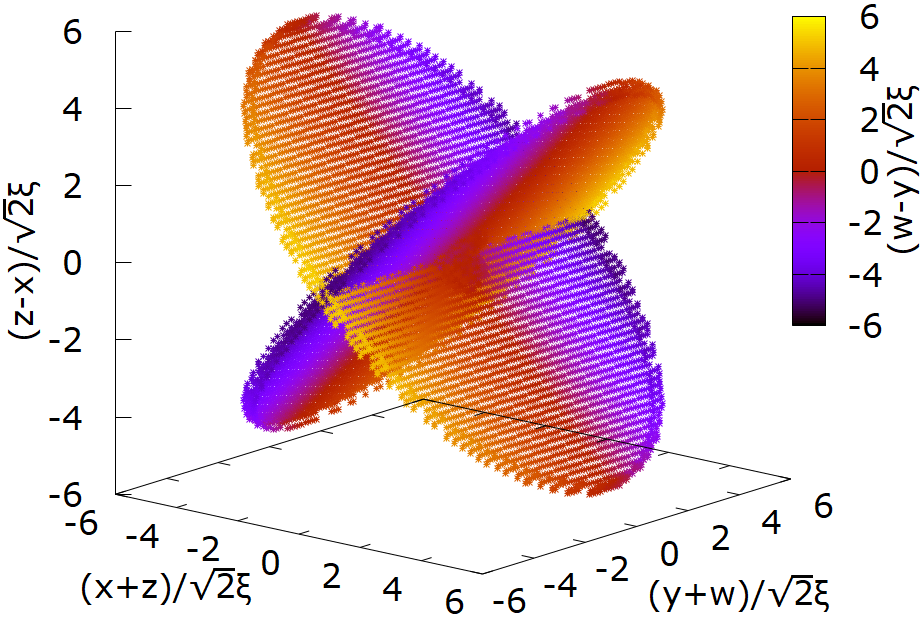}
    \caption{Vortex core of orthogonal intersecting planes in the final state of the ITEM with parameters \(\Omega_{xy} =\Omega_{zw} = 1.5\Omega_c\), and \(\Delta x = 0.2\xi\), giving \(R\approx 8.3\xi\). Note that the coordinates we are plotting against are rotated relative to those used in the numerics, in order to show both planes at once.}
    \label{fig:OrthoTilt}
\end{figure}

\subsection{Numerical Methods and Results}
\label{sec:Numerics}

In this section we will briefly describe the numerical methods which are then used to support our analytical results throughout the rest of this paper. We shall then illustrate these methods with an example of a structure with orthogonal vortex planes, so as to complement the results previously presented in Ref.~\cite{mccanna2021} and to provide a basis for comparison with later sections of this paper. 

As in our earlier work~\cite{mccanna2021}, the imaginary time evolution method (ITEM) is used to find solutions of the 4D GPE with double rotation [Eq.~\eqref{eq:GPE4DR}]. We use second order finite differences in space and a first order explicit discretisation in time. All calculations are performed on a Cartesian grid within a 4D hypersphere of radius roughly equal to \(N_{\text{grid}}\approx41\) gridpoints, with a hardwall boundary condition imposed on the boundary points (defined as the points with fewer than 8 nearest neighbours). This then corresponds to a total number of gridpoints roughly equal to \(1.4\times10^{7}\). The spatial step size for most calculations is set to \(\Delta x = 0.5\xi\), which ensures a large system of radius \(R\approx21\xi\) to reduce the importance of boundary effects.

We calculate the predicted critical frequency \(\Omega_c\) from Eq.~\eqref{eq:4Dcrit} with \(R\) set to \(N_{\text{grid}}\Delta x - \xi\), subtracting one healing length in order to approximately account for the boundary region. As we shall see later, our numerical results suggest that, for \(\Delta x=0.5\xi\), a more accurate value for the critical frequency of \(0.9\Omega_c\). This is likely due to a combination of finite size effects and the fact that Eq.~\eqref{eq:OrthoHydro} is an approximate result based on fitting to the function form of Eq.~\eqref{eq:HollowOrthoHydro}.

Our initial states are constructed in terms of a density profile and phase profile, with a degree of noise (up to $20\%$ of the background value) then added to the real and imaginary parts of \(\psi\). The initial density profile is chosen to be homogeneous except at the boundary where it smoothly goes to zero, while the initial phase factor is determined by the vortex configuration we expect to see at low energy for the chosen parameters. The ITEM is deemed to have converged once the relative variations in the particle number, \(N\) (calculated as the sum of \(\abs{\psi}^2\Delta x^4\)), and chemical potential (the sum of the LHS of Eq.~\eqref{eq:GPE4DR} multiplied by \(\psi^*\Delta x^4/N\)), from one iteration to the next reach below \(10^{-10}\). Once the ITEM reaches this threshold accuracy level we output the state and calculate the energy (using a finite difference version of Eqs~\eqref{eq:Kinetic} and \eqref{eq:Interaction}). We also determine the coordinates of all points making up the vortex core and separately output these, where we deem a point to be in the core if it is more than one healing length from the boundary and if the modulus of the order parameter at that location is less than the spatial resolution \(\Delta x/\xi\). This latter criterion is motivated by the fact that the order parameter goes to zero linearly as one approaches a singly charged vortex core~\cite{pitaevskii2003}. In order to then plot the vortex core, we supplement a 3D scatter plot (showing the \(x\), \(y\), and \(z\) coordinates) with colour (representing the \(w\) coordinate). 

To illustrate this numerical method, in Fig~\ref{fig:OrthoTilt} we show the vortex core structure obtained from ITEM under equal-frequency double rotation with parameters \(\Omega_{xy} =\Omega_{zw} = 1.5\Omega_c\), and \(\Delta x = 0.2\xi\), giving \(R\approx 8.3\xi\). Here, the initial state (before adding noise) was chosen to have a density profile that was homogeneous within the system away from the boundaries, and a phase profile given by $\arctan2(y,x) + \arctan2(w,z)$, corresponding to the phase-winding expected from Eq.~\eqref{eq:OrthogonalAnsatz} with $k_1=k_2=1$.
Note that both the Cartesian grid and the initial phase profile numerically break the hyperspherical symmetry, and the symmetry associated with isoclinic rotation. However, we expect the grid effects to be small, and the phase profile simply picks out one of several degenerate states for equal frequency rotation.
% , and our numerical results do contain vortex cores that are tilted and curved off of the grid axes so this appears to be the case.
As in Ref.~\cite{mccanna2021}, the resulting stationary state is found to contain a vortex core structure consisting of a pair of intersecting completely orthogonal planes (here, corresponding to the $x\en y$ and $z\en w$ planes). Note that in plotting Fig~\ref{fig:OrthoTilt}, we have rotated our coordinates according to
\begin{equation}
    \begin{pmatrix}
        x \\ y \\ z \\ w
    \end{pmatrix} \to
    \frac{1}{\sqrt{2}}
    \begin{pmatrix}
        1 & 0 & 1 & 0 \\
        0 & 1 & 0 & 1 \\
        -1 & 0 & 1 & 0 \\
        0 & -1 & 0 & 1
    \end{pmatrix}
    \begin{pmatrix}
        x \\ y \\ z \\ w
    \end{pmatrix}
    \label{eq:CrossedRotation}
\end{equation}
in order to better depict both planes at the same time. This visualization of the vortex-core structure complements the results previously presented in Ref.~\cite{mccanna2021} for the phase and density profile of the stationary state, and will serve as a useful basis for comparison to results obtained in later sections of this paper under other parameters and initial conditions.

\section{Non-orthogonal vortex planes}
\label{sec:NonOrtho}

Having introduced vortex structures composed of completely orthogonal vortex planes in the previous section, we shall now consider the possibility of a pair of non-orthogonal vortex planes in a 4D superfluid. As we shall show later in this paper, such non-orthogonal vortex planes are a natural candidate for the low energy configuration of a 4D superfluid doubly rotating at unequal frequencies. In preparation, we shall therefore derive in this section the total hydrodynamic energy of a pair of non-orthogonal vortex planes.

As in Section~\ref{sec:vortexreview} and Section~\ref{sec:4Dsuper}, we shall neglect the contributions from density variations and from interparticle interactions, such that the kinetic energy can be approximated as
\begin{equation}
    E_{\text{h}} \approx  \frac{m}{2} \int \rho {\bf v}^2 \diff^4 r = \frac{m}{2} \int \rho ( {\bf v}_1 + {\bf v}_2 )^2 \diff^4 r ,
\end{equation}
where we have again assumed that the velocity field can be decomposed as a sum of the velocity fields associated with each of the (now non-orthogonal) vortex planes separately. From this it can be seen that in general, we can split the total hydrodynamic energy up into a sum of the energies of each individual vortex plane [c.f. Section~\ref{sec:OrthoHydro}] together with the hydrodynamic vortex-vortex interaction, which is given by
\begin{equation}
   E_{\text{vv}} = m  \int \rho
    \mathbf{v}_1\cdot\mathbf{v}_2 \diff^4 r.
    \label{eq:Evv}
\end{equation}
As we discussed above and in Ref.~\cite{mccanna2021}, the velocity fields \(\mathbf{v}_{1,2}\) induced by two orthogonal planes were themselves everywhere orthogonal, \(\mathbf{v}_1\cdot\mathbf{v}_2 = 0\), meaning that this hydrodynamic vortex-vortex interaction between the planes  vanished. However, as we shall now show, this is not true for non-orthogonal vortex planes, meaning that the vortex-vortex interaction term is non-zero in general.  

In order to find $E_{\text{vv}}$, we will start from the assumption that a pair of non-orthogonal vortex planes can be described by the dimensionless ansatz
\begin{align}
    &\psi = r_1^{|k_1|}{r_2'}^{|k_2|}e^{i\left(k_1\ta_1+k_2\ta'_2\right)}g(r_1^2,r_2'^2),
    \nonumber \\
    &= \left(x+\sigma_1iy\right)^{|k_1|} \left(z'+ \sigma_2iw'\right)^{|k_2|} g(x^2+y^2,z'^2+w'^2), \qquad
    \label{eq:skewVortices}
\end{align}
where the primed and unprimed coordinates are related by a double rotation given by a matrix \(\tiltR\) defined below, such that \(\mathbf{r}' = \tiltR\mathbf{r}\), and \(k_{1,2}\) are the winding numbers of the two vortices while \(\sigma_j=\sgn(k_j)\). The function \(g\) ensures that the density returns to the homogeneous value when both \(r_1\) and \(r_2'\) are large compared to the healing length, and is given by \(g(r_1^2,r_2'^2)=\text{const} \times \ f_{k_1,k_2}(r_1,r_2')/r_1^{|k_1|}{r_2'}^{|k_2|}\), where \(f_{k1,k2}\) is the dimensionless profile associated with the ansatz for orthogonal vortex planes in Eq.~\eqref{eq:OrthogonalAnsatz}. In particular, \(g\) is always positive, and must satisfy the asymptotic relations
\begin{align}
    g(r_1^2,{r_2'}^2) \sim \text{const} \times\ \frac{f_{k_1}(r_1)}{r_1^{|k_1|}} \quad \text{as} \quad r_2'\to \infty, \\
    g(r_1^2,r_2'^2) \sim \text{const}  \times\ \frac{f_{k_2}(r_2')}{r_2'^{|k_2|}} \quad \text{as} \quad r_1\to \infty, 
\end{align}
where \(f_k\) is the dimensionless 2D vortex profile described in Sec~\ref{sec:2DVortex}. These asymptotics physically are the requirement that far from one of the vortex planes, the density profile is determined purely by the remaining one. 

Concretely, the state in Eq.~\eqref{eq:skewVortices} contains vortex planes along \(x=y=0\) and \(z'=w'=0\) respectively, which intersect at the origin. In order to keep this ansatz general while minimizing the number of parameters, we will refer to appendix \ref{app:TiltWLOG}, where we derive the general form of \(\tiltR\) required to describe the tilting of a plane in \(\mathbb{R}^4\). The result is that, without loss of generality, we may choose the following form
\begin{equation}
    \tiltR
    =
    \begin{pmatrix}
        \cos\alpha_1 & 0 & -\sin\alpha_1 & 0 \\
        0 & \cos\alpha_2 & 0 & -\sin\alpha_2 \\
        \sin\alpha_1 & 0 & \cos\alpha_1 & 0 \\
        0 & \sin\alpha_2 & 0 & \cos\alpha_2
    \end{pmatrix},
\end{equation}
with \(\alpha_{1,2} \in [0,\pi/2)\), such that
\begin{equation}
    z' = \sin\alpha_1x + \cos\alpha_1z, \qquad
    w' = \sin\alpha_2y + \cos\alpha_2w.
    \label{eq:primedZW}
\end{equation}
 To use this result in describing our skewed vortex planes we must assume that the vortices exist within a spherically symmetric 4D superfluid of radius \(R\), such that \(r_1^2+r_2^2=r_1'^2+r_2'^2\leq R\). Note that having a pair of orthogonal vortex planes, as discussed in the previous section, corresponds to taking \(\alpha_1=\alpha_2=0\), such that \(\tiltR\) becomes an identity matrix and ${\bf r}' = {\bf r}$. Given the spherical geometry we will assume, in analogy with Section~\ref{sec:4Dsuper}, that the velocity fields induced by each vortex have the following simple forms
\begin{equation}
    \mathbf{v}_1 = k_1 \vDimens \frac{\bfhat{\ta}_1}{r_1} \qquad \mathbf{v}_2' = k_2 \vDimens \frac{\bfhat{\ta}_2'}{r_2'}.
    \label{eq:4DVelocity}
\end{equation}
Let's first consider the special case where the matrix \(\tiltR\) is a simple rotation, meaning one of the angles \(\alpha_{1,2}\) is equal to zero. In this case it is easier to use the Cartesian representation of Eq.~\eqref{eq:4DVelocity}, which is
\begin{equation}
    \mathbf{v}_1 = k_1 \vDimens \frac{x\bfhat{y}-y\bfhat{x}}{x^2+y^2}
    \qquad \mathbf{v}_2' = k_2 \vDimens \frac{z'\bfhat{w}'-w'\bfhat{z}'}{z'^2+w'^2}.
    \label{eq:4DVeloCart}
\end{equation}
Without loss of generality we can choose \(\alpha_2=0\) such that we have \(w'=w\) and \(\bfhat{w}'=\bfhat{w}\), which is of course orthogonal to both \(\bfhat{x}\) and \(\bfhat{y}\). Therefore, the dot product between the velocity fields is given by
\begin{align}
    \mathbf{v}_1\cdot\mathbf{v}_2' = k_1k_2 \left(\vDimens\right)^2 \frac{w}{z'^2+w^2}\frac{y\bfhat{z}'\cdot\bfhat{x} - x\bfhat{z}'\cdot\bfhat{y} }{x^2+y^2}.
\end{align}
Using Eq.~\eqref{eq:primedZW} we have that \(\bfhat{z}' = \sin\alpha_1\bfhat{x}+\cos\alpha_1\bfhat{z}\), and therefore the interaction energy is given by 
\begin{align}
    E_{\text{vv}} = k_1k_2 \sin\alpha_1 \frac{\hbar^2}{m} \int \frac{wy \rho(x^2+y^2,z'^2+w^2)}{(z'^2+w^2)(x^2+y^2)} \diff^4 r.
\end{align}
Keeping in mind that only \(x\) and \(z\) appear in \(z'\), we can see that the above integrand is an odd function of both \(y\) and \(w\). This integral therefore vanishes, since our chosen geometry is symmetric with respect to both of these coordinates. In order to get a non-zero interaction potential we would need the superfluid to occupy a region that is asymmetric in both the \(y\) and \(w\) directions. This may be an interesting avenue for future work but is beyond the scope of this paper.

We shall now derive the form of the hydrodynamic interaction energy [Eq.~\eqref{eq:Evv}] in the special cases where \(\tiltR\) is an isoclinic rotation. There are two main reasons for this choice of rotation: firstly, an isoclinic tilt allows us to derive an analytic form for the interaction using the integral transform into non-orthogonal double polar coordinates derived in Appendix~\ref{app:IntSkew}; and secondly, in Sec~\ref{sec:Unequal} we will use the results we derive here to investigate possible low-energy vortex configurations in a superfluid doubly rotating at unequal frequencies, and we obtain predictions that agree closely with numerics when the frequencies are not too high. Isoclinic rotation here corresponds to the condition \(\alpha_2 = \nu\alpha_1\), with \(\nu = \pm1\) denoting whether \(\tiltR\) is left (\(-\)) or right (\(+\)) isoclinic. Let us therefore define \(\eta\equiv\alpha_1\) for simplicity and proceed. Using Eq.~\eqref{eq:primedZW}, we see that the primed coordinates now take the form
\begin{equation}
    z'+iw' = c(z+iw) + s(x+\nu iy),
    \label{eq:isoTilt}
\end{equation}
where we have applied the shorthand \(c=\cos\eta\), \(s=\sin\eta\). The equations of the plane $z'=w'=0$ then become
\begin{align}
    z = -x\tan\eta,  \\
    w = -y\nu\tan\eta.
\end{align}

This is tilted away from the plane $z=w=0$ by an angle $\eta$, so the angular separation of the planes $z'=w'=0$, and $x=y=0$ is $\pi/2 - \eta$.

Alternatively, we can derive the angle between the planes as follows. First, note that the plane \(z'=w'=0\) is spanned by the unit vectors \(\bfhat{x}'\) and \(\bfhat{y}'\), and that these are related to the unprimed basis vectors by
\begin{align}
\bfhat{x}' &= c\bfhat{x} - s\bfhat{z}, \\
\bfhat{y}' &= c\bfhat{y} - s\bfhat{w}.   
\end{align}
This lets us define arbitrary unit vectors in the \(x'\en y'\) plane and \(z\en w\) plane as follows
\begin{align}
    \bfhat{u}' &= \cos\phi' \bfhat{x}' + \sin\phi' \bfhat{y}' \\
    \bfhat{u} &= \cos\phi \bfhat{z} + \sin\phi \bfhat{w},
\end{align}
where \(\phi'\) and \(\phi\) are arbitrary angles between \(0\) and \(2\pi\). We can then find the angle between these vectors in the usual way using the dot product
\begin{align}
    \bfhat{u}' \cdot \bfhat{u} &= \left( \cos\phi' \bfhat{x}' + \sin\phi' \bfhat{y}' \right) \cdot \left(\cos\phi \bfhat{z} + \sin\phi \bfhat{w} \right) \nonumber\\
    &= \cos\phi' \cos\phi \bfhat{x}' \cdot \bfhat{z} + \sin\phi' \sin\phi \bfhat{y}' \cdot \bfhat{w} \nonumber \\
    &= -s \left( \cos\phi' \cos\phi + \sin\phi' \sin\phi \right) \nonumber\\
    &= -\sin\eta \cos\left( \phi' - \phi \right).
\end{align}
When \(\phi' - \phi = \pi/2\) we obtain the minimum angle between the two planes, which is given by \(\arccos(\sin\eta) = \pi/2 - \eta\). In the rest of the paper, we will sometimes refer to the angle \(\eta\) as the ``skewness" between two planes, since it measures how far from orthogonal those two planes are; when \(\eta = 0\), the two planes are orthogonal, and when \(\eta\) reaches its maximum value of \(\pi / 2\) the two planes coincide.

Continuing, we substitute Eq.~\eqref{eq:isoTilt} into Eq.~\eqref{eq:skewVortices} and find that our ansatz is given by
\begin{align}
    \psi = (x+\sigma_1iy)^{|k_1|}\left[c(z+\sigma_2iw) + s(x+\nu\sigma_2 iy)\right]^{|k_2|}g,
\end{align}
where we have suppressed the arguments of \(g\) for brevity. Note that if we have \(\nu=\sigma_1\sigma_2\), then \(x+\sigma_1iy = x+\nu\sigma_2iy\), and the planes are skewed in such a way that they are beginning to perfectly align, while \(\nu=-\sigma_1\sigma_2\)
corresponds to pure anti-aligning. We therefore expect that \(\nu=\sgn(k_1k_2)\) will give rise to a repulsive interaction between the planes, while \(\nu=-\sgn(k_1k_2)\) will lead to a attractive interaction [c.f. Section~\ref{sec:Multiple2D}]. 

In order to compute the hydrodynamic vortex-vortex interaction energy, we will first rewrite Eq.~\eqref{eq:isoTilt} in double polar coordinates as
\begin{equation}
    r_2'e^{i\ta'_2} = c\cpolar{2} + sr_1e^{i\nu\ta_1}.
    \label{eq:radIsoTilt}
\end{equation}
As the hydrodynamic interaction energy density depends on \(\rho\mathbf{v}_1\cdot\mathbf{v}_2'\) [c.f. Eq.~\eqref{eq:Evv}], we must find an expression for the dot product \(\bfhat{\ta}_1\cdot\bfhat{\ta}_2'\) under the assumption of velocity fields of the form in Eq.~\eqref{eq:4DVelocity}. To do this, we will start by taking the vector gradient of Eq.~\eqref{eq:radIsoTilt} as
\begin{equation}
    (\bfhat{r}_2'+i\bfhat{\ta}_2')e^{i\ta'_2} = c(\bfhat{r}_2+i\bfhat{\ta}_2)e^{i\ta_2} + s(\bfhat{r}_1+i\nu\bfhat{\ta}_1)e^{i\nu\ta_1}.
\end{equation}
where we have used the primed coordinate system on the LHS and the unprimed coordinate system on the RHS. Then taking the dot product of both sides with \(\bfhat{\ta}_1\) gives
\begin{equation}
    (\bfhat{r}_2'\cdot\bfhat{\ta}_1+i\bfhat{\ta}_2'\cdot\bfhat{\ta}_1)e^{i\ta'_2} = i\nu se^{i\nu\ta_1}.
\end{equation}
Dividing through by \(ie^{i\ta'_2}\) and then taking the real part of both sides gives
\begin{equation}
    \bfhat{\ta}_2'\cdot\bfhat{\ta}_1 = \nu s\cos(\ta'_2-\nu\ta_1),
\end{equation}
such that \(\mathbf{v}_1\cdot\mathbf{v}_2'= \nu k_1k_2(\hbar/m)^2 s\cos(\ta'_2-\nu\ta_1) / (r_1 r_2')\), and the vortex-vortex interaction is given by
\begin{equation}
    E_{\text{vv}} = \nu k_1k_2 \frac{\hbar^2}{m} s \dint_{B^4(R)}\rho(r_1,r_2')\frac{\cos(\ta'_2-\nu\ta_1)}{r_1r_2'}\diff^4 r,
\end{equation}
where \(B^d(R)\) denotes the $d$-dimensional ball of radius \(R\) centred at the origin, which is our chosen geometry.
\begin{figure} [t!]
    \centering
    \begin{tikzpicture}[line width=0.5pt]

    \pgfdeclarelayer{below}
    \pgfdeclarelayer{above}
    \pgfsetlayers{below,main,above}

    \colorlet{myPurple}{Red!50!Blue}
    \newlength{\axisLength};
    \setlength{\axisLength}{0.35\linewidth};
    \coordinate (O) at (0,0);
    \node (re) at (\axisLength,0) {Re};
    \node (im) at (0,\axisLength) {Im};
    \draw [-{Latex[length=2mm]}] (-0.25\axisLength,0) -- (re);
    \draw [-{Latex[length=2mm]}](0,-0.25\axisLength) -- (im);
    
    \pgfmathsetmacro\scale{1.2}
    \pgfmathsetmacro\cosPhi{0.55}
    \pgfmathsetmacro\sinPhi{sqrt{1-\cosPhi*\cosPhi}}

    \pgfmathsetmacro\rX{0.3}
    \pgfmathsetmacro\rY{0.4}
    \pgfmathsetmacro\redx{\rX*\scale*\sinPhi}
    \pgfmathsetmacro\redy{\rY*\scale*\sinPhi}

    \pgfmathsetmacro\bX{0.96}
    \pgfmathsetmacro\bY{0.28}
    \pgfmathsetmacro\rTwo{sqrt{1-\rX*\rX - \rY*\rY}}
    \pgfmathsetmacro\bluex{\bX*\scale*\cosPhi*\rTwo}
    \pgfmathsetmacro\bluey{\bY*\scale*\cosPhi*\rTwo}
    
    \coordinate (red) at (\redx\axisLength,\redy\axisLength);
    \begin{pgfonlayer}{above}
        \draw [red, thick, ->] (O) -- (red) node[above left = -0.2em] {\( sr_1e^{i\nu\theta_1}\)};
    \end{pgfonlayer}
    
    \coordinate (blue) at ($(red) + (\bluex\axisLength,\bluey\axisLength)$);
    \draw [blue, thick, ->] (red) -- (blue) node[above right = -0.4em and -0.1em] {\( cr_2e^{i\theta_2}\)};
    \node (c) at (red) [draw, dashed, blue, circle through=(blue)] {};
    
    \begin{pgfonlayer}{below}
        \draw[myPurple, thin, dashed, ->] (O) -- (intersection 1 of c and O--red) node[above right = -0.4em] {$r_+e^{i\nu\ta_1}$};
    \end{pgfonlayer}
    
    \draw[myPurple, thick] (O) -- (blue) node[midway, right] {\( r_2'e^{i\theta_2'}\)};

\end{tikzpicture}
    \begin{tikzpicture}[line width=0.5pt]

    \pgfdeclarelayer{below}
    \pgfdeclarelayer{above}
    \pgfsetlayers{below,main,above}

    \colorlet{myPurple}{Red!50!Blue}
    % \newlength{\axisLength};
    \setlength{\axisLength}{0.4\linewidth};
    \coordinate (O) at (0,0);
    \node (re) at (\axisLength,0) {Re};
    \node (im) at (0,\axisLength) {Im};
    \draw [-{Latex[length=2mm]}] (-0.1\axisLength,0) -- (re);
    \draw [-{Latex[length=2mm]}](0,-0.1\axisLength) -- (im);
    
    \pgfmathsetmacro\scale{1.8}
    \pgfmathsetmacro\cosPhi{0.75}
    \pgfmathsetmacro\sinPhi{sqrt{1-\cosPhi*\cosPhi}}
    \pgfmathsetmacro\rX{0.6}
    \pgfmathsetmacro\rY{0.6}
    \pgfmathsetmacro\redx{\rX*\scale*\sinPhi}
    \pgfmathsetmacro\redy{\rY*\scale*\sinPhi}

    \pgfmathsetmacro\bX{0.96}
    \pgfmathsetmacro\bY{0.28}
    \pgfmathsetmacro\rTwo{sqrt{1-\rX*\rX - \rY*\rY}}
    \pgfmathsetmacro\bluex{\bX*\scale*\cosPhi*\rTwo}
    \pgfmathsetmacro\bluey{\bY*\scale*\cosPhi*\rTwo}

    \coordinate (red) at (\redx\axisLength,\redy\axisLength);
    \coordinate (blue) at ($(red) + (\bluex\axisLength,\bluey\axisLength)$);

    \node (c) at (red) [draw, dashed, blue, circle through=(blue)] {};
    \draw [blue, thick, ->] (red) -- (blue) node[midway, below] {\( cr_2e^{i\theta_2}\)};
    
    \begin{pgfonlayer}{below}
        \draw[myPurple, dashed, thin, ->] (O) -- (intersection 1 of c and O--red) node[above right = -0.4em] {$ r_+e^{i\nu\ta_1}$};
    \end{pgfonlayer}

    \draw [red, thick, ->] (O) -- (red) node[left] {\( sr_1e^{i\nu\theta_1} \)};

    \begin{pgfonlayer}{above}
        \draw[myPurple, dashed, thick, ->] (O) -- (intersection 1 of c and red--O) node[right = 0.1em] {$ r_-e^{i\nu\ta_1}$};
    \end{pgfonlayer}

    % \draw[myPurple] (O) -- (blue) node[midway, right] {\( r_1'e^{i\theta_1'}\)};

    \draw[myPurple, dashed, -] (O) -- (tangent cs:node=c, point={(O)}, solution=1) node (thetam) {};
    \draw[myPurple, dashed, -] (O) -- (tangent cs:node=c, point={(O)}, solution=2) node (thetap) {};
    
    \coordinate (A) at ($(O)!.25!(thetap)$);
    \coordinate (B) at ($(O)!.25!(red)$);
    
    \node at ($(A)!.5!(B)$) [myPurple, above right = -0.3em and -0.6em] {$\theta_*$};
    \path[draw] 
      let
      \p1=( $ (O) - (A) $ )
      in
      pic[draw, myPurple, angle radius={veclen(\x1,\y1)}] {angle = B--O--A};
\end{tikzpicture}
    \caption{A geometric/algebraic description of the two terms in Eq.~\eqref{eq:EvvIntegrals}. (Left) The first term corresponds to the region where \(sr_1<cr_2\), meaning that the complex coordinate \(\cpolar[']{2}\) encircles the origin. This means that \(r_2'\) always reaches a minimum value of \(0\), and \(\ta_2'\) spans a full period, i.e. \(r_2'\in[0,r_+]\), and \(\ta_2'\in[-\pi,\pi)\). (Right) The second term corresponds to \(sr_1\leq cr_2\), which means that \(\cpolar[']{2}\) no longer winds around the origin, so \(r_2'\) has some minimum value \(r_-\) which can be positive, and \(\ta_2'\) no longer spans a full period. This in turn means that \(\ta_2'\) takes values in the interval \([\ta_1-\ta_*,\ta_1+\ta_*]\), where \(\ta_*\leq\pi/2\) with equality occuring when \(sr_1=cr_2\) and the blue dotted circle passes through the origin. The limits on \(r_1\) are derived in Appendix~\ref{app:IntSkew}, and come from combining the inequalities on \(sr_1\) and \(cr_2\) with the spherical geometry \(r_1^2+r_2^2\leq R^2\).}
    \label{fig:regions}
\end{figure}
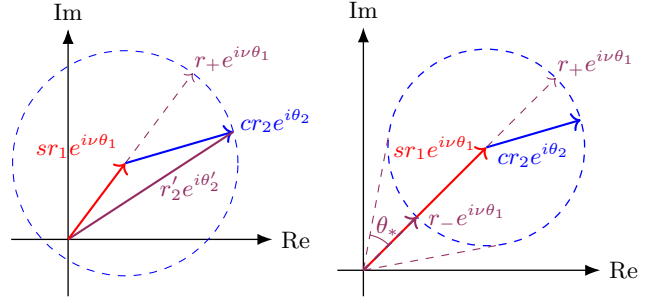

This integral will be as difficult to compute in the primed coordinate system as the unprimed one; however, we can greatly simplify the integrand by using the non orthogonal coordinate system defined by \((r_1,\ta_1,r_2',\ta'_2)\), at the cost of complicating the integration limits. In Appendix \ref{app:IntSkew} we derive the integral transformation into this non-orthogonal coordinate system for the region \(B^4(1)\). To use this result we must make the substitution \(\mathbf{r}\to R\mathbf{r}\) so that the integral is over \(B^4(1)\) rather than \(B^4(R)\), and we make the substitution \(\ta_2' = \intTheta + \nu\ta_1\) to simplify the cosine. Altogether, we then have
\begin{align}
    E_{\text{vv}} &= A \frac{s}{c^2} \dint_0^c \diff{r_1} \cintm\diff{\ta_1} \cintm\diff{\intTheta} \dint_0^{r_+}\diff{r_2'} \rho \cos\intTheta \\
    &+ A \frac{s}{c^2} \dint_c^1 \diff{r_1} \cintm\diff{\ta_1} \dint_{-\intTLim}^{\intTLim} \diff{\intTheta} \dint_{r_-}^{r_+}\diff{r_2'} \rho \cos\intTheta,
    \label{eq:EvvIntegrals}
\end{align}
% \begin{widetext}
%     \begin{align}
%         E_{\text{vv}} &= \nu k_1k_2 R^2 \frac{s}{c^2} \left( \dint_0^c \diff{r_1} \cintm\diff{\ta_1} \cintm\diff{\intTheta} \dint_0^{r_+}\diff{r_2'} \rho \cos\intTheta +
%         \dint_c^1 \diff{r_1} \cintm\diff{\ta_1} \dint_{-\intTLim}^{\intTLim} \diff{\intTheta} \dint_{r_-}^{r_+}\diff{r_2'} \rho \cos\intTheta \right),
%         \label{eq:EvvIntegrals}
%     \end{align}
% \end{widetext}
where the prefactor \(A=\nu k_1k_2 R^2\hbar^2/m\), and the limits \(\intTLim\) and \(r_\pm\) are given by
\begin{align}
    \intTLim &= \arcsin\left[c\left(1-r_1^2\right)^{ \frac{1}{2} }/sr_1\right], \ \text{and} \\
    r_\pm &= sr_1\cos\intTheta \pm \left[c^2\left(1 - r_1^2\right) - s^2r_1^2\sin^2\intTheta\right]^{ \frac{1}{2} }.
\end{align}
While this integral transform is derived in great detail in Appendix~\ref{app:IntSkew}, a quick pictoral explanation for the form taken by Eq.~\eqref{eq:EvvIntegrals} is given in Fig~\ref{fig:regions}. In particular, the fact that there are two distinct terms in this equation with different integration limits is directly related to whether \(\cpolar[']{2}\) encircles the origin (first term) or not (second term).

Now that we are using the natural coordinates for this problem, we can proceed to evaluate the integrals. Similar to Section~\ref{sec:2DVortex}, we will ignore the vortex core, and approximate the density as constant \(\rho = N/V\), where \(N\) is the particle number and \(V=\pi^2R^4/2\) is the 4D volume of the system. This approximation works provided the angle \(\eta\) is not too close to \(\pi/2\) as will be discussed later. The vortex-vortex interaction energy is then given as
\begin{align}
    E_{\text{vv}} &= A' \frac{s}{c^2}\dint_0^c \diff{r_1} \cintm\diff{\intTheta} \dint_0^{r_+}\diff{r_2'} \cos\intTheta \\
    &+ A' \frac{s}{c^2}\dint_c^1 \diff{r_1} \dint_{-\intTLim}^{\intTLim} \diff{\intTheta} \dint_{r_-}^{r_+}\diff{r_2'} \cos\intTheta,
\end{align}    
where the new prefactor \(A' = 4\nu k_1k_2N\hbar^2/\pi mR^2\). After further algebraic steps detailed in Appendix~\ref{app:IntSkew}, we then obtain the final result
\begin{align}
    E_{\text{vv}} =  -4k_1k_2\nu N\frac{\hbar^2}{mR^2} \ln\left(\cos\eta\right).
    \label{eq:Eint1}
\end{align}
We now see that our expectation regarding the sign of \(E_{\text{vv}}\) was correct: the overall sign is given by \(\sgn(\nu k_1k_2)\) such that the interaction is positive (i.e. repulsive) when the planes are skewed in an aligning sense (\(\nu=\sgn(k_1k_2)\)), and negative (i.e. attractive) when they are anti-aligning (\(\nu=-\sgn(k_1k_2\))).

Combining this with the result from Sec~\ref{sec:OrthoHydro}, we have that the total hydrodynamic energy of the non-orthogonal vortex plane state is
\begin{align}
    E_{\text{h}} =  2N\frac{\hbar^2}{mR^2} \left[ \left(k_1^2+k_2^2\right) \ln\left(\frac{R}{\xi}\right) - 2\nu k_1k_2 \ln\left(\cos\eta\right)\right]. \quad
    \label{eq:hydroTilt}
\end{align}
Note, however, that we just derived the interaction term under the constant density approximation, while the individual hydrodynamic energies of the vortices was calculated using a hollow core model. This hollow core was needed to remove the unphysical \(1/r^2\) singularity around each vortex that gives a divergent contribution to the energy, which in a mathematical sense is why vortices have cores. In contrast, the interaction energy density only goes as \(1/r_1r_2'\) which is not singular once integrated. 

Recall, as per the discussion in Sec~\ref{sec:Multiple2D}, that the same is true of point vortices in 2D: their interaction can be approximated with a constant density, but this fails to give a finite answer for the individual energies. As previously discussed, the correct answer can still be obtained if we take the vortices to be combined once their separation is of the order of \(\xi\) or below.

The question arises whether we can recover the expression for vortex combination in this 4D case. Here we have an angle between the vortex planes, given by \(\pi/2-\eta\), and we see that the interaction energy \(E_{\text{vv}}\) diverges in the limit \(\eta\to\pi/2\) under the constant density approximation. This also occurred when using this approximation for point vortices in 2D as their separation distance \(\Delta r\) approached zero, so we see that the angular separation \(\pi/2-\eta\) is playing a similar role here in 4D as \(\Delta r\) did in 2D. In contrast to 2D, however, we do not have a unique value for the separation distance between the vortices, so coming up with a criterion for when they have combined seems difficult. 

We will identify a natural separation as follows: each plane makes a circle of intersection with the boundary of the hypersphere, and we argue that the maximum reasonable value for the distance between the planes should be given by the minimum distance between these circles.  This distance is the length of the most direct straight line between the two planes at the boundary, and --- as we derive in Appendix~\ref{app:circles} --- is given by \(\sqrt{2}R\left(1-\sin\eta\right)^{1/2}\). This result can be obtained by naively applying the cosine rule in an analogy to lines in 2D. Setting this distance less than or equal to \(\xi\), we have
\begin{align}
    &\sqrt{2}R\left(1-\sin\eta\right)^{\frac{1}{2}} \leq \xi,
\end{align}
which rearranges to
\begin{align}
    &\sin\eta \geq 1 - \frac{\xi^2}{2R^2}.
\end{align}
We want this inequality in terms of \(\cos\eta\), since this is what appears in the interaction. Therefore we square both sides, which is safe as they are each non-negative, giving
\begin{align}
    &1-\cos^2\eta \geq \left( 1- \frac{\xi^2}{2R^2} \right)^2,
\end{align}
which rearranges to
\begin{align}
    &\cos\eta \leq \frac{\xi}{R}\left( 1 - \frac{\xi^2}{4R^2} \right)^{\frac{1}{2}}.
\end{align}
To leading order in \(\xi/R\), we can therefore say that the vortices are combined once \(\cos\eta = \xi/R\) or less. Substituting this into Eq.~\eqref{eq:hydroTilt} (using that \(\nu^2=1\)) we obtain
\begin{align}
    E_{\text{h}} =  \frac{2N\hbar^2}{mR^2}\left(k_1 + \nu k_2\right)^2 \ln\left(\frac{R}{\xi}\right),
\end{align}
which is the correct result for a vortex plane with winding number \(k_1 +\nu k_2\).This concludes our discussion of the energetic properties of non-orthogonal vortex planes, the results of which we will shall now use to construct a model of superfluids doubly rotating at unequal frequencies.

\section{Unequal frequency double-rotation}
\label{sec:Unequal}

In this section, we will consider the behaviour of a 4D superfluid undergoing constant double rotation with unequal frequencies, given by \(\Omega_{xy} = \Omega_{zw} + \Delta\Omega\) in the lab (\(x,y,z,w\)) frame. We will take \(\Delta\Omega >0\) without loss of generality, and we will also assume the superfluid occupies a hyperspherical (4D ball) region for simplicity. Note that this breaks the isoclinic SU(2) symmetry [c.f. Sec.\ref{sec:4DRot}], which is associated with the manifold of different rotation planes when the frequencies are equal. We will begin in Section~\ref{sec:Unequal:OnePlane} by considering what happens to a single vortex plane in this set-up, before discussing the case of two vortex planes in Section~\ref{sec:Unequal:TwoPlanes}. We shall then present our numerical results in Section~\ref{sec:Unequal:TwoPlanes:Results}. 

\subsection{A single vortex plane under unequal frequency double rotation}
\label{sec:Unequal:OnePlane}

In this section, we shall develop our intuition by considering the simple case of a single vortex plane in a system with unequal frequency double rotation. We will assume that this plane remains rigid but allow it to arbitrarily tilt in order to optimise its energy. The energy of the superfluid in the rotating frame is reduced by the amount
\begin{align}
    E_{\text{rot}}  = \Omega_{xy} \expected{\hat{L}_{xy}} + \Omega_{zw} \expected{\hat{L}_{zw}}
    \label{eq:Erot}
\end{align}
relative to an inertial frame. Since \(\Omega_{xy} > \Omega_{zw}\) it is natural to presume that the lowest energy occurs when the vortex plane lies along the \(z\en w\) plane --- thereby inducing rotation in the \(x\en y\) plane --- such that \(\expected{\hat{L}_{xy}} = N\hbar\) and \(\expected{\hat{L}_{zw}} = 0\). The converse case of a vortex plane spanning the \(x\en y\) plane would certainly be higher in energy, as this state would have \(\expected{\hat{L}_{xy}} = 0\) and \(\expected{\hat{L}_{zw}} = N\hbar\); however, it is not obvious that the superfluid energy decreases monotonically as the vortex plane is tilted from the \(x\en y\) plane to the \(z\en w\) plane. That is, the lowest energy overall could occur when the vortex plane is oriented somewhere between these two limits. Such a state would have positive values of both \(\expected{\hat{L}_{xy}}\) and \(\expected{\hat{L}_{zw}}\), and would be given as follows
\begin{align}
    \psi &= (x'+iy')g(x'^2+y'^2),
\end{align}
where the primed coordinates are to be defined shortly, and the function \(g\) is given by \(g(r^2) = const \times f_1(r/\xi)/r\), with \(f_1\) the dimensionless density profile of a point vortex in 2D (see Sec~\ref{sec:2DVortex}). The primed coordinates are defined such that the vortex plane is given by \(x'=y'=0\), so --- as derived in appendix~\ref{app:TiltWLOG} --- we may assume without loss of generality that the primed coordinates are given by
\begin{align}
    \begin{pmatrix}
        x' \\ y' \\ z' \\ w'
    \end{pmatrix}
    &= 
    \begin{pmatrix}
        \cos\alpha_1 & 0 & \sin\alpha_1 & 0 \\
        0 & \cos\alpha_2 & 0 & \sin\alpha_2 \\
        -\sin\alpha_1 & 0 & \cos\alpha_1 & 0 \\
        0 & -\sin\alpha_2 & 0 & \cos\alpha_2
    \end{pmatrix}
    \begin{pmatrix}
        x \\ y \\ z \\ w
    \end{pmatrix}.
\end{align}
A more useful form of the order parameter for this state is given in polar coordinates, \(\cpolar[']{1} \equiv x' + i y'\), as follows
\begin{align}
    \psi = \cpolar[']{1}g(r_1'^2).
\end{align}
In order to use this, we must express the transformation into the primed coordinate system in polar coordinates as well. Taking the combination \(x'+iy'\) gives
\begin{align}
    \cpolar[']{1} &= \cos\alpha_1x + \sin\alpha_1z + i\cos\alpha_2y + i\sin\alpha_2w,
\end{align}
and switching to polar coordinates and rearranging gives 
% substituting for \(x,y,z,w\) in terms of \(\cpolar{1,2}\) gives
\begin{align}
    % &= \cos\alpha_1r_1\frac{e^{i\ta_1}+e^{-i\ta_1}}{2} + \sin\alpha_1r_2\frac{e^{i\ta_2}+e^{-i\ta_2}}{2} \\
    % &\hphantom{=} + \cos\alpha_2r_1\frac{e^{i\ta_1}-e^{-i\ta_1}}{2} + \sin\alpha_2r_2\frac{e^{i\ta_2}-e^{-i\ta_2}}{2} \\ 
    \cpolar[']{1} &= \frac{\cos\alpha_1+\cos\alpha_2}{2}\cpolar{1} + \frac{\sin\alpha_1+\sin\alpha_2}{2}\cpolar{2} \\
    &\hphantom{=} + \frac{\cos\alpha_1-\cos\alpha_2}{2}\ccpolar{1} + \frac{\sin\alpha_1-\sin\alpha_2}{2}\ccpolar{2}.
\end{align}
The terms on the second line are proportional to \(e^{-i\ta_j}\), and therefore generate angular momentum counter to the external rotation (since \(\Omega_{xy}L_{xy}=-i\hbar\Omega_{xy}\partial_{\ta_1}\) and similarly for \(zw\)). To maximize the energy  reduction from rotation we should eliminate these terms. We therefore set \(\alpha_1=\alpha_2\equiv\eta\) to obtain the following
\begin{align}
    \cpolar[']{1} &= \cos\eta \, \cpolar{1} + \sin\eta \, \cpolar{2}.
\end{align}

An expression of this same exact form [Eq.~\eqref{eq:RotationPlane}] was derived in Sec~\ref{sec:IsoPlanes} in the context of finding the rotation planes of any left isoclinic rotation in the \(x\en y\) and \(z \en w\) planes, i.e. a rotation \(M_L\) generated by \(\hat{L}_{+} = \hat{L}_{xy} + \hat{L}_{zw}\). Here, however, \(\cpolar[']{1} = 0\) encodes the vortex plane, so we see that the vortex plane always lies in a rotation plane of \(M_L\), regardless of the value of \(\eta\). This vortex generates angular momentum in the plane orthogonal to itself, which is also a rotation plane of \(M_L\). This is all essentially summed up by the (easily verifiable) fact that \(\cpolar[']{1}\) is an eigenfunction of the sum of the angular momenta, \(\hat{L}_{+}\), despite not being an eigenfunction of either component. The superfluid containing this tilted vortex therefore has the same value of \(\expected{\hat{L}_{+}} = N\hbar\) for every value of \(\eta\). We can exploit this knowledge by rewriting the rotational energy [Eq.~\eqref{eq:Erot}] as 
\begin{align}
    E_{\text{rot}} = \Omega_{zw}\expected{\hat{L}_+} + \Delta\Omega\expected{\hat{L}_{xy}}.
    \label{eq:Erot2}
\end{align}

Since the first term is constant with respect to \(\eta\) we therefore maximize \(E_{\text{rot}}\) by simply maximizing the second term, proportional to \(\expected{\hat{L}_{xy}}\). This clearly occurs at \(\eta=0\), where \(\cpolar[']{1} = \cpolar{1}\), and so the initial intuition was correct: a single perfectly rigid vortex plane will always want to fully align with the higher frequency.

\subsection{Two vortex planes under unequal frequency double rotation}
\label{sec:Unequal:TwoPlanes}

Armed with this knowledge we now seek to find the optimal configuration of two rigid vortex planes in a doubly rotating superfluid with unequal rotation frequencies. Each vortex will want to align its angular momentum with the \(x\en y\) plane as much as possible to gain from the larger rotation frequency in this plane. However, as we showed in Sec~\ref{sec:NonOrtho}, vortex planes will interact with each other hydrodynamically once they are not orthogonal. This interaction will limit how close together (in orientation) each vortex can be and the competition between this effect and the rotational energy will determine the optimal orientation of each plane respectively. This is a simplified model of the situation, and it is worth briefly discussing the approximations we are making. 

We are going to again assume a constant density profile given by \(\rho=N/V\), thereby ignoring the vortex core. As previously discussed in Section~\ref{sec:2DVortex}, one needs to account for the core to avoid a divergent hydrodynamic energy cost of a vortex --- see, for example, Eq.~\eqref{eq:HollowKinetic} with \(\xi\) taken to zero. However, in this case we are only interested in how the energy varies with the orientation of each vortex plane. This means we can ignore any terms which do not vary as the planes tilt. If we denote the velocity field induced by each plane by \(\mathbf{v}_j\), with \(j=1,2\) respectively, then the hydrodynamic energy can be expanded as in Section~\ref{sec:NonOrtho} as
\begin{align}
    \frac{1}{2}\int\rho {\bf v}^2 \diff^4 r = \frac{1}{2}\int\rho \left({\bf v}_1^2 + {\bf v}_2^2\right) \diff^4 r + \int\rho \mathbf{v}_1\cdot\mathbf{v}_2 \diff^4 r.
\end{align}
The first term is the individual hydrodynamic cost of each vortex, which diverges if we ignore the core by assuming a constant density. However, this term does not vary with orientation due to the spherical symmetry of the boundary. On the other hand, the second term, which is the hydrodynamic interaction between the planes, depends on their relative orientation but does not diverge in a constant density approximation, as explained in Sec~\ref{sec:NonOrtho}.

We therefore ignore the constant first term, keeping only the second term which can safely be approximated using a constant density. This constant density approximation also allows us again to ignore the energy contributions from quantum pressure (the first term in Eq.~\eqref{eq:Kinetic}), and the bosonic interaction (Eq.~\eqref{eq:Interaction}) as is also done in Section~\ref{sec:NonOrtho}.

As we are assuming a constant density \(n=N/V\), we may take the order parameter for this configuration to be
\begin{align}
    \psi = n^{\frac{1}{2}}e^{i(\act_1+\grt_2)},
    \label{eq:Unequal:skewAnsatz}
\end{align}
where the acute (\(\acute{r}\)) and grave (\(\grave{r}\)) coordinate systems are to be defined shortly.  Note that this assumes the vortex planes remain flat and intersecting at the origin. In our numerical results, we do find some curvature and an avoided crossing near the origin (see Section~\ref{sec:Unequal:TwoPlanes:Results}), but these seem to have only a small effect on the energies. We will investigate the phenomena of curved vortex cores with avoided crossings in more in~\cite{mccannaequal}. 

Recall from Sec~\ref{sec:Unequal:OnePlane} that the vortex planes will want to stay on one of the rotation planes of left isoclinic rotations, \(M_L\), generated by \(\hat{L}_+\). We will therefore use our result [Eq.~\eqref{eq:RotPlaneTransformation}] from Sec~\ref{sec:IsoPlanes} for the general form of such rotation planes relative to a fixed basis (in this case the lab basis, \(x,y,z,w\)). Using this result for each of the acute and grave coordinates, we have
\begin{align}
    \begin{pmatrix}
        \acr_1e^{i\act_1} \\
        \acr_2e^{i\act_2}
    \end{pmatrix}
    &= 
    \begin{pmatrix}
        \cos\eta_1 & e^{i\varphi_1}\sin\eta_1 \\
        -e^{-i\varphi_1}\sin\eta_1 & \cos\eta_1
    \end{pmatrix}
    \begin{pmatrix}
        \cpolar{1} \\
        \cpolar{2}
    \end{pmatrix}, \quad
    \label{eq:acute}
    \\
    \begin{pmatrix}
        \grr_1e^{i\grt_1} \\
        \grr_2e^{i\grt_2}
    \end{pmatrix}
    &= 
    \begin{pmatrix}
        \cos\eta_2 & e^{i\varphi_2}\sin\eta_2 \\
        -e^{-i\varphi_2}\sin\eta_2 & \cos\eta_2
    \end{pmatrix}
    \begin{pmatrix}
        \cpolar{1} \\
        \cpolar{2}
    \end{pmatrix}. \quad
    \label{eq:grave}
\end{align}
where \(\eta_{1,2}\in[0,\pi/2]\) and \(\varphi_{1,2}\in[0,2\pi)\), with \(\varphi_j\) undefined when \(\eta_j=0\) or \(\pi/2\). The location of each vortex plane is then given by \(\acr_1 e^{i\act_1}=0\) and \(\grr_2 e^{i\grt_2}=0\), respectively. The parameters \(\eta_{1,2}\) denote the angle that each plane makes with the \(x\en y\) (resp. \(z\en w\)) plane, while \(\varphi_{1,2}\) denote the direction of this tilt. 

 The vortex at \(\acr_1 e^{i\act_1}=0\) is tilted by an angle \(\eta_1\) away from the \(x\en y\) plane, while the vortex at \(\grr_2 e^{i\grt_2}=0\) is tilted by \(\pi/2-\eta_2\) off of the same plane. Since the two vortex planes are indistinguishable, we can define the acute and grave coordinates such that the former vortex is closer to the \(x\en y\) plane than the latter, which translates to the following constraint on the angles
\begin{align}
    \eta_1 \leq \frac{\pi}{2} - \eta_2.
    \label{eq:EtaRestriction}
\end{align}
Note that \(\eta_1=\eta_2=0\) corresponds to the configuration that we have previously studied \cite{mccanna2021} --- a completely orthogonal pair of vortex planes spanning the rotation planes of the superfluid: the \(x\en y\) and \(z\en w\) planes, respectively. A change of basis in either of these planes redefines the \(\varphi_j\) variables as follows
\begin{align}
    &\begin{pmatrix}
        e^{-i\alpha} & 0 \\
        0 & e^{-i\beta}
    \end{pmatrix}
    \begin{pmatrix}
        \cos\eta_j & e^{i\varphi_j}\sin\eta_j \\
        -e^{-i\varphi_j}\sin\eta_j & \cos\eta_j
   \end{pmatrix}
    \begin{pmatrix}
        e^{i\alpha} & 0 \\
        0 & e^{i\beta}
    \end{pmatrix} \nonumber \\
    &\qquad= 
    \begin{pmatrix}
        \cos\eta_j & e^{i(\varphi_j-\alpha+\beta)}\sin\eta_j \\
        -e^{-i(\varphi_j-\alpha+\beta)}\sin\eta_j & \cos\eta_j
    \end{pmatrix}
\end{align}
We will choose a basis in which \(-\varphi_2=\varphi_1\equiv\varphi\), leaving us with three free parameters, (\(\eta_1,\eta_2,\varphi\)), describing the orientation of the two vortex planes relative to the lab frame. In terms of these parameters the two planes are defined by the zeroes of the following complex coordinates
\begin{align}
    \acr_1e^{i\act_1} &= \cos\eta_1\cpolar{1} + e^{i\varphi}\sin\eta_1\cpolar{2},
    \label{eq:acute1} \\
    \grr_2e^{i\grt_2} &= \cos\eta_2\cpolar{2} - e^{i\varphi}\sin\eta_2\cpolar{1},
    \label{eq:grave2}
\end{align}
restated here for clarity. We will now find and then minimise the sum of the rotational and hydrodynamic energies of the superfluid with respect to these three variables. 

\subsubsection{Rotational Energy}
\label{sec:Unequal:TwoPlanes:Rotation}

Firstly, we will calculate the rotational energy, which is the expectation value of \(-\Omega_{xy} \hat{L}_{xy}-\Omega_{zw} \hat{L}_{zw}  \) in the state \(\psi\) [c.f. Eq.~\eqref{eq:Erot}]. Since this is a first order differential operator we may use the product rule on Eq.~\eqref{eq:Unequal:skewAnsatz}, e.g. for each angular momentum component as
\begin{align}
   \Omega_{xy} \hat{L}_{xy}\psi &= n^{\frac{1}{2}} \left( e^{i\grt_2}   \Omega_{xy} \hat{L}_{xy} e^{i\act_1} + e^{i\act_1}   \Omega_{xy} \hat{L}_{xy} e^{i\grt_2} \right),  \\
    \psi^*\Omega_{xy} \hat{L}_{xy}\psi &= n \left( e^{-i\act_1}   \Omega_{xy} \hat{L}_{xy} e^{i\act_1} + e^{-i\grt_2}   \Omega_{xy} \hat{L}_{xy} e^{i\grt_2} \right),
\end{align}
such that the rotational energy density is simply the sum of contributions from each vortex independently. From Section~\ref{sec:Unequal:OnePlane}, recall that the favourable possible orientations of the two planes are limited to those which are planes of rotation of the isoclinic rotation generated by \(\hat{L}_+\). This means that -- as in the single vortex case we just considered -- we can rewrite the rotational energy density as
\begin{align}
    \psi^*&\odl\psi = \nonumber \\ &\qquad 2n\hbar \Omega_{zw}
    + n \Delta\Omega \left( \frac{\hat{L}_{xy} e^{i\act_1}}{e^{i\act_1}} + \frac{\hat{L}_{xy} e^{i\grt_2}}{e^{i\grt_2}} \right),
\end{align}
where the extra factor of $2$ in the first term on the RHS arises because we now have two vortex planes instead of one [c.f. Eq.~\eqref{eq:Erot2}]. This means that to proceed we simply have to evaluate the angular momentum of each vortex in the \(x\en y\) plane. In other words, we must compute the following integral
\begin{align}
    \dint_{B^4(R)} \frac{\hat{L}_{xy} e^{i\act_1}}{e^{i\act_1}} \diff^4 r;
    \label{eq:integrand}
\end{align}
this is carried out in Appendix~\ref{app:angular}, with the final result that
\begin{align}
    \dint_{B^4(R)} \diff^4 r \frac{\hat{L}_{xy} e^{i\act_1}}{e^{i\act_1}} = \hbar\frac{\pi^2}{2}R^4 \cos^2\eta_1.
\end{align}
The calculation for the \(e^{i\grt_2}\) term follows identical logic and so we simply state the result, which is
\begin{align}
    \dint_{B^4(R)} \diff^4 r \frac{\hat{L}_{xy} e^{i\grt_2}}{e^{i\grt_2}} &= \hbar\frac{\pi^2}{2}R^4 \sin^2\eta_2.  
\end{align}
Putting these together, remembering that \(n=N/V\) and \(V=\pi^2R^4/2\), we have
\begin{align}
 E_{\text{rot}}&= 2N \hbar \Omega_{zw} + N \hbar \Delta\Omega\left(\cos^2\eta_1+\sin^2\eta_2\right), 
    \label{eq:ODLExpectation}
\end{align}
for the rotational energy of two rigid, intersecting vortex planes under unequal frequency double rotation, with $\eta_{1}$ (resp. $\eta_{2}$) denoting the angle that the first (resp. second) plane is tilted compared to the \(x\en y\) (resp. \(z\en w\)) rotation plane. 

\subsubsection{Vortex-Vortex Interaction Energy}
\label{sec:Unequal:TwoPlanes:Interaction}

Secondly, we consider the hydrodynamic vortex plane interaction previously derived in Sec~\ref{sec:NonOrtho}, which, as we state again here, is calculated from
\begin{equation}
   E_{\text{vv}} = m  \int \rho
    \mathbf{v}_1\cdot\mathbf{v}_2 \diff^4 r .
\end{equation}
As this depends on the dot product between the velocity fields of each individual vortex, this energy is entirely dependent on the skewness, \(\eta\), of the two planes, that measures how far from being mutually orthogonal the vortex planes are (see Sec~\ref{sec:NonOrtho} for details). The most direct way to find \(\eta\) is to take the acute and grave coordinates and define a rotation transforming between them. Remembering that each of the vortex planes is a rotation plane of a left isoclinic rotation \(M_L\), we can use our result from Sec~\ref{sec:IsoPlanes} that the transformation between them has the following form [c.f. Eq.~\eqref{eq:SU(2)}]
\newcommand{\pharb}{\phi}
\begin{align}
    \begin{pmatrix}
        \grr_1e^{i\grt_1} \\
        \grr_2e^{i\grt_2}
    \end{pmatrix}
    &= 
    \begin{pmatrix}
        e^{i\pharb_1}\cos\eta & e^{i\pharb_2}\sin\eta \\
        -e^{-i\pharb_2}\sin\eta & e^{-i\pharb_1}\cos\eta
    \end{pmatrix}
    \begin{pmatrix}
        \acr_1e^{i\act_1} \\
        \acr_2e^{i\act_2}
    \end{pmatrix},
\end{align}
where \(\eta\in[0,\pi/2]\) and \(\phi_{1,2}\in[0,2\pi)\). Substituting Eq.~\eqref{eq:grave} into Eq.~\eqref{eq:acute} gives a relation of this form; specifically, examining the top-left entry of the matrix allows us to relate \(\eta\) to \(\eta_1, \eta_2,\) and \(\varphi\) as follows
% \begin{align}
%     &\begin{pmatrix}
%         \cos\eta_2 & e^{-i\varphi}\sin\eta_2 \\
%         -e^{i\varphi}\sin\eta_2 & \cos\eta_2
%     \end{pmatrix}
%     \begin{pmatrix}
%         \cos\eta_1 & -e^{i\varphi}\sin\eta_1 \\
%         e^{-i\varphi}\sin\eta_1 & \cos\eta_1
%     \end{pmatrix} \\
%     &= 
%     \begin{pmatrix}
%         \cos\eta_2\cos\eta_1 \\
%         +e^{-2i\varphi}\sin\eta_2\sin\eta_1 & e^{-i\varphi}\sin\eta_2 \\ \\
%         -e^{i\varphi}\sin\eta_2 & \cos\eta_2\cos\eta_1 \\ & +e^{2i\varphi}\sin\eta_2\sin\eta_1
%     \end{pmatrix}
% \end{align}
\begin{align}
    e^{i\pharb_1}\cos\eta &= \cos\eta_1\cos\eta_2+e^{-2i\varphi}\sin\eta_1\sin\eta_2.
    \label{eq:EtaComplex}
\end{align}
Taking particular values of \(\varphi\) in the above equation will give us intuition for how this parameter corresponds to the direction of the tilt, as mentioned previously, and hence allow us to deduce the form of the vortex-vortex interaction. In particular, we will examine the cases in which  \(\varphi\) is a multiple of \(\pi/2\), as this renders the RHS of Eq.~\eqref{eq:EtaComplex} real and non-negative, allowing us to simplify this equation by choosing \(\pharb_1=0\). Specifically, when \(\varphi = 0\) or \(\pi\) we have
\begin{align}
    \cos\eta &= \cos(\eta_1-\eta_2), \quad \rightarrow \quad
    \eta = \abs{\eta_1-\eta_2},
    \label{eq:EtaDiff}
\end{align}

while \(\varphi = \pi/2\) or \(3\pi/2\) gives
\begin{align}
    \cos\eta &= \cos(\eta_1+\eta_2), \quad \rightarrow \quad
    \eta = \eta_1+\eta_2.
    \label{eq:EtaSum}
\end{align}
Note that \(\cos(\eta_1+\eta_2)\) is non-negative due to the constraint on \(\eta_{1,2}\) [Eq.~\eqref{eq:EtaRestriction}].

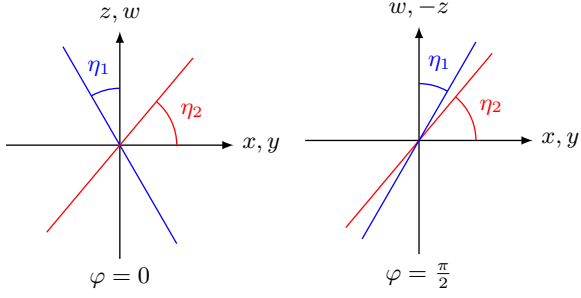
\begin{figure}
    \centering
    \begin{tikzpicture}[line width=0.5pt]

    \pgfdeclarelayer{below}
    \pgfdeclarelayer{above}
    \pgfsetlayers{below,main,above}

    \colorlet{myPurple}{Red!50!Blue}
    \setlength{\axisLength}{0.35\linewidth};
    \coordinate (O) at (0,0);
    
    \pgfmathsetmacro\etaOne{30}
    \pgfmathsetmacro\etaTwo{50}
    \draw[-latex] (-0.5\axisLength,0) -- (0.5\axisLength,0) node[right] (xy) {\(x,y\)};
    \draw[-latex] (0,-0.5\axisLength) -- (0,0.5\axisLength) node[above] (zw) {\(z,w\)};
    
    \node[below] at (0,-0.5\axisLength) {\(\varphi=0\)};
    
    \draw[blue, rotate=\etaOne] (0,-0.5\axisLength) -- (0,0.5\axisLength);
    \node[blue] at ($(O) + ({(90+\etaOne/2)}:0.35\axisLength) $) {\(\eta_1\)};
    \draw[blue] (0,0.25\axisLength) arc (90:90+\etaOne:0.25\axisLength);
    
    \draw[red, rotate=\etaTwo] (-0.5\axisLength,0) -- (0.5\axisLength,0);
    \node[red] at ($(O) + ({\etaTwo/2}:0.35\axisLength)$) {\(\eta_2\)};
    \draw[red] (0.25\axisLength,0) arc (0:\etaTwo:0.25\axisLength);
    
\end{tikzpicture}
    \begin{tikzpicture}[line width=0.5pt]

    \pgfdeclarelayer{below}
    \pgfdeclarelayer{above}
    \pgfsetlayers{below,main,above}

    \colorlet{myPurple}{Red!50!Blue}
    \setlength{\axisLength}{0.35\linewidth};
    \coordinate (O) at (0,0);
    
    \pgfmathsetmacro\etaOne{50}
    \pgfmathsetmacro\etaTwo{-30}
    \draw[-latex] (-0.5\axisLength,0) -- (0.5\axisLength,0) node[right] (xy) {\(x,y\)};
    \draw[-latex] (0,-0.5\axisLength) -- (0,0.5\axisLength) node[above] (zw) {\(w,-z\)};
    
    \node[below] at (0,-0.5\axisLength) {\(\varphi=\frac{\pi}{2}\)};
    
    \draw[red, rotate=\etaOne] (-0.5\axisLength,0) -- (0.5\axisLength,0);
    \node[red] at ($(O) + ({\etaOne/2}:0.35\axisLength)$) {\(\eta_2\)};
    \draw[red] (0.25\axisLength,0) arc (0:\etaOne:0.25\axisLength);
    
    \draw[blue, rotate=\etaTwo] (0,-0.5\axisLength) -- (0,0.5\axisLength);
    \node[blue] at ($(O) + ({(90+\etaTwo/2)}:0.35\axisLength) $) {\(\eta_1\)};
    \draw[blue] (0,0.25\axisLength) arc (90:90+\etaTwo:0.25\axisLength);
    
\end{tikzpicture}
    \caption{A possible configuration of tilted vortex planes relative to the \(x\en y\) and \(z\en w\) planes, visualised as lines in 2D, for special values of \(\varphi\). When \(\varphi=0\) the planes are given by Eqs~\eqref{eq:samePlane1} and \eqref{eq:samePlane2}, which describe a pair of lines in \(x,z\) space and an identical pair of lines in \(y,w\) space, plotted simultaneously in the left panel. When \(\varphi=\pi/2\) we have instead Eqs~\eqref{eq:oppositePlane1} and \eqref{eq:oppositePlane2}, describing a pair of lines in \(x,w\) space and their reflection about the horizontal axis in \(y,z\) space, which we plot on the same graph here by transforming \(z,w\to w,-z\). Equivalent pictures for \(\varphi=\pi\) and \(3\pi/2\) respectively can be found by reflection about the vertical axis.}
    \label{fig:TiltPlanesLines}
\end{figure}

To understand these special cases let's look at the equations of the vortex planes directly. Substituting \(\varphi = 0\) into Eqs~\eqref{eq:acute1} and \eqref{eq:grave2} gives
\begin{align}
    \acr_1e^{i\act_1} &= \cos\eta_1(x+iy) + \sin\eta_1(z+iw), \\
    \grr_2e^{i\grt_2} &= \cos\eta_2(z+iw) - \sin\eta_2(x+iy),
\end{align}
such that the two planes are defined by
\begin{align}
    \acr_1 e^{i\act_1} &= 0: \hspace{0.5em} x = - \tan\eta_1z \hspace{0.5em} \& \hspace{0.5em} y = - \tan\eta_1w,
    \label{eq:samePlane1} \\
    \grr_2 e^{i\grt_2} &=0: \hspace{0.5em} z = \tan\eta_2x \hphantom{-}\hspace{0.5em} \& \hspace{0.5em} w = \tan\eta_2y,
    \label{eq:samePlane2}
\end{align}
while the same procedure for \(\varphi = \pi/2\) gives
\begin{align}
    \acr_1 e^{i\act_1} &= 0: \hspace{0.5em} x = \tan\eta_1w \hphantom{-}\hspace{0.5em} \& \hspace{0.5em} y = - \tan\eta_1z,
    \label{eq:oppositePlane1} \\
    \grr_2 e^{i\grt_2} &=0: \hspace{0.5em} z = - \tan\eta_2y \hspace{0.5em} \& \hspace{0.5em} w = \tan\eta_2x.
    \label{eq:oppositePlane2}
\end{align}
In the \(\varphi=0\) case, Eqs~\eqref{eq:samePlane1} and \eqref{eq:samePlane2} describe a pair of lines in the 2D \((x,z)\) subspace, and identical lines found by taking the first lines and sending \((x,z)\to (y,w)\). Similarly, when \(\varphi=\pi/2\) we have a pair of lines in the \((x,w)\) subspace, and a pair in the \((y,z)\) space which are related to the first pair by \((x,w)\to(y,-z)\). For \(\varphi=\pi,3\pi/2\) note that \(\varphi\to\varphi+\pi\) transforms the equations for the planes simply by \(z\to-z\) and \(w\to-w\). 

Since the equations separate into lines in 2D subspaces this way, we can visualise the planes by simply plotting these lines, as shown in Fig~\ref{fig:TiltPlanesLines}. From the left panel of this figure it is clear that when \(\varphi=0\) (or \(\pi\)) the vortex planes are tilted away from the coordinate planes in the same direction. Similarly, the right panel shows that when \(\varphi=\pi/2\) or \(3\pi/2\) the vortices are tilted in opposite directions, and hence towards each other. Other values of \(\varphi\) interpolate between these two scenarios such that the two vortex planes are not tilted along a common direction. This visual understanding also agrees with the two expressions for \(\eta\), given in Eqs~\eqref{eq:EtaDiff} and \eqref{eq:EtaSum}.

Now we can use physical intuition to deduce -- without any further calculation -- the most energetically favourable value for \(\varphi\) for any fixed values of the parameters \(\eta_{1,2}\). Recalling that the rotational energy was independent of \(\varphi\), we need only consider the interaction potential between the two vortices, given by
\begin{align}
  E_{\text{vv}} &= -4\mu N\frac{\xi^2}{R^2}\ln\cos\eta,
\end{align}
which is positive and therefore repulsive. This result was derived in Sec~\ref{sec:NonOrtho} in the case that \(\phi_{1,2}=0\), however these angles do not affect the interaction since they can be absorbed into the definition of \(\act_{1,2}\) and \(\grt_{1,2}\). Since this interaction is repulsive we can clearly see that it is maximised when \(\varphi=\pi/2\) or \(3\pi/2\) as the vortex planes are tilted directly toward one another. Equally, in the other case where \(\varphi=0\) or \(\pi\) the vortices are tilted in the same direction and the interaction energy cost is minimised. Therefore, we can set \(\eta=\abs{\eta_1-\eta_2}\) and proceed with finding the minimum energy as a function of the remaining parameters \(\eta_{1,2}\). For concreteness we will also set \(\varphi=0\), but note that \(\varphi=\pi\) provides an equivalent solution with the same energy.

\subsubsection{Finding the minimum}
\label{sec:Unequal:TwoPlanes:Minimum}

Our final step is to add the rotational and the vortex-vortex interactions energies together and to minimise the resulting sum. Firstly we will define quantities that will make the calculation simpler. Let \(E_{\text{rot}}^{\perp}=N \hbar (\Omega_{xy}+\Omega_{zw})\) denote the reduction in energy due to rotation of the state with orthogonal vortex planes along the \(x\en y\) and \(z\en w\) planes [c.f. Eq.~\eqref{eq:orthrot}]. We then define a dimensionless energy density relative to \(E_{\text{rot}}^{\perp}\), given by
\begin{align}
    \varepsilon = \frac{R^2}{2\xi^2\mu N}(-E_{\text{rot}} + E_{\text{rot}}^{\perp} + E_{\text{vv}}), 
\end{align}
and a dimensionless frequency difference \(\omega = R^2\hbar\Delta\Omega/2\xi^2\mu\). Note that in units of the critical frequency \(\Omega_c\) [Eq.~\eqref{eq:4Dcrit}], this dimensionless frequency is given by \(\omega = \ln(2.07R/\xi)\Delta\Omega/\Omega_c\). We then must find the minimum of the following
\begin{align}
    \varepsilon &= \omega \left( 1 - \cos^2\eta_1 - \sin^2\eta_2 \right) - 2\ln\cos(\eta_1-\eta_2). \label{eq:tiltEnergy}
\end{align}
Note that we have not needed to include the absolute value on the RHS of Eq.~\eqref{eq:EtaDiff}, since \(\cos\abs{x}=\cos{x}\). Additionally, the logarithmic divergences as \(\eta_1-\eta_2\to\pi/2\) are unphysical as there the vortex planes coincide and the constant density approximation that we took in Sec~\ref{sec:NonOrtho} fails.

\begin{figure}[t!]
    \centering
    \includegraphics[width=0.45\textwidth]{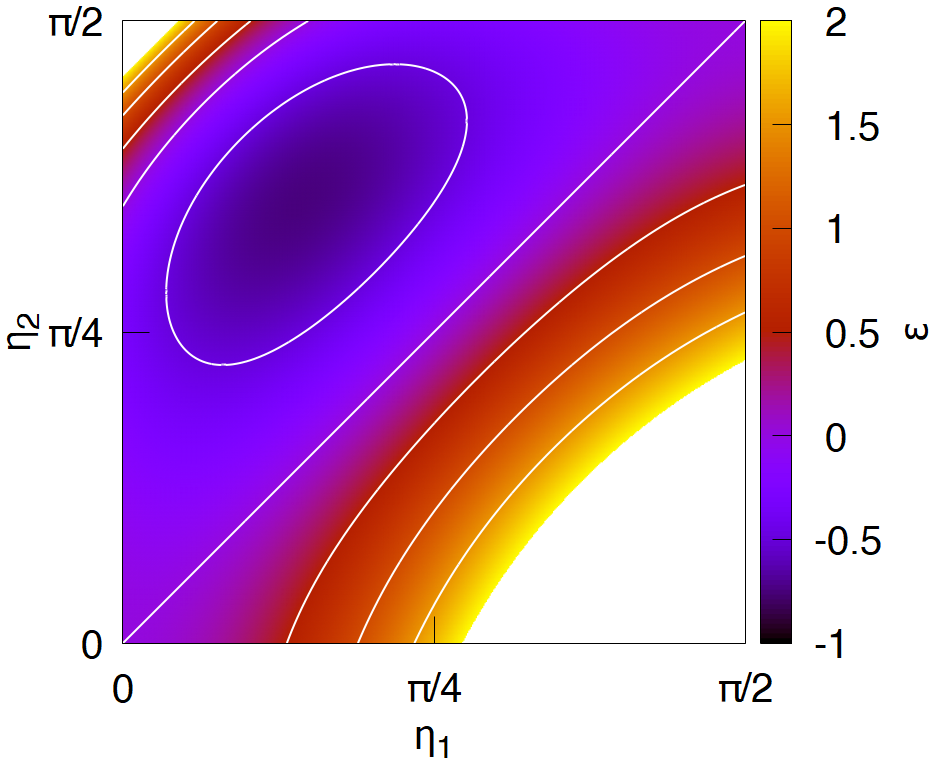}
    \caption{Difference in dimensionless energy density [Eq.~\eqref{eq:tiltEnergy}] between the skew vortex plane ansatz and the orthogonal configuration as a function of the tilt angles \(\eta_{1,2}\). The dimensionless frequency difference has been set to \(\omega=2\), for this value the minimum occurs at a skewness of \(\eta\approx 38\degree\) as calculated from Eq.~\eqref{eq:EtaOfOmega}. Contour lines are included as guides to the eye to highlight the degeneracy along the line \(\eta_1=\eta_2\), and the minimum in the upper left. We have omitted values of \(\varepsilon>2\) to avoid the logarithmic divergences at \((0,\pi/2)\) and \((\pi/2,0)\).}
    \label{fig:TiltEnergyAnalytic}
\end{figure}

Setting the derivatives of this energy to zero gives us the following simultaneous equations
\begin{align}
    \frac{\partial\varepsilon}{\partial\eta_1} &= \omega\sin2\eta_1 + 2\tan(\eta_1-\eta_2) = 0, \label{eq:eta1Deriv} \\
    \frac{\partial\varepsilon}{\partial\eta_2} &= -\omega\sin2\eta_2 - 2\tan(\eta_1-\eta_2) = 0. \label{eq:eta2Deriv}
\end{align}

Firstly, examining the sign of the terms in each of these equations (recalling that \(\eta_{1,2}<\pi/2\)), we must have that 
\begin{align}
    \eta_1\leq\eta_2.
    \label{eq:etaOrder}
\end{align}
Physically, this is because if \(\eta_1\) is greater than \(\eta_2\) then the force from the repulsive interaction acts in the same direction as the force from the rotational energy. Therefore we can eliminate the absolute value in Eq.~\eqref{eq:EtaDiff}, such that
\begin{align}
    \eta=\eta_2-\eta_1.
    \label{eq:eta}
\end{align}

Secondly, Eqs~\eqref{eq:eta1Deriv} and \eqref{eq:eta2Deriv} together imply \(\sin2\eta_1 = \sin2\eta_2\). There are two ways to satisfy this, i.e. by taking 
\begin{align}
    \eta_1 &= \eta_2, \label{eq:orthogonalBranch} \\
  \text{or}\quad  \eta_1 &= \frac{\pi}{2} - \eta_2. \label{eq:symmetricBranch}
\end{align}

The former case is precisely the condition that the two planes are orthogonal, which eliminates the interaction term. Substituting Eq.~\eqref{eq:orthogonalBranch} into Eqs~\eqref{eq:eta1Deriv} and \eqref{eq:eta2Deriv} then leads to the result \(\eta_1=\eta_2=0\), the state we have previously studied \cite{mccanna2021}. This state has an energy of \(\varepsilon=0\) by definition, since \(\varepsilon\) was defined relative to this state. Moreover, we also note that if we substitute Eq.~\eqref{eq:orthogonalBranch} into Eq.~\eqref{eq:tiltEnergy}, we see that {\it any} state with \(\eta_1=\eta_2\) has energy \(\varepsilon=0\). Interestingly, these orthogonal states are still all degenerate despite the isoclinic symmetry being broken when \(\omega\neq 0\). The stationary point at \((\eta_1, \eta_2) =(0,0)\) is therefore a saddle point, since it has this line of constant energy passing through it.

\begin{figure*}[t!]
    \centering
    \includegraphics[width=\linewidth]{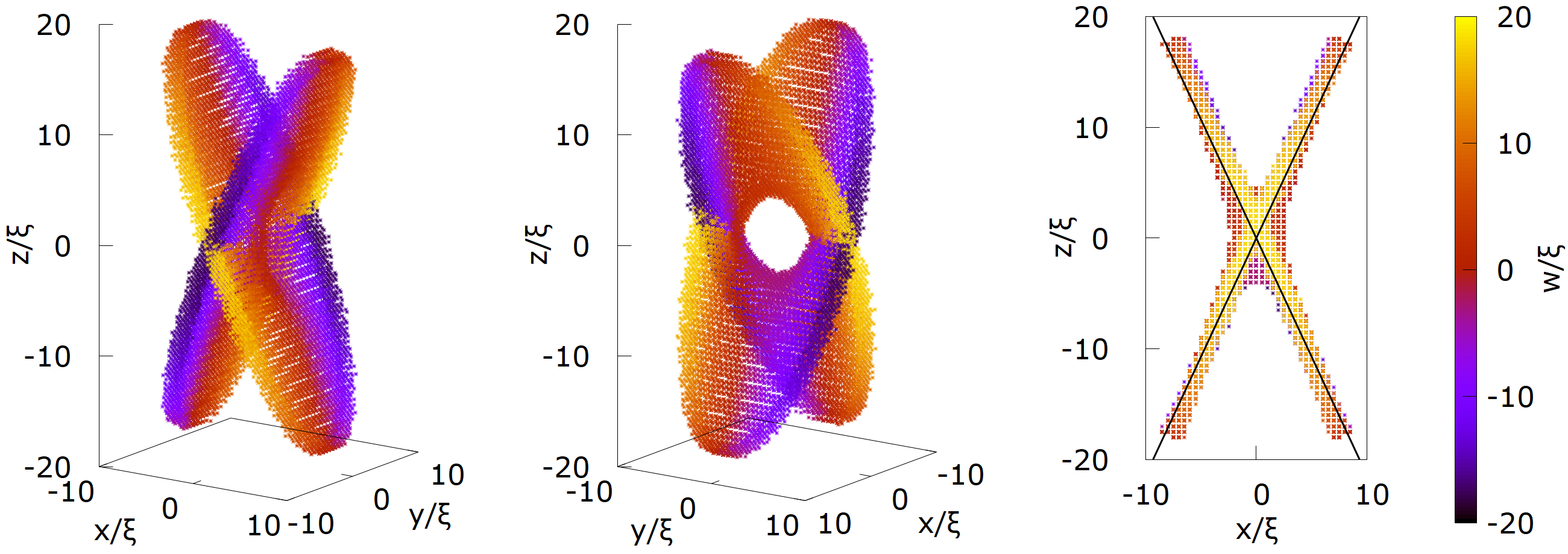}
    \caption{Numerical vortex core in the final state of the ITEM-evolved 4D GPE under double rotation [Eq.~\eqref{eq:GPE4DR}]. The spatial step size was \(\Delta x=0.5\xi\), corresponding to \(R\approx20.6\xi\). Rotation frequencies were \(\Omega_{zw}=0.85\Omega_c\), and \(\Omega_{xy} \approx1.43\Omega_c\), chosen such that the predicted skewness was precisely \(\eta=40\degree\). The initial phase profile was given by that of the ansatz described in Sec~\ref{sec:Unequal:TwoPlanes}, with this predicted value of \(\eta\) and added noise. The first two panels show two different views of the core in \((x,y,z)\) space, with points coloured according to their \(w\) value. The overall structure resembles the skew planes of the ansatz but the second panel clearly shows how the core curves away from these planes to form an avoided crossing~\cite{mccannaequal}. The final panel shows the points projected down into \((x,z)\) space --- again with \(w\) shown as colour --- as well as the lines \(z=\pm x\tan(\pi/4+\eta/2)\), which the theory predicts that the core points should lie along from this perspective. The agreement between these analytical lines and numerical points is very good.}
    \label{fig:40Skew}
\end{figure*}

The latter case is much more interesting as it arises from competition between the interaction and rotational energies. Note that the relation between the tilt angles [Eq.~\eqref{eq:symmetricBranch}] ensures that the two planes are symmetrically tilted with respect to the rotation planes of the superfluid; what we mean by this is that each vortex makes the same angle with the \(x\en y\) plane, and also with the \(z\en w\) plane. This can be seen by considering Fig~\ref{fig:TiltPlanesLines} and noting that the vortices are each tilted away from the \(x\en y\) plane by an angle of \(\pi/2-\eta_1\), and \(\eta_2\) respectively. When \(\eta_1=\pi/2-\eta_2\) these two angles are equal, and the same is of course true with the angles the vortices make with the \(z\en w\) plane. Using Eqs~\eqref{eq:eta} and \eqref{eq:symmetricBranch} we can write both the angles \(\eta_{1,2}\) in terms of the skewness as follows
\begin{align}
    \eta_1 &= \frac{\pi}{4} - \frac{\eta}{2}, 
    \label{eq:eta1} \\
    \eta_2 &= \frac{\pi}{4} + \frac{\eta}{2},
    \label{eq:eta2}
\end{align}
which makes it clear that \(\eta_1\leq\pi/4\) and \(\eta_2\geq\pi/4\). At this point it is worth restating our ansatz, since it now only depends on \(\eta\). Recall that the order parameter is defined as [Eq.~\eqref{eq:Unequal:skewAnsatz}] \(\psi=n^{1/2} e^{i\act_1}e^{i\grt_2}\), with these angles defined by Eqs~\eqref{eq:acute1} and \eqref{eq:grave2}. Substituting \(\varphi=0\) and the above equations for \(\eta_{1,2}\), Eqs~\eqref{eq:acute1} and \eqref{eq:grave2} become 

\begin{align}
    \acr_1e^{i\act_1}
    &= \sin\left(\frac{\pi}{4}+\frac{\eta}{2}\right) \cpolar{2} + \sin\left(\frac{\pi}{4}-\frac{\eta}{2}\right) \cpolar{1}, \quad
    \label{eq:AcuteFinal} \\
    \grr_2e^{i\grt_2}
    &= \sin\left(\frac{\pi}{4}+\frac{\eta}{2}\right) \cpolar{2} - \sin\left(\frac{\pi}{4}-\frac{\eta}{2}\right) \cpolar{1}, \quad
    \label{eq:GraveFinal}
\end{align}
where we have used that \(\cos(\pi/4\pm\eta/2) = \sin(\pi/4\mp\eta/2)\). Note that we can now clearly see that two planes are arranged symmetrically, in the sense that after a rotation of angle \(\pi\) in the \(x\en y\) plane (\(\ta_1\to\ta_1+\pi\)) the two equations Eq.~\eqref{eq:AcuteFinal} and \eqref{eq:GraveFinal} swap and hence the two vortex planes swap (note that this is also true for a \(\pi\) rotation in the \(z\en w\) plane, up to a shift in the angles \(\act_1\) and \(\grt_2\)). This symmetry can also be seen in the the equations for the vortex planes, Eqs~\eqref{eq:samePlane1} and \eqref{eq:samePlane2} which are now given by
\begin{align}
    \hspace{0.5em} z = \pm\tan\left(\frac{\pi}{4}+\frac{\eta}{2}\right)x \hspace{0.5em} \& \hspace{0.5em} w = \pm\tan\left(\frac{\pi}{4}+\frac{\eta}{2}\right)y
    \label{eq:symPlanes}
\end{align}
where \(+\) and \(-\) refer to the planes given by \(\grr_2e^{i\grt_2}=0\) and \(\acr_1e^{i\act_1}=0\), respectively. From these equations we can actually see that this configuration is invariant under a \(\pi\) rotation in any one of the six coordinate planes.

We also now see from these equations that both vortices are closer in angle to the \(z\en w\) plane than they are to the \(x\en y\) plane. An interesting consequence of this is that when \(\eta=0\) the orthogonal state we get does not consist of vortices spanning the \(x\en y\) and \(z\en w\) planes. Instead the vortices occupy a pair of diagonal (in terms of the lab frame) planes, given by \(z=\pm x\) and \(w=\pm y\). This doesn't matter when the frequency difference is zero, as then the rotation is isoclinic and these diagonal planes are also rotation planes [c.f. Section~\ref{sec:4DRot}], but for any other value of \(\Delta\Omega\) the only rotation planes are the \(x\en y\) and \(z\en w\) planes so it is perhaps surprising that none of these states ever occupy them.

Substituting Eqs~\eqref{eq:eta1} and \eqref{eq:eta2} into Eq.~\eqref{eq:eta1Deriv} gives the following relation between \(\omega\) and the optimal skewness
\begin{align}
    \omega = \frac{2\tan\eta}{\cos\eta},
    \label{eq:omega}
\end{align}
which rearranges to the following quadratic equation for \(\sin\eta\)
\begin{align}
    \sin^2\eta  + \frac{2}{\omega}\sin\eta - 1 = 0.
\end{align}
This has only one solution for \(\sin\eta\) in the interval \([-1,1]\), given by
\begin{align}
    \sin\eta = \frac{\left( 1 + \omega^2 \right)^{\frac{1}{2}} - 1}{\omega}
    \label{eq:EtaOfOmega}
\end{align}
Physically this means that the optimal skewness vanishes in the limit that $\omega$ (i.e. the frequency difference in rotation) goes to zero, corresponding to the situation where the two vortex planes become completely orthogonal, as expected. In the opposite limit that $\omega$ becomes very large, this formula instead predicts that $\sin \eta \rightarrow 1$ and hence $\eta \rightarrow \pi/2$, meaning that the angle between the two planes goes to zero. This corresponds physically to the two planes both aligning with the $z\!-\!w$ plane so as to maximise the energetic reduction due to the higher rotation frequency $\Omega_{xy} > \Omega_{zw}$. However, this limit should also be treated with caution, as at high enough frequencies, we expect that it will become energetically favourable to introduce more vortices and/or more complicated vortex structures, as briefly discussed in Appendix~\ref{app:numerical}. We also expect there will be other contributions to the energy, which we have neglected here; for example, our assumption of a constant density profile will break down when the two vortex planes become very close together.

Now we can find the energy of this optimally skewed state; using Eqs~\eqref{eq:eta}, \eqref{eq:symmetricBranch}, and \eqref{eq:eta2} the energy [Eq.~\eqref{eq:tiltEnergy}] becomes
\begin{align}
    \varepsilon &= \omega(1-2\sin^2\eta_2) -2\ln\cos\eta, \\
    &= \omega\cos\left(\frac{\pi}{2}+\eta\right) - \ln\cos^2\eta.
\end{align}
Rearranging Eq.~\eqref{eq:omega} we can quickly find that \(\cos^2\eta = 2\sin\eta/\omega\), and after using \(\cos(\pi/2+\eta)=-\sin\eta\) we have everything in terms of \(\sin\eta\). Substituting Eq.~\eqref{eq:EtaOfOmega} then gives the energy density for the optimal skewed statesin terms of \(\omega\) as
\begin{align}
    \varepsilon &= -\left[ \left(1+\omega^2\right)^{\frac{1}{2}} - 1 \right] + \ln\left\{ \frac{\omega^2}{2\left[\left(1+\omega^2\right)^{\frac{1}{2}} - 1\right]} \right\}.
    \label{eq:EofOmega}
\end{align}
For small and large \(\omega\) we have the following asymptotics
\begin{align}
    \varepsilon &= -\frac{\omega^2}{2} + o(\omega^2) \ \text{as} \ \omega\to0, \\
    \varepsilon &= -\omega + o(\omega) \ \text{as} \ \omega\to\infty.
\end{align}
Recall that \(\omega \propto R^2\Delta\Omega\), therefore these limits can be reached by decreasing (resp. increasing) either \(\Delta\Omega\) or the radius \(R\).

This energy [Eq.~\eqref{eq:EofOmega}] is also negative for all \(\omega>0\), which means our simplified model has predicted that this tilted vortex plane state is lower energy than the orthogonal state for any frequency difference \(\Delta\Omega>0\). Fig~\ref{fig:TiltEnergyAnalytic} shows the energy landscape as a function of both tilt angles for dimensionless frequency difference \(\omega=2\). As expected, we see a line of constant energy along \(\eta_1=\eta_2\), a minimum energy along the line \(\eta_1=\pi/2-\eta_2\), and a range of tilt angles for which the energy is negative.

\subsection{Numerical Results}
\label{sec:Unequal:TwoPlanes:Results}

We will now compare the above analytical predictions for tilted vortex planes to numerical results obtained using the methods described in Sec~\ref{sec:Numerics}. We choose an initial phase profile identical to that of our non-orthogonal vortex ansatz [Eq.~\eqref{eq:Unequal:skewAnsatz}], where the acute and grave coordinates are given by Eqs~\eqref{eq:AcuteFinal} and \eqref{eq:GraveFinal}, respectively, with a chosen value for the skewness, \(\eta\). We then use Eq.~\eqref{eq:omega} to calculate the frequency difference, \(\Delta\Omega\), that will energetically favour the chosen value of \(\eta\) if the model is accurate, and then we run the ITEM with this value of \(\Delta\Omega\) on our initial state with added noise. 

Once the ITEM is converged we compare the geometry of the vortex core in the numerical final state with that of our predictions. Fig~\ref{fig:40Skew} shows the numerical vortex core for \(\Delta x=0.5\xi\), which corresponds to a system radius of \(R\approx20.6\xi\), and with frequencies \(\Omega_{zw}=0.85\Omega_c\), and \(\Omega_{xy}\approx1.43\Omega_c\), corresponding to a predicted skewness of \(\eta=40\degree\). The first two panels show two different rotations of the core in \((x,y,z)\) space, with the points coloured according to their \(w\) value (see the colourbar on the far right). Already we can see that, at large distances from the origin, the vortex cores look like the predicted tilted planes, symmetrically arranged with respect to the rotation planes of the superfluid. The third panel in Fig~\ref{fig:40Skew} shows a side-on view where the vortex core appears approximately as a pair of lines, just as in Fig~\ref{fig:TiltPlanesLines}. On top of these data points we have plotted the lines \(z=\pm\tan(\pi/4+\eta/2)x\) [c.f. Eq.~\eqref{eq:symPlanes}] which are the predicted lines on which the numerical data should lie. As can be seen in the figure, there is excellent agreement between these numerical final states and our analytic predictions. However, note that near the origin we see --- most prominently in the second panel of the figure --- an avoided crossing structure~\cite{mccannaequal}, as also discussed further below. 

\begin{figure}
    \centering
    \includegraphics[width=0.45\textwidth]{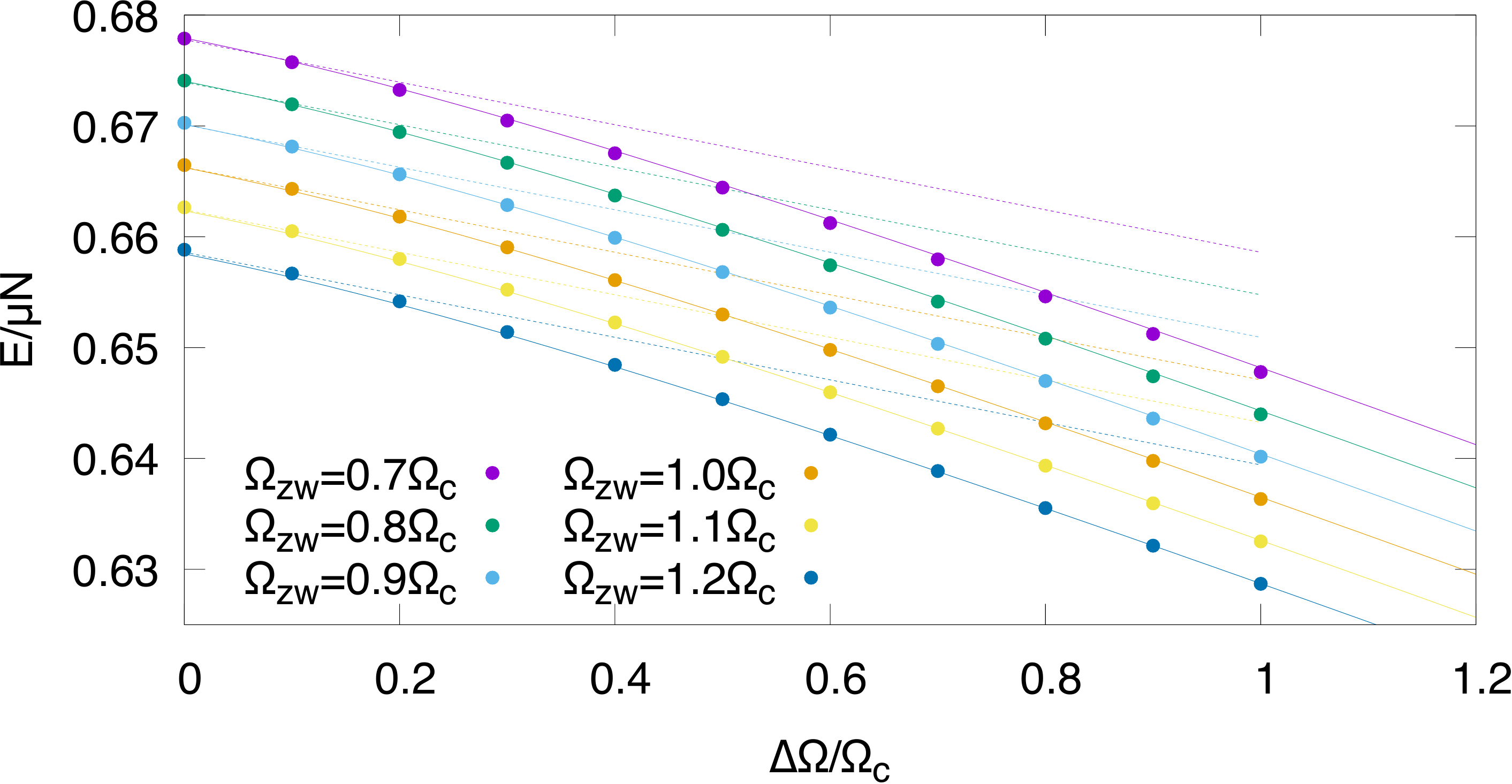}
    \caption{Numerical (points) and analytical (solid lines) results for the energy of aligning non-orthogonal vortex states --- as a function of \(\Omega_{zw}\) and \(\Delta\Omega\) ---  described in this section, as well as numerical results for the energy of the orthogonal vortex state (dotted lines). The resolution was set to \(\Delta x=0.5\xi\), giving a system radius of \(R\approx20.6\xi\). All of the analytical lines were generated by a single fit of the numerical points to Eq.~\eqref{eq:EofOmegaDimensions}, with \(E_0\) and \(R\) as fitting parameters. The fit produced \(E_0\approx 0.71\mu N\) and \(R\approx 19.5\xi\), and agrees excellently with the numerical data. The numerical zero-vortex ground state in this system has an energy of \(E_\text{hom}\approx0.67\mu N\), which suggests that the numerical critical frequency in this system is very close to \(0.9\Omega_c\), where \(\Omega_c\) is our predicted value [Eq.~\eqref{eq:4Dcrit}]. }
    \label{fig:TiltEnergyNumerics}
\end{figure}

In addition to the above qualitative comparisons of the numerical and predicted vortex cores, we have made a quantitative analysis of the accuracy of our predicted energy [Eq.~\eqref{eq:EofOmega}]. To do this we performed the ITEM for a range of different frequencies \(\Omega_{zw}\) and \(\Delta\Omega\), again using our ansatz to determine the initial phase, and then calculating the energy of each final state. The procedure for these calculations was as follows; we fixed a value of \(\Omega_{zw}\), then ran the ITEM with \(\Delta\Omega=\Omega_c\) on our prescribed initial state. From the final state we calculated the energy, and then to speed up calculations we used this final state as the initial state for the next ITEM run with \(\Delta\Omega=0.9\Omega_c\). This process was repeated down to the isoclinic point, \(\Delta\Omega=0\). Finally, many of these loops were run at once with different values of \(\Omega_{zw}\), so that we could explore an area in frequency space rather than just a line. The results for these energies are shown in Fig~\ref{fig:TiltEnergyNumerics}, with each value of \(\Omega_{zw}\) corresponding to a different colour. On top of these points we have plotted lines given by performing a single fit of this data over the area in frequency space to a redimensionalized version of Eq.~\eqref{eq:EofOmega} given by
\begin{widetext}
    \begin{align}
        E &= E_0 -N\hbar\left(\Delta\Omega + 2\Omega_{zw}\right) + 2\mu N\frac{\xi^2}{R^2} \left[ 1 - \sqrt{1 + \frac{R^4\hbar^2\Delta\Omega^2}{4\xi^4\mu^2}} + \ln\left( \frac{R^4\hbar^2\Delta\Omega^2}{4\xi^4\mu^2} \right) -  \ln\left(\sqrt{1+\frac{R^4\hbar^2\Delta\Omega^2}{4\xi^4\mu^2}} - 1 \right)\right], \quad
        \label{eq:EofOmegaDimensions}
    \end{align}
\end{widetext}
where \(E_0\) is simply the energy of the system when the vortex planes are orthogonal and there is no external rotation. Since \(\xi\) and \(\mu\) are both external parameters in the numerics, the only parameters in the fit were \(R\) and \(E_0\), the energy at \(\Omega_{zw}=\Omega_{xy}=0\). Furthermore, \(R\) is not truly a free parameter as it is determined up to boundary effects by the radius of the simulated region, which is roughly \(20.6\xi\). The fit produced a value of \(R\approx19.5\xi\), which is consistent if we estimate the size of the boundary region to be roughly equal to \(\xi\). Additionally, the agreement between the numerical points and lines from the fit is excellent. We also attempted to track the energy of the theoretically predicted skew plane branch at a frequency of \(\Omega_{zw}=1.3\Omega_c\), by performing two of these iterative ITEM runs from \(\Delta\Omega=\Omega_c\) down to \(\Delta\Omega=0\). Interestingly, as we decreased \(\Delta\Omega\) for these runs the states we obtained increasingly diverged from the skew plane states, even down to the isoclinic point at \(\Delta\Omega=0\). For plots of these states at the isoclinic point, see Appendix~\ref{app:numerical}.

The dotted line shows the numerical energy of the orthogonal state~\cite{mccanna2021} as a function of the frequencies. This was found by running the ITEM once, with \(\Omega_{zw}=\Omega_{xy}=1.2\Omega_c\), and then calculating the energy of this fixed final state for different values of \(\Omega_{zw}\) and \(\Delta\Omega\). As shown in Fig~\ref{fig:TiltEnergyNumerics}, the energies of the orthogonal state form dotted straight lines that meet the fit lines tangentially at \(\Delta\Omega=0\), which is exactly as predicted since the skewness of the tilted state is approaching zero in this limit. Finally, note that the numerical zero-vortex ground state in this system has an energy of \(E_\text{hom}\approx0.67\mu N\), which is roughly equal to the energy of the skew and orthogonal vortex states when \(\Omega_{zw}=0.9\Omega_c\), and \(\Delta\Omega=0\). This means that the value of the numerical critical frequency in this system is very close to \(0.9\Omega_c\), where \(\Omega_c\) is our predicted value [Eq.~\eqref{eq:4Dcrit}].

\begin{figure}[h]
    \centering
    \includegraphics[width=0.45\textwidth]{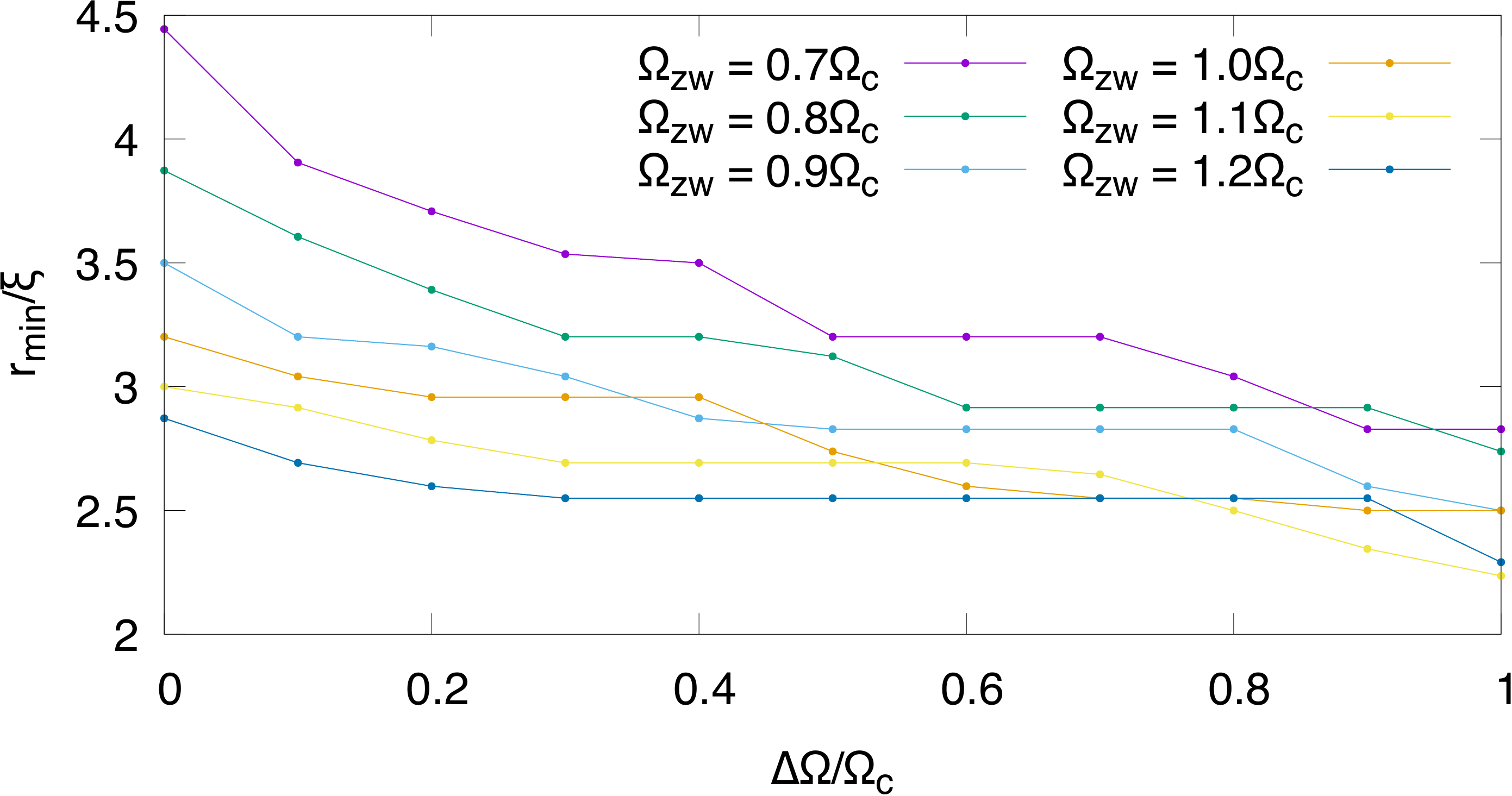}
    \caption{Minimum distance \(r_{\text{min}}\) of the vortex core from the origin --- as a function of \(\Omega_{zw}\) and \(\Delta\Omega\) --- for the non-orthogonal numerical states whose energy is shown in Fig~\ref{fig:TiltEnergyNumerics}. This quantity captures the size of the avoided crossing region seen in Figs~\ref{fig:40Skew} and similar final states. As shown above, this region generally shrinks as either frequency increases. These data are not particularly smooth, which may be  due to some sampling error from the discretisation of the Cartesian grid.}
    \label{fig:TiltDistPlot}
\end{figure}

To get an idea of the size of the avoided crossing as a function of the frequencies, we have taken each state represented in Fig~\ref{fig:TiltEnergyNumerics} and calculated the minimum distance between its vortex core and the origin, which we denote \(r_{\text{min}}\). This is then plotted in Fig~\ref{fig:TiltDistPlot}, which shows that, in general, the avoided crossing decreases in size with both \(\Omega_{zw}\) and \(\Delta\Omega\). Note that these lines are not perfectly smooth and also change their ordering as \(\Delta\Omega\) changes, suggesting that there are multiple metastable branches being sequentially followed by our numerical states. Nevertheless, these must be very close together in energy, since the fit in Fig~\ref{fig:TiltEnergyNumerics} is very good, and the long range core structure has good agreement with the predicted state.

Lastly, we have further tested our analytical results by using a different initial phase profile in the numerics. We still use the tilted plane ansatz [Eq.~\eqref{eq:Unequal:skewAnsatz}], but with different values of \(\eta_1\) and \(\eta_2\) than those predicted. Instead of the predicted values, \(\eta_1 + \eta_2 = \pi/2\) [Eq.~\eqref{eq:symmetricBranch}], we chose \(\eta_1 = \delta\), \(\eta_2 = \eta + \delta\), where \(\eta\) is the skewness of the theoretically predicted configuration, and \(\delta\) is a small angle added to ensure all symmetries are broken. The corresponding planes still have skewness given by \(\eta\), but are now asymmetric with respect to the planes of the external rotation. This initial state (with added noise) converges to the same final state as Fig~\ref{fig:40Skew} under the same value of all parameters (\(\Delta x=0.5\xi\), corresponding to \(R\approx20.6\xi\) \(\Omega_{zw}=0.85\Omega_c\), \(\Omega_{xy}\approx1.43\Omega_c\), corresponding to \(\eta=40\degree\)), showing that this predicted state is likely the ground state in this regime. However, at a higher value of \(\Omega_{zw}=1.25\Omega_c\), we find very different final states of the ITEM depending on which initial phase profile is used. Specifically, using the predicted phase profile (with added noise) we find the same final states as before, with just as good agreement to the analytics. Using the asymmetric phase profile (with added noise) described above, we find very different vortex core structures, with slightly higher energies than the theoretical states (see Appendix~\ref{app:numerical}).

\section{Conclusions}
\label{sec:concl}

In the first part of this section, we will focus on summarizing the main conclusions of our paper and highlighting the open questions that directly follow on from our study. In the second part of this section, we will then briefly discuss the more general future outlook for research into topological excitations in 4D superfluids, looking beyond the physics of the minimal model studied in this paper. 

\subsection{Summary}

In this paper, we have demonstrated that stationary states of the 4D GPE under unequal-frequency double rotation can host complicated vortex core structures consisting of skew planes and curved surfaces. This work generalises 
Ref.~\cite{mccanna2021}, which focused on completely orthogonal and rigid vortex planes, and lays the groundwork for Ref.~\cite{mccannaequal}, which will explore whether a different configuration of tilted vortex planes can be energetically favoured under equal-frequency double rotation.

In more detail, we firstly showed in Sec.~\ref{sec:NonOrtho} that non-orthogonal vortex planes interact hydrodynamically, in contrast to the case of orthogonal planes, and we analytically derived the form of this interaction in the special cases of planes related by an isoclinic or simple rotation. Understanding this interaction potential in the general double rotation case would be an interesting topic for future work, allowing our analytics to extend to the full configuration space of non-orthogonal vortex planes.

Secondly, we approached the problem of a 4D hyperspherical superfluid doubly rotating at unequal frequencies, under the assumption that the vortices remain rigid planes. In Sec.~\ref{sec:Unequal:OnePlane} we showed that a single vortex plane in this system will always want to fully align with the higher of the two rotation frequencies. Then, using this result, we tackled the case of an intersecting vortex pair in Sec.~\ref{sec:Unequal:TwoPlanes}, proposing a non-orthogonal such pair as an ansatz for the ground state. This was based on the observation that for both planes to benefit from the higher frequency they had to be skewed in a purely aligning sense, thereby inducing repulsive vortex-vortex interaction from our result in Sec.~\ref{sec:NonOrtho}. We built an analytic model based simply on the balance between these two energies, which predicted that a skew configuration of vortex planes could indeed have lower energy than an orthogonal one. With this model we were able to find which of these configurations was optimal, and calculated the predicted tilt-angles and energy. Comparing these results with numerics, we found excellent agreement, despite the fact that we did not account for the avoided crossing of these states that was seen numerically, but which will be discussed further in Ref.~\cite{mccannaequal}. At high frequencies we also found that more exotic states with highly curved vortex surfaces could appear (see Appendix~\ref{app:numerical}), suggesting the ground state of a doubly rotating superfluid is in general very complex.

The results presented in this paper show that the physics of vortex surfaces in 4D can be incredibly rich, even in the absence of dynamics. The fact that curved and tilted vortex surfaces can be stable and exist at low energies in such a minimal model is a dramatic departure from the physics of lower dimensions under rotation, suggesting a vast configuration space to explore and investigate in the future.

\subsection{General Outlook}

In this paper, we have focused on a minimal theoretical model for a 4D superfluid based on the 4D GPE under rotation. This is motivated as the simplest extension of textbook 2D and 3D superfluids [c.f. Section~\ref{sec:vortexreview}] into higher spatial dimensions~\cite{mccanna2021}. However, in the future, it will be both interesting and relevant to go beyond this simple model to study more realistic systems with the aim of making an experimental proposal, and to explore the even richer vortex physics that will likely emerge. 

Recent interest in higher spatial dimensions has been sparked by various theoretical and experimental works aimed at exploring signatures of single-particle physics in artificial 4D systems, based e.g. on topological pumping~\cite{Thouless1983,verbin2013,Kraus2012Adiabatic,Kraus2012Fibonacci,Kraus2013,verbin2015,Lohse2016,Nakajima2016,Lohse2018,Zilberberg2018,ChengW2021,Chen2021}, ``synthetic dimensions"~\cite{Schreiber2010,Regensburger2011,Boada2012,Celi2014,Stuhl2015,Mancini2015,Luo2015,Gadway2015, wang2015,Livi2016,Meier2016,Ozawa2016,Yuan2016,Kolkowitz2017,Cardano2017,Ozawa2017,Wimmer2017,Martin2017,Signoles2017, Sundar2018,Wang2018,Chen2018,Baum2018,Peng2018,salerno2019quantized, viebahn2019matter,barbiero2019bose, Price2020,chalopin2020exploring,ChengD2021,Kanungo2022, ozawa2017synthetic, lustig2019photonic, Yuan2018,Yuan2019, yuan2020creating, Dutt2020, baum2018setting, chen2018experimental, cai2019experimental, wimmer2021superfluidity, price2019synthetic,crowley2019half,boyers2020exploring,lienhard2020realization, kang2020creutz,balvcytis2021synthetic,chen2021real,oliver2021bloch,englebert2021bloch, bouhiron2022realization}, artificial parameter spaces~\cite{sugawa2018second,lu2018topological, kolodrubetz2016measuring, wang2020exceptional,zhu2020four,palumbo2018revealing, chen2022synthetic} and the connectivity of classical electrical circuits~\cite{wang2020circuit,Price2018, yu2019genuine,li2019emergence,ezawa2019electric}. Of these schemes that of ``synthetic dimensions", in particular, may provide a way in the future to experimentally explore the physics of a 4D superfluid. 
In this general approach, a set of states or internal degrees of freedom are externally coupled together and then re-interpreted as lattice sites along an extra spatial dimension~\cite{Boada2012}. Such a synthetic dimension can then be combined with other real or synthetic dimensions to allow particles to explore a system with the desired effective dimensionality, such as e.g. four dimensions. Interest in this approach has grown dramatically in recent years, with significant theoretical and experimental progress in implementing synthetic dimensions across ultracold atoms~\cite{Stuhl2015,Mancini2015,Gadway2015,Meier2016,Livi2016,Price2017,Kolkowitz2017,An_2017,viebahn2019matter,chalopin2020exploring,bouhiron2022realization}, photonics~\cite{lustig2019photonic,Dutt2020,balvcytis2021synthetic,chen2021real, Schwartz2013,Luo2015,Ozawa2016,Yuan2016,Bell2017,Zhou2017,Cardano2017,Wang2018,Yuan2018photonics,Yuan2018rings,price2022roadmap}, and other systems~\cite{baum2018setting, price2019synthetic,crowley2019half,boyers2020exploring}. Within ultracold atoms, for example, synthetic dimension schemes have so far been realised based on using internal atomic states~\cite{Boada2012, Celi2014,Stuhl2015,Mancini2015,Chalopin2020, bouhiron2022realization}, momentum states~\cite{Gadway2015,Gadway2016,  viebahn2019matter}, harmonic trap states~\cite{Price2017, salerno2019quantized,oliver2021bloch}, orbital states~\cite{kang2020creutz}, superradiant states~\cite{cai2019experimental}, and Rydberg states~\cite{lienhard2020realization,Kanungo2022} amongst others. Of particular note, a recent experiment has realized a 4D atomic quantum Hall system made up of two synthetic dimensions of internal states as well as of two real dimensions~\cite{bouhiron2022realization}. Combining such a scheme with the inter-particle interactions necessary for superfluidity may open the way for the experimental investigation of 4D interacting states, such as 4D vortices, in the future.     

However, as we discussed in our previous paper~\cite{mccanna2021}, the 4D GPE that we have considered [Eq.~\eqref{eq:GPER}] is a toy model lacking elements which are necessary for experimental relevancy to current synthetic dimension approaches. For example, we have considered a purely hypothetical four-dimensional space that is isotropic and continuous, and we have chosen a hyperspherical hard-wall boundary to preserve the rotational symmetry. However, the motion,  inter-particle interactions, and boundary conditions along any synthetic dimensions can differ from that in real space. In practice, it is likely that an experiment may contain both real and synthetic dimensions, which would break \(SO(4)\) rotational symmetry. This will affect the behaviour of the tilted and curved vortex planes that we have studied, adding in additional physics that will compete with the rotational and hydrodynamic energies that we have considered. Additionally, most synthetic dimension implementations are discrete with hard-wall boundary conditions, and hence are best described by tight binding models on a lattice. It is then important to consider how many synthetic lattice sites are spanned by the typical length scales of the problem. If the answer is many, then a continuum approximation can be appropriate in the mean-field regime. If not, then a tight binding model must be used and rich physics can be expected to arise from competition between these length scales and the synthetic lattice spacing.

Moreover, synthetic dimensions can also have features that are rarely seen in typical tight binding models, and which in themselves warrant further research. These can include nonuniform hoppings, limited system sizes, non-equilibrium effects from external driving and long-range interactions along the synthetic dimension~\cite{Celi2014,Mancini2015}. All of these are details that should be considered to make this work more experimentally relevant but they also depend strongly on the experiment in question. For this reason, and for simplicity, we have studied a minimal extension of 3D superfluid physics into 4D, in order to begin investigating what is possible in higher dimensions. 

An obvious direction of future work is therefore to connect these results to experiment, by studying more complex models that take experimental details into account. We hope that such research can build upon our work by using similar techniques and ansatzes, and that more physical models will yield even richer behaviour. One simple modification that could still have interesting effects is to keep the continuous, isotropic 4D GPE model but to change the geometry to one which breaks the rotational symmetry and better reflects the boundary conditions in a synthetic dimension. A possible choice would be to pick out one or two directions as ``synthetic" and give them independent hard wall boundaries (i.e. \(w\in[-L,L]\), while retaining a rotationally symmetric geometry in the remaining coordinates.

There are many other interesting avenues for extending our research, aside from making the model more relevant to experiment. Our numerical stationary states with curved vortex surfaces raises the interesting possibility that other stationary states under rotation could contain closed vortex surfaces that do not meet the boundary of the system. These would be the four dimensional generalisation of vortex loops (including links and knots) in 3D~\cite{Proment2012,Proment2014,Scheeler2014,villois2017universal,villois2020,Proment2020}.
% , although vortex loops are not known to form stationary states.
Additionally, there is a far richer classification of closed surfaces~\cite{gallier2013} than of closed loops, suggesting there could be more possible closed vortex configurations in 4D.

It would also be interesting to study vortex surface configurations for even higher rotation frequencies. The presence of intersection, curvature, and avoided crossings in our vortex core results suggests that vortices can lose their individual character in 4D. It is therefore not entirely clear, even in some low energy stationary states, whether we can meaningfully assign an integer to the number of vortices in the system. In lower dimensions the number of vortices becomes very large in the rapidly rotating limit, where the vortices form an Abrikosov lattice~\cite{abo2001observation}. Investigating the limit of high frequency in one or both planes of rotation in 4D is therefore an interesting and open problem, due to the more malleable nature of the vortex core(s).

This work can also be extended to consider more interesting order parameters in 4D. Certain phases of spinor condensates in 3D are known to host non-Abelian vortices~\cite{Kawaguchi,Machon}, which have more interesting behaviour when they intersect and reconnect. Given that intersection and reconnection are also relevant for the behaviour of vortex planes, it is natural to ask what phenomena would arise for non-Abelian vortices in 4D. Finally, this work also represents a small step towards the strongly interacting fractional quantum Hall effect in 4D~\cite{Zhang2001, karabali2002quantum}, thanks to the analogy between a rotating superfluid and a quantum Hall system~\cite{cooper2008rapidly}.\\

{\it Acknowledgements:} We thank Tomoki Ozawa, Mike Gunn, Iacopo Carusotto, Mark Dennis, Davide Proment and Russell Bisset for helpful discussions. 
This work is supported by the Royal Society via grants UF160112, RGF\textbackslash{}EA\textbackslash{}180121 and RGF\textbackslash{}R1\textbackslash{}180071 and by the Engineering and Physical Sciences Research Council [grant number EP/W016141/1].

\appendix

\section{General rotation of a plane in 4D}
\label{app:TiltWLOG}

We want to derive the simplest rotation to describe a plane tilting in 4D without loss of generality. Consider the plane \(P\) defined in 4D Cartesian coordinates as the set of solutions to \(x=y=0\), and another plane \(P'\) as the image of \(P\) under a double rotation. We will represent \(P'\) as the set of solutions to \(x'=y'=0\), where the primed coordinates are related to the original coordinates by double rotation with matrix \(M\), that is, \(\mathbf{r}'=M\mathbf{r}\). It will be useful to write this in a block form such that
\begin{align}
    \begin{pmatrix}
         x' \\ y' \\ z' \\ w'
    \end{pmatrix}
    &=
    \begin{stretchPmatrix}{2}
        A & B \\
        C & D
    \end{stretchPmatrix}
    \begin{pmatrix}
         x \\ y \\ z \\ w
    \end{pmatrix},
    \label{eq:M}
\end{align}
where \(A, B, C, D\) are the \(2\times2\) blocks of \(M\). Rotations in 4D generally have six free parameters, but we can reduce this down to two for the matrix \(M\) by exploiting the symmetry of \(P\) under certain rotations, and by using our freedom to choose a basis. Firstly, using the following shorthand for a 2D rotation matrix
\begin{align*}
    R(\phi) =
    \rot{\phi},
\end{align*}
note that, for arbitrary \(\phi_{1,2}\), we can redefine \(M\) to be
\begin{equation}
    M =
    \begin{stretchPmatrix}{2}
        A & B \\
        C & D
    \end{stretchPmatrix}
    \begin{stretchPmatrix}{2}
        \Rp{1} & 0 \\
        0 & \Rp{2}
    \end{stretchPmatrix},
    \label{eq:M2}
\end{equation}
without changing \(P'\). The reason for this is that the initial rotation we have added is a double rotation in the \(x\en y\) and \(z\en w\) planes [c.f. Section~\ref{sec:4DRot}], which leaves the plane \(P\) invariant, such that the combined transformation results in the same transformed plane \(P'\). Secondly, we will use another double rotation in the \(x\en y\) and \(z\en w\) planes to change basis, as follows
\begin{align}
    &\begin{pmatrix}
         x \\ y \\ z \\ w
    \end{pmatrix}
    \to
    \begin{stretchPmatrix}{2}
        \Rp{3} & 0 \\
        0 & \Rp{4}
    \end{stretchPmatrix}
    \begin{pmatrix}
         x \\ y \\ z \\ w
    \end{pmatrix}.
    \label{eq:blockRot}
\end{align}
Denoting this matrix as \(R(\phi_3,\phi_4)\), we have that \(M \to R(\phi_3,\phi_4) M R(-\phi_3,-\phi_4)\) under this transformation. Combining this with the redefinition from Eq.~\eqref{eq:M2} we can write, for arbitrary \(\phi_j\), \(j=1,2,3,4\)
\begin{equation}
    M = 
    \begin{stretchPmatrix}{2}
        \Rp{3} & 0 \\
        0 & \Rp{4}
    \end{stretchPmatrix}
    \begin{stretchPmatrix}{2}
        A & B \\
        C & D
    \end{stretchPmatrix}
    \begin{stretchPmatrix}{2}
        \Rp{1} & 0 \\
        0 & \Rp{2}
    \end{stretchPmatrix},
    \label{eq:M3}
\end{equation}
without any loss of generality. Note that we have made the shifts \(\phi_1 \to \phi_1 + \phi_3\), and \(\phi_2 \to \phi_2 + \phi_4\) for simplicity. We can use these four free parameters to transform the upper left (\(A\)) and lower right (\(D\)) blocks into diagonal \(2\times 2\) matrices. To see this, start by expanding the product in Eq.~\eqref{eq:M3}
\begin{align}
    M &= {}
    \begin{stretchPmatrix}{2}
        \Rp{3} A \Rp{1} & \Rp{3} B \Rp{2} \\
        \Rp{4} C \Rp{1} & \Rp{4} D \Rp{2}
    \end{stretchPmatrix}.
\end{align}
Denoting the elements of \(A\) in the standard fashion
\begin{align}
    &A =
    \begin{pmatrix}
        a_{11} & a_{12} \\
        a_{21} & a_{22}
    \end{pmatrix},
\end{align}
and employing the shorthand $s_j = \sin\phi_j,$ $c_j = \cos\phi_j$, the off-diagonal elements of \(A' = \Rp{3} A \Rp{1}\) are given by
\begin{align}
    \left[A'\right]_{12} &= -a_{11}s_1c_3 + a_{12}c_1c_3 + a_{21}s_1s_3 - a_{22}c_1s_3, \\
    \left[A'\right]_{21} &= \hphantom{-}a_{11}c_1s_3 + a_{12}s_1s_3 + a_{21}c_1c_3 + a_{22}s_1c_3.
\end{align}
Setting these both to zero and taking the sum and difference of the two gives the following simultaneous equations
\begin{align*}
    (a_{11}-a_{22})\sin(\phi_3-\phi_1) + (a_{21}+a_{12})\cos(\phi_3-\phi_1) &= 0, \\
    (a_{11}+a_{22})\sin(\phi_3+\phi_1) + (a_{21}-a_{12})\cos(\phi_3+\phi_1) &= 0,
\end{align*}
which always have solutions for $\phi_{1,3}$. Similarly $R(\phi_4)DR(\phi_2)$ can be made diagonal by choosing particular values for $\phi_{2,4}$. We now look at the full transformed matrix \(M\) to see what form the off-diagonal blocks must take. The matrix now reads
\begin{align}
    M =
    \begin{pmatrix}
        a_1 & 0 & b_{11} & b_{12} \\
        0 & a_2 & b_{21} & b_{22} \\
        c_{11} & c_{12} & d_1 & 0 \\
        c_{21} & c_{22} & 0 & d_2,
    \end{pmatrix}
\end{align}
where $a_{1,2}$ and $d_{1,2}$ now denote the only non-zero elements of the upper left and lower right blocks after these blocks have been made diagonal. 
To proceed further, we will first focus on the upper right $B$ block. Normalization of the first two rows of \(M\) can be ensured, without loss of generality, by the following form
\begin{align}
    M = {}
    \begin{pmatrix}
        \cos\alpha_1 & 0 & -\sin\alpha_1\cos\beta_1 & -\sin\alpha_1\sin\beta_1 \\
        0 & \cos\alpha_2 & -\sin\alpha_2\cos\beta_2 & -\sin\alpha_2\sin\beta_2 \\
        c_{11} & c_{12} & d_1 & 0 \\
        c_{21} & c_{22} & 0 & d_2
    \end{pmatrix},
\end{align}
such that orthogonality of the first two rows now implies
\begin{align}
    \sin\alpha_1\sin\alpha_2\cos(\beta_1-\beta_2)=0.
\end{align}
This has $\sin\alpha_1=0$ or $\sin\alpha_2=0$ as special cases, which we ignore for now since these each lead to a simple rotation of the plane $P$ [c.f. Section~\ref{sec:4DRot}]. What we will derive instead is the general case for double rotation by requiring $\beta_2 = \beta_1 + \pi/2$, and this general case will actually include the simple rotation as a special case. Proceeding, we have
\begin{align}
    M = {}
    \begin{pmatrix}
        \cos\alpha_1 & 0 & -\sin\alpha_1\cos\beta_1 & -\sin\alpha_1\sin\beta_1 \\
        0 & \cos\alpha_2 & \hphantom{-}\sin\alpha_2\sin\beta_1 & -\sin\alpha_2\cos\beta_1 \\
        c_{11} & c_{12} & d_1 & 0 \\
        c_{21} & c_{22} & 0 & d_2
    \end{pmatrix}.
\end{align}
Orthogonality of the last two columns gives
\begin{align}
    \left(\sin^2\alpha_1-\sin^2\alpha_2\right)\cos\beta_1\sin\beta_1=0.
\end{align}
Again, we have a special case, given by $\alpha_2=\alpha_1$, which will give an isoclinic rotation of the plane $P$ [c.f. Section~\ref{sec:4DRot}]. We will ignore this solution for now, and again find that it can be found as a particular case of the remaining solution. We therefore require either $\cos\beta_1=0$ or $\sin\beta_1=0$. This leads to the following two forms
\begin{align}
    M = {}
    \begin{pmatrix}
        \cos\alpha_1 & 0 & 0 & -\sin\alpha_1 \\
        0 & \cos\alpha_2 & \hphantom{-}\sin\alpha_2 & 0 \\
        c_{11} & c_{12} & d_1 & 0 \\
        c_{21} & c_{22} & 0 & d_2
    \end{pmatrix}, \\
    M = {}
    \begin{pmatrix}
        \cos\alpha_1 & 0 & -\sin\alpha_1 & 0 \\
        0 & \cos\alpha_2 & 0 & -\sin\alpha_2 \\
        c_{11} & c_{12} & d_1 & 0 \\
        c_{21} & c_{22} & 0 & d_2
    \end{pmatrix},
\end{align}
respectively, up to an unimportant common sign in the upper right block which can be absorbed in to the definition of \(\alpha_1\) and \(\alpha_2\). Furthermore, these two forms are related to each other by a change of basis and redefinition of parameters. We therefore choose the second form without loss of generality. Orthonormality of the columns and the last two rows now allows us to determine the remaining unknowns, such that we finally have
\begin{align}
    \begin{pmatrix}
        \cos\alpha_1 & 0 & -\sin\alpha_1 & 0 \\
        0 & \cos\alpha_2 & 0 & -\sin\alpha_2 \\
        \hphantom{-}\sin\alpha_1 & 0 & \cos\alpha_1 & 0 \\
        0 & \hphantom{-}\sin\alpha_2 & 0 & \cos\alpha_2
    \end{pmatrix},
    \label{eq:TiltWLOG}
\end{align}
as is used in the main text.

\begin{figure*}[t!]
% \begin{figure*}
    \centering
    \includegraphics[width=0.9\textwidth]{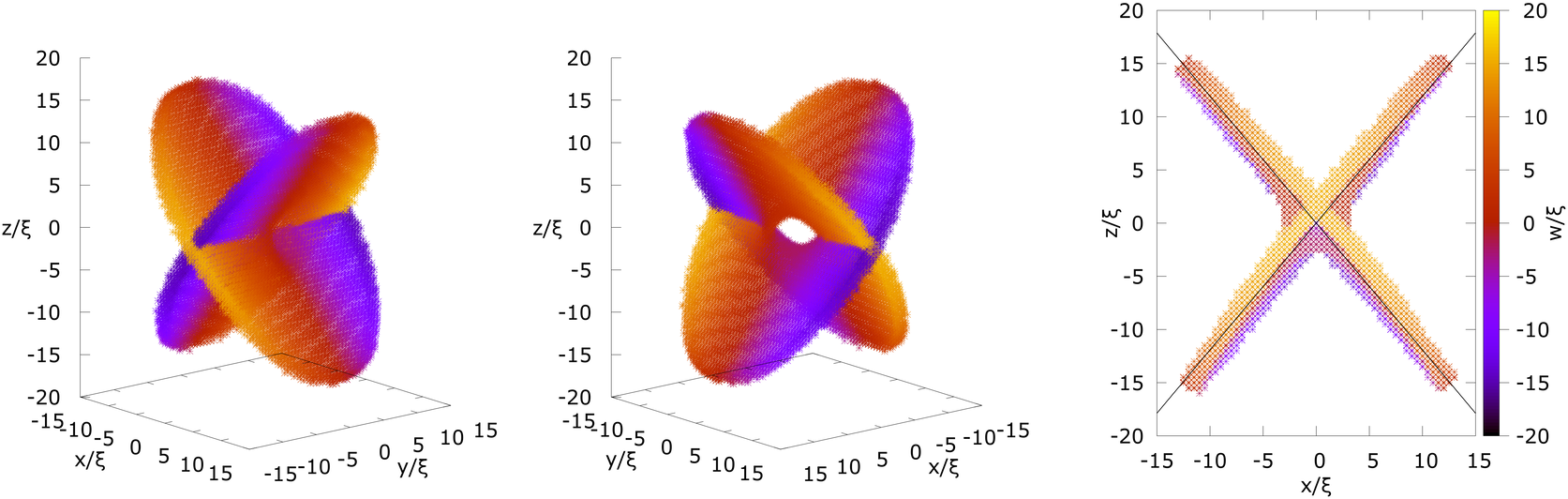}
    \caption{As in Fig~\ref{fig:40Skew}, except with rotation frequencies of \(\Omega_{zw}=1.25\Omega_c\) and \(\Omega_{xy}\approx1.35\Omega_c\), corresponding to a predicted skewness of \(\eta=10\degree\). The overall structure is essentially the same as in Fig~\ref{fig:40Skew}, and the agreement with theory is still very good as shown in panel three. However, this panel also shows that the data has a slight mirror asymmetry in the vertical axis that is not visibly present in Fig~\ref{fig:40Skew} or accounted for in the theory.}
    \label{fig:10Skew}
\end{figure*}

\section{Integration in skew double polar coordinates}
\label{app:IntSkew}
\renewcommand{\cpt}{\cos\left(\ta'_2-\nu\ta_1\right)}
\renewcommand{\spt}{\sin\left(\ta'_2-\nu\ta_1\right)}
In this appendix we derive an integral transformation from 4D Cartesian coordinates (\(x, y, z, w\)) into a non-orthogonal coordinate system given by (\(x, y, z', w')\), where the primed coordinates form another Cartesian framed related to the unprimed one by a double rotation and apply this to the vortex-vortex interaction energy. 

If we are only interested in preserving the relationship between the two planes defined by \(x=y=0\) and \(z'=w'=0\) respectively, then without loss of generality we can choose this double rotation to have the form
\begin{align}
    \begin{pmatrix}
        x' \\ y' \\ z' \\ w'
    \end{pmatrix} = 
    \begin{pmatrix}
        \cos\alpha_1 & 0 & -\sin\alpha_1 & 0 \\
        0 & \cos\alpha_2 & 0 & -\sin\alpha_2 \\
        \sin\alpha_1 & 0 & \cos\alpha_1 & 0 \\
        0 & \sin\alpha_2 & 0 & \cos\alpha_2
    \end{pmatrix}
    \begin{pmatrix}
        x \\ y \\ z \\ w
    \end{pmatrix},
\end{align}
as derived in Appendix~\ref{app:TiltWLOG}. As discussed in the main text, here we will only deal with the special case where this double rotation is isoclinic, such that \(\alpha_2 = \nu\alpha_1\), with \(\nu = \pm 1\). From here on we will employ the shorthand \(c=\cos\alpha_1\), \(s=\sin\alpha_1\). We will derive this integration over non-orthogonal coordinates for the case of a 4D ball of unit radius, since this geometry preserves the symmetry between the primed and unprimed coordinates. The primed coordinates can be introduced into the integral using Dirac deltas as follows
\begin{widetext}
\begin{align}
    \dint_{B^4(1)}\diff{x}\diff{y}\diff{z}\diff{w} &= \dint_{B^4(1)}\diff{z}\diff{w}\diff{x}\diff{y}
    \dint_{\mathbb{R}^2}\diff{z'}\diff{w'}\delta(z'-cz-sx)\delta(w'-cw-\nu sy) \nonumber \\
    &= \frac{1}{c^2} \dint_{B^2(1)}\diff{x}\diff{y} \dint_{\mathbb{R}^2}\diff{z'}\diff{w'} \dint_{B^2(R)}\diff{z}\diff{w}  \delta\left(z-\frac{z'-sx}{c}\right) \delta\left(w-\frac{w'-\nu sy}{c}\right),
    \label{eq:deltas}
\end{align}
\end{widetext}
where \(B^d(R)\) is the ball of radius \(R\) centred at the origin in \(d\) dimensions, and here \(R^2=1^2-x^2-y^2\).

Our goal now is to eliminate \(z\) and \(w\) by evaluating the corresponding integrals. This in turn will define the limits of integration for their primed counterparts. However, this is more easily accomplished in double polar coordinates, whereby
\begin{align}
    \dint_{B^2(1)}\diff{x}\diff{y} = &\dint_0^{1}r_1\diff{r_1}\cintm\diff{\ta_1}, \
    \dint_{\mathbb{R}^2}\diff{z'}\diff{w'} = \dint_0^\infty r_2'\diff{r_2'}\cintm\diff{\ta'_2}, \\
    \dint_{B^2(R)}&\diff{z}\diff{w} \delta\left(z-z_0\right)\delta\left(w-w_0\right) = \nonumber \\ 
    &\dint_0^{R} r_2\diff{r_2}\cintm\diff{\ta_2}\frac{1}{r_2}\delta\left(r_2-r_0\right) \delta\left(\ta_2-\ta_0\right),
\end{align}
where we have defined \(z_0=(z'-sx)/c\), \(w_0=(w'-\nu sy)/c\), \(r_0^2=z_0^2+w_0^2\), and \(\ta_0=\arctan(z_0,w_0)\).
% \begin{align}
%     \delta\left(z-z_0\right)&\delta\left(w-w_0\right) = 
%       \frac{1}{r_2}\delta\left(r_2-r_0\right) \delta\left(\ta_2-\ta_0\right),  
%     % \frac{1}{r_2}\delta\left[r_2-\left(z_0^2+w_0^2\right)^{\frac{1}{2}} \right] \delta\left[\ta_2-\arctan(z_0,w_0)\vphantom{\left(z_0^2+w_0^2\right)^{\frac{1}{2}}} \right],
% \end{align}
We now integrate out \(r_2\) and \(\ta_2\) as follows
\begin{align}
    \label{eq:Theta}
    &\dint_0^{R}\diff{r_2} \delta\left(r_2-r_0 \right) = \Ta\left(R^2 - r_0^2 \right), \\
    &\cintm\diff{\ta_2} \delta\left(\ta_2-\ta_0\right) = 1,
\end{align}
where we have used that \(\Ta(R-r_0) = \Ta(R^2-r_0^2)\) since both \(R\) and \(r_0\)  are non-negative. In double polar coordinates we have that \(R^2 = 1 - r_1^2\), and \(r_0^2 = [r_2'^2 + s^2r_1^2 - 2sr_1r_2'\cpt]/c^2\). Substituting this into the Theta function on the RHS of Eq.~\eqref{eq:Theta} gives
\begin{align}
    \Ta(R^2-r_0^2) = \Ta\left[c^2-r_1^2-r_2'^2+2sr_1r_2'\cpt\right],
\end{align}
where we have used that \(\Ta(c^2\ \cdot)=\Ta(\cdot)\). Altogether this gives
\begin{widetext}
\begin{align}
  \dint_{B^4(1)}\diff{x}\diff{y}\diff{z}\diff{w} = \frac{1}{c^2}\dint_0^{1} r_1\diff{r_1} \cintm\diff{\ta_1} \dint_0^1 r_2'\diff{r_2'} \cintm \diff{\ta'_2} \Ta\left[c^2-r_1^2-r_2'^2+2sr_1r_2'\cpt\right],
  \label{eq:IntSkew}
\end{align}
\end{widetext}
where we have also used the spherical symmetry to restrict the upper limit of \(r_2'\) to \(1\), by comparison to that of \(r_1\). (This is unnecessary, since the step function will ultimately control the limits of whichever radius is integrated over first, but it makes the equivalence between the primed and unprimed coordinates fully clear.)

From now we will assume that the primed coordinates will be integrated over first, so let us make the substitution \(\ta_2' = \intTheta + \nu\ta_1\), treating \(\ta_1\) as a constant within the \(\ta_2'\) integral, in order to simplify the cosine. The limits of the \(\intTheta\) integral will be \((-\nu\ta_1-\pi,-\nu\ta_1+\pi)\), but this is arbitrary since we are integrating over a full circle, so we can just as easily write \(\intTheta\in(-\pi,\pi)\). In order to figure out exactly how the step function translates into integration limits, consider the inequality it enforces
\begin{align}
    c^2 - r_1^2 - r_2'^2 + 2sr_1r_2'\cos\intTheta > 0.
    \label{inqeq:Heaviside}
\end{align}
This form is ideal for integrating over \(\intTheta\) first, but it will actually be easier to integrate over \(r_2'\) first. For this reason we will rewrite Eq.~\eqref{inqeq:Heaviside} by completing the square for \(r_2'\) as
\begin{align}
    \left(r_2' - sr_1\cos\intTheta\right)^2 < c^2\left(1 - r_1^2\right) - s^2r_1^2\sin^2\intTheta
    \label{ineq:r_2'}
\end{align}
This inequality has no solutions for \(r_2'\) where the RHS is negative, so we immediately obtain
\begin{equation}
    \abs{\sin\intTheta} < \frac{ c\left(1-r_1^2\right)^{\frac{1}{2}} }{sr_1},
    \label{ineq:ta_-}
\end{equation}
as a constraint for \(\intTheta\). Note that this constraint is trivially satisfied whenever \(c\left(1-r_1^2\right)^{1/2} > sr_1\), which occurs when \(r_1 < c\).  Given this condition for \(\intTheta\), we can then satisfy the inequality~\eqref{ineq:r_2'} when \(r_2'\in(r_-,r_+)\), where
\begin{align}
    r_\pm = sr_1\cos\intTheta \pm \left[c^2\left(1 - r_1^2\right) - s^2r_1^2\sin^2\intTheta\right]^{\frac{1}{2}}.
    \label{eq:r_pm}
\end{align}
The last step is to enforce the constraint \(r_\pm \geq 0\), since \(r_2'\) cannot be negative. Rearranging each inequality gives
\begin{align}
    r_- \geq 0 \quad &\iff \quad \hphantom{-}\cos\intTheta \geq 
    \left(\thetaFrac^2-\sin^2\intTheta\right)^{\frac{1}{2}}, \\
    r_+ \geq 0 \quad &\iff \quad -\cos\intTheta \leq \left(\thetaFrac^2-\sin^2\intTheta\right)^{\frac{1}{2}},
\end{align}
% so \(\ta \in [-\pi/2,\pi/2]\) is a necessary condition for \(r_- \geq 0\), and a sufficient one for \(r_+ \leq 0\).
where \(\thetaFrac=c\left(1-r_1^2\right)^{1/2}/sr_1\). Note that the quantity on the RHS of both of these inequalities is always non-negative, so \(r_- \geq 0\) requires \(\ta \in [-\pi/2,\pi/2]\), while \(r_+ \geq 0\) is automatically satisfied in this same region. With this consideration of the sign of the LHS in mind, we can square both inequalities and rearrange to find
\begin{align}
    r_- \geq 0&: \quad 1 \geq \thetaFrac^2 \onlyif r_1^2 \geq c^2, \\
    r_+ \geq 0&: \quad 1 \leq \thetaFrac^2 \onlyif r_1^2 \leq c^2,
    \label{eq:r_pmNonNeg}
\end{align}
Combining all of this with the inequality~\eqref{ineq:ta_-}, gives us two separate integration regions. We have
\begin{align}
    r_1 \in (0,c), &\quad r_2'\in(0,r_+), \quad \intTheta\in(-\pi,\pi),
\end{align}
and
\begin{align}
    r_1 \in (c,1), &\quad r_2'\in(r_-,r_+) \quad \intTheta\in(-\intTLim,\intTLim),
\end{align}
where \(\intTLim = \arcsin{\thetaFrac}\). Finally, we can write the full result as
\begin{widetext}
\begin{align}
    \dint_{B^4(1)}\diff{x}\diff{y}\diff{z}\diff{w} = \frac{1}{c^2} \dint_0^c r_1\diff{r_1} \cintm\diff{\ta_1} \cintm\diff{\intTheta} \dint_0^{r_+}r_2'\diff{r_2'} + \frac{1}{c^2}\dint_c^1 r_1\diff{r_1} \cintm\diff{\ta_1} \dint_{-\intTLim}^{\intTLim} \diff{\intTheta} \dint_{r_-}^{r_+}r_2'\diff{r_2'}.
\end{align}
\end{widetext}
As stated in the main text, the vortex-vortex interaction energy is then given as
\begin{align}
    E_{\text{vv}} &= A' \frac{s}{c^2}\dint_0^c \diff{r_1} \cintm\diff{\intTheta} \dint_0^{r_+}\diff{r_2'} \cos\intTheta \nonumber \\
    &+ A' \frac{s}{c^2}\dint_c^1 \diff{r_1} \dint_{-\intTLim}^{\intTLim} \diff{\intTheta} \dint_{r_-}^{r_+}\diff{r_2'} \cos\intTheta \\
    &= A' \left(J_1 + J_2\right),
    \label{eq:E_int}
\end{align}
where we have introduced $J_1$ and $J_2$ as shorthand to denote the two integrals. 
We will now deal with each of these integrals separately; focusing on the first term, we have
\begin{align}
    % J_1 &= \dint_0^c \diff{r_1} \cintm\diff{\ta_1} \cintm\diff{\intTheta}\cos\intTheta \dint_0^{r_+}\diff{r_2'} \\
    J_1 \equiv& \frac{s}{c^2}\dint_0^c \diff{r_1} \cintm\diff{\intTheta}\cos\intTheta r_+ = \frac{s}{c^2}\dint_0^c \diff{r_1} \cintm\diff{\intTheta} sr_1\cos^2\intTheta  \nonumber \\
    &\hphantom{=} + \frac{s}{c^2} \dint_0^c \diff{r_1} \cintm\diff{\intTheta}\cos\intTheta\left[c^2\left(1 - r_1^2\right) - s^2r_1^2\sin^2\intTheta\right]^{\frac{1}{2}},
\end{align}
where we have carried out the integral over $r_2'$.
The second integral on the RHS of this equation can be shown to vanish as follows
% \begin{align}
%     &\cintm \diff{\intTheta} \cos\intTheta F(\sin^2\intTheta) \\
%     &= \left( \dint_{-\pi}^{-\pi/2} + \dint_{-\pi/2}^{\pi/2} + \dint_{\pi/2}^{\pi} \right) \diff{\intTheta}\cos\intTheta F(\sin^2\intTheta), \\
%     &= \dint_{0}^{\pi/2} \diff{\ta}\cos(\intTheta-\pi) F\left[ \sin^2\left(\intTheta+\pi\right) \right], \\
%     &\hphantom{=} + \dint_{-\pi/2}^{\pi/2} \diff{\ta}\cos(\intTheta) F(\sin^2\intTheta) \\
%     &\hphantom{=} + \dint_{-\pi/2}^{0} \diff{\ta}\cos(\intTheta+\pi) F\left[ \sin^2\left(\intTheta+\pi\right) \right] \\
%     &= \left( -\dint_{0}^{\pi/2} + \dint_{-\pi/2}^{\pi/2} - \dint_{-\pi/2}^{0} \right) \diff{\intTheta}\cos\intTheta F(\sin^2\intTheta), \\
%     &= 0,
% \end{align}
\begin{widetext}
\begin{align}
    &\cintm \diff{\intTheta} \cos\intTheta F(\sin^2\intTheta) = \left( \dint_{-\pi}^{-\frac{\pi}{2}} + \dint_{-\frac{\pi}{2}}^{\frac{\pi}{2}} + \dint_{\frac{\pi}{2}}^{\pi} \right) \diff{\intTheta}\cos\intTheta F(\sin^2\intTheta), = \dint_{0}^{\frac{\pi}{2}} \diff{\ta}\cos(\intTheta-\pi) F\left[ \sin^2\left(\intTheta-\pi\right) \right] \nonumber \\
    &+ \dint_{-\frac{\pi}{2}}^{\frac{\pi}{2}} \diff{\ta}\cos(\intTheta) F(\sin^2\intTheta) + \dint_{-\frac{\pi}{2}}^{0} \diff{\ta}\cos(\intTheta+\pi) F\left[ \sin^2\left(\intTheta+\pi\right) \right] = \left( -\dint_{0}^{\frac{\pi}{2}} + \dint_{-\frac{\pi}{2}}^{\frac{\pi}{2}} - \dint_{-\frac{\pi}{2}}^{0} \right) \diff{\intTheta}\cos\intTheta F(\sin^2\intTheta)= 0,
\end{align}
\end{widetext}
which works for any arbitrary function \(F\). This leaves us with
\begin{align}
    J_1 &= \frac{s^2}{c^2}\dint_0^c r_1 \diff{r_1} \cintm\diff{\intTheta} \cos^2\intTheta = \frac{\pi}{2}s^2.
\end{align}
We now turn to the second term of Eq.~\eqref{eq:E_int}, which depends on
\begin{align}
    % J_2 &= 2\pi\dint_c^1 \diff{r_1} \dint_{-\intTLim}^{\intTLim} \diff{\intTheta} \cos\ta \dint_{r_-}^{r_+}\diff{r_2'}, \\
    J_2 &= \frac{s}{c^2} \dint_c^1 \diff{r_1} \dint_{-\intTLim}^{\intTLim}\diff{\intTheta} \cos\intTheta\left(r_+ - r_-\right), \nonumber \\
    &= \frac{2s}{c^2} \dint_c^1 \diff{r_1} \dint_{-\intTLim}^{\intTLim}\diff{\intTheta} \cos\intTheta\left[c^2\left(1 - r_1^2\right) - s^2r_1^2\sin^2\intTheta\right]^{\frac{1}{2}}.
\end{align}
This has the form of the vanishing term in \(J_1\), except that the \(\intTheta\) limits now do not cover a full period. In fact, the limits do not cover even half a period since \(\intTLim\leq\pi/2\) (consider Fig~\ref{fig:regions} with the blue dotted circle passing through the origin), with the consequence being that this term now contributes. To compute it we will apply the substitution \(sr_1\sin\intTheta=c\left(1-r_1^2\right)^{1/2}\sin{u}\), to give
\begin{align}
    J_2 &= 2 \dint_c^1 \diff{r_1}\frac{\left(1-r_1^2\right)}{r_1} \dint_{-\pi/2}^{\pi/2}\diff{u} \cos^2u \nonumber \\
    &= -\pi\ln c - \frac{\pi}{2}s^2.
\end{align}
Combining these results then gives the final result
\begin{align}
    E_{\text{vv}} =  -4k_1k_2\nu N\frac{\hbar^2}{mR^2} \ln\left(\cos\eta\right).
    \label{eq:Eint2}
\end{align}
as stated in the main text. 

\begin{figure*}[t!]
% \begin{figure*}
    \centering
    \includegraphics[width=0.9\textwidth]{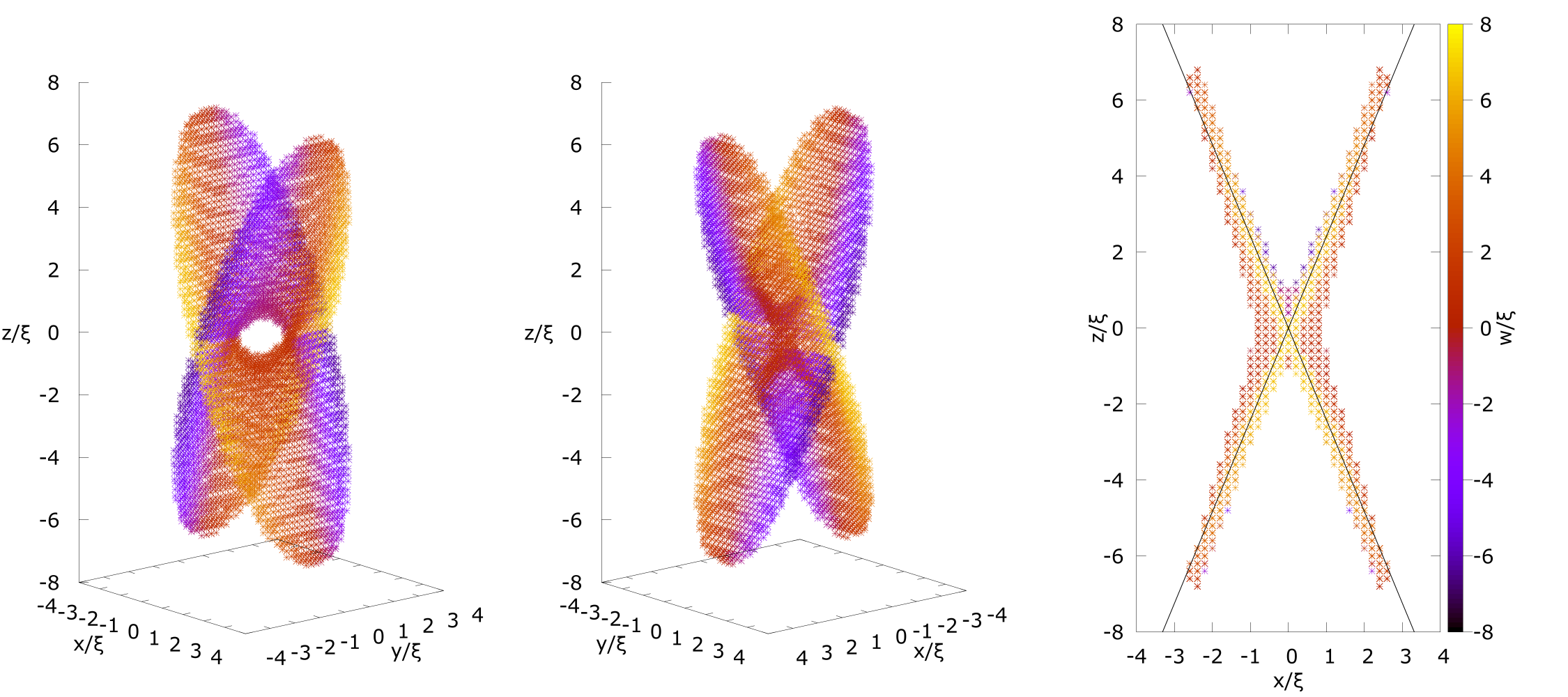}
    \caption{As in Fig~\ref{fig:10Skew}, except with a higher resolution of \(\Delta x=0.2\xi\), and with \(\Omega_{xy}\approx2.25\Omega_c\), corresponding to a predicted skewness of \(\eta=45\degree\). The agreement with the theoretical planes is still excellent, as seen in the third panel, and there is no visible asymmetry about the vertical axis. Interestingly, the avoided crossing at the origin has a different orientation to that seen in Figs~\ref{fig:40Skew} and \ref{fig:10Skew}, and appears to be smaller.}
    \label{fig:HiResSkew}
\end{figure*}

\section{The minimum distance between two circles of common centre and radius in 4D}
\label{app:circles}

    Consider a pair of circles \((C_1,C_2)\) in \(\mathbb{R}^4\) with the same radius and centre, but occupying different planes, and let these two planes be related by an isoclinic rotation. In this appendix we will derive an expression for the minimum distance between two such circles. Without loss of generality, we may encode the two circles in the following vector equations:
    \begin{align}
        \bm{r}_{C_1} &= R\left(\cos{\ta_1}\bfhat{x} + \sin{\ta_1}\bfhat{y}\right) \\
        \bm{r}_{C_2} &= R\left[\cos{\ta_2'}\left(c\bfhat{z}-s\bfhat{x}\right) + \sin{\ta_2'}\left(c\bfhat{w}-\nu s\bfhat{y}\right)\right].
    \end{align}
    The vector between an arbitrary point on \(C_1\) and an arbitrary point on \(C_2\) is given by \(\bm{d} = \bm{r}_{C_1} - \bm{r}_{C_2}\). All we have to do is compute the length of this vector and minimize it with respect to \(\ta_1\) and \(\ta_2'\). Evaluating the modulus squared of \(d\), we have
    \begin{align}
        % d^2 &= R^2\left[ \left(\cos\ta_1+s\cos\ta_2'\right)^2 + \left(\sin\ta_1+\nu s\sin\ta_2'\right)^2 + c^2\cos\ta_2' + c^2\sin\ta_2' \right] \nonumber \\
        d^2 &= 2R^2\left[ 1 + s\left(\cos\ta_1\cos\ta_2' + \nu\sin\ta_1\sin\ta_2' \right) \right] \nonumber \\
        d &= \sqrt{2}R\left[ 1 + s\cos\left(\ta_1-\nu\ta_2'\right)\right]^{\frac{1}{2}}
    \end{align}
    The minimum value of \(d\) is therefore \(\sqrt{2}R(1-s)^{1/2}\), which occurs when \(\ta_1 = \nu\ta_2' + \pi\).

\section{Evaluation of rotational energy integrals}
\label{app:angular}

As stated in the main text, in order to calculate the rotational energy of two vortex planes under unequal frequency double rotation, we must compute the integral (Eq.~\eqref{eq:integrand}):
\begin{align}
    \dint_{B^4(R)} \frac{\hat{L}_{xy} e^{i\act_1}}{e^{i\act_1}} \diff^4 r,
\end{align}
as well as the corresponding integral for \(e^{i\grt_2}\). Here, we will only show the direct calculation of the first integral, since the second follows identical logic. To begin, we will consider acting with $\hat{L}_{xy} \equiv - i\hbar\partial_{\theta_1}$ on Eq.~\eqref{eq:acute1} as
\begin{align}
    \frac{\hat{L}_{xy}\acr_1e^{i\act_1}}{\acr_1e^{i\act_1}} &= \frac{\hbar\cos\eta_1\cpolar{1}}{\cos\eta_1\cpolar{1} + e^{i\varphi}\sin\eta_1\cpolar{2}},  \end{align}
where we have also divided through by  Eq.~\eqref{eq:acute1}.    
Then by using the product rule, we can see that the desired integrand in Eq.~\eqref{eq:integrand} can be expressed as %   
\begin{align}    
    \frac{\hat{L}_{xy} e^{i\act_1}}{e^{i\act_1}} &= \frac{\hbar r_1e^{i(\ta_1-\ta_2-\varphi)}}{r_1e^{i(\ta_1-\ta_2-\varphi)} + \tan\eta_1r_2} - \frac{\hat{L}_{xy}\acr_1}{\acr_1}.
\end{align}
%
% \begin{align}
%     \grr_2e^{i\grt_2} &= -e^{i\varphi}\sin\eta_2\cpolar{1} + \cos\eta_2\cpolar{2} \\
%     \frac{L_{xy}\grr_2e^{i\grt_2}}{\grr_2e^{i\grt_2}} &= \frac{-e^{i\varphi}\sin\eta_2\cpolar{1}}{-e^{i\varphi}\sin\eta_2\cpolar{1} + \cos\eta_2\cpolar{2}} \\
%      e^{-i\grt_2}L_{xy}e^{i\grt_2} &= \frac{-e^{i\varphi}\tan\eta_2\cpolar{1}}{-e^{i\varphi}\tan\eta_2\cpolar{1} + \cpolar{2}} - \frac{L_{xy}\grr_2}{\grr_2}
% \end{align}
The second term can be shown to integrate to zero as follows
\begin{align}
    \cintm\diff\ta_1 \frac{\hat{L}_{xy}\acr_1}{\acr_1} &= -i\hbar\cintm\diff\ta_1 \acr_1^{-1}\frac{\partial\acr_1}{\partial\ta_1}  \nonumber \\
    &= -i\hbar\Big[\ln\acr_1\Big]_{-\pi}^{\pi} = 0.
\end{align}
In terms of the integral over $\theta_1$, we are then left with the following
\begin{align}
    \cintm \diff\ta_1 \frac{\hat{L}_{xy} e^{i\act_1}}{e^{i\act_1}} &= \hbar\cintm\diff{\ta_1} \frac{r_1e^{i(\ta_1-\ta_2-\varphi)}}{r_1e^{i(\ta_1-\ta_2-\varphi)} + \tan\eta_1r_2},
\end{align}
which can be evaluated as a contour integral in the complex plane. Setting \(\zeta = e^{i\left(\ta_1-\ta_2-\varphi\right)}\), such that \(\diff\zeta = ie^{i\left(\ta_1-\ta_2-\varphi\right)}\diff\ta_1\), we have
\begin{align}
    \cintm \diff\ta_1 \frac{\hat{L}_{xy} e^{i\act_1}}{e^{i\act_1}} &= \hbar\dint_{\abs{\zeta}=1} \frac{-ir_1\diff{\zeta}}{r_1\zeta + \tan\eta_1r_2}\nonumber \\
    &= 2\pi\hbar\Ta\left(r_1-\tan\eta_1r_2\right).
\end{align}
The integral over \(\ta_2\) then simply gives another factor of \(2\pi\). What is left is a fairly straightforward double integral over the two polar radii
\begin{align}
    \dint_{B^4(R)} \diff^4 r \frac{\hat{L}_{xy} e^{i\act_1}}{e^{i\act_1}} &= 4\hbar\pi^2\dint_0^R r_2\diff{r_2} \smashoperator{\dint_0^{\sqrt{R^2-r_2^2}}} r_1\diff{r_1} \Ta(r_1-\tan\eta_1r_2),
\end{align}
where the limits reflect that the 4D hypersphere is bounded by $r_1^2+r_2^2 = R^2$. 
Since both \(r_j\) are non-negative we can safely rewrite the step function as \(\Ta(r_1^2-\tan^2\eta_1r_2^2)\), which allows us to make the substitutions \(u_j = r_j^2/R^2\), such that the integral then becomes
\begin{align}
    \dint_{B^4(R)} \diff^4 r \frac{\hat{L}_{xy} e^{i\act_1}}{e^{i\act_1}} &=     \hbar\pi^2R^4\dint_0^1 \diff{u_2} \smashoperator{\dint_0^{1-u_2}} \diff{u_1} \Ta(u_1-\tan^2\eta_1u_2).
\end{align}

\begin{figure*}[t!]
% \begin{figure*}
    \centering
    \includegraphics[width=0.9\textwidth]{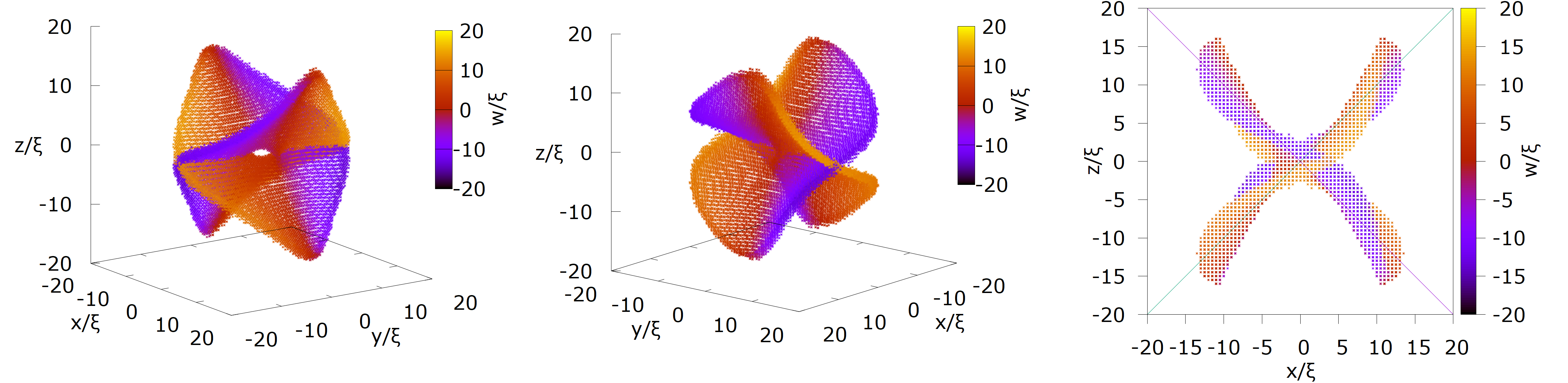}
    \includegraphics[width=0.9\textwidth]{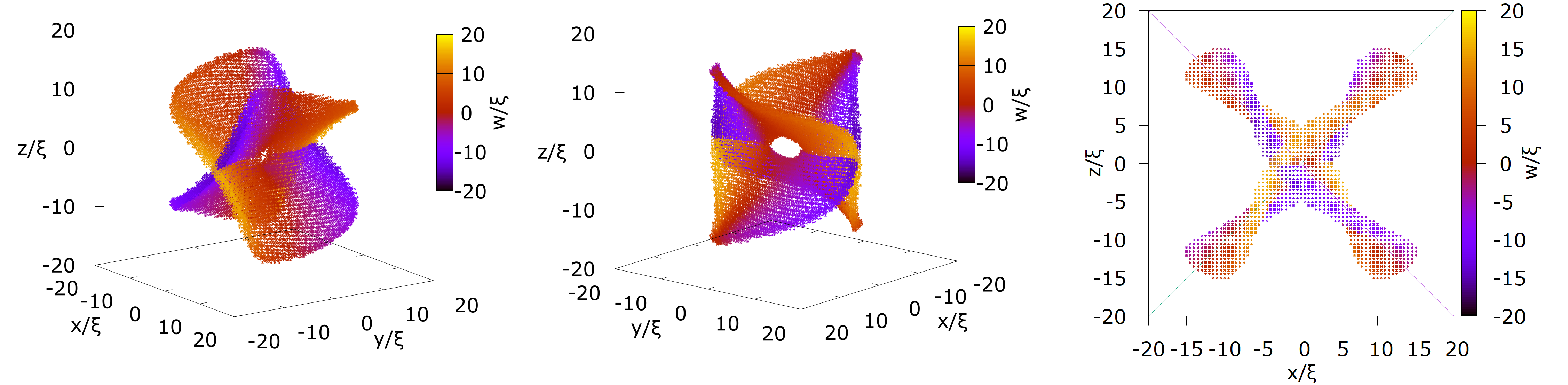}
    \caption{(Top and bottom) Two different unusual vortex core structures observed in final states of ITEM runs with the same parameters. The frequencies were set at the isoclinic point, \(\Omega_{zw}=\Omega_{xy}=1.3\Omega_c\), and the spatial resolution was equal to \(\Delta x=0.5\xi\), corresponding to  system radius of \(R\approx20.6\xi\). These simulations were the final iteration in a series of ITEM runs similar to those that generated Fig~\ref{fig:TiltEnergyNumerics}. That is, we fixed \(\Omega_{zw}\), and varied \(\Delta\Omega\) starting from \(\Omega_c\) and stepping down to the isoclinic point in units of \(0.1\Omega_c\). We used the final state of each run as the initial state of the next. These two states started out very close to the predicted skew planes at high frequency difference, however as we decreased  \(\Delta\Omega\) these planes began to increasingly curve, up to the maximum amount shown in this figure at \(\Delta\Omega=0\). These states both have energy given by \(E\approx0.65\mu N\), which is approximately equal to the orthogonal state energy at these frequencies, so we would need higher precision to tell if these are degenerate with the orthogonal state or if they are metastable excited states.}
    \label{fig:WeirdThings}
\end{figure*}

It is now much easier to compare the step function to the integration limits, as we have that \(u_1 > \tan^2\eta_1u_2\) and \(0<u_1<1-u_2\), whilst \(0<u_2<1\). Clearly \(\tan^2\eta_1u_2\) is non-negative, so we can make this value the new lower limit for \(u_1\) provided it is not greater than the upper limit of \(1-u_2\). This will be true for a certain range of \(u_2\) values which satisfy the inequality
\begin{align}
    \tan^2\eta_1u_2 &\leq 1-u_2, \\
    u_2 &\leq \cos^2\eta_1,
\end{align}
in which case the above integral including the step function is equivalent to
\begin{align}
    \dint_{B^4(R)} \diff^4 r \frac{\hat{L}_{xy} e^{i\act_1}}{e^{i\act_1}} &=     \hbar\pi^2R^4\smashoperator{\dint_0^{c^2}} \diff{u_2} \smashoperator{\dint_{t^2u_2}^{1-u_2}} \diff{u_1},
\end{align}
where \(c\) and \(t\) are shorthand for \(\cos\eta_1\) and \(\tan\eta_1\), respectively. Evaluating this, we obtain
\begin{align}
    \dint_{B^4(R)} \diff^4 r \frac{\hat{L}_{xy} e^{i\act_1}}{e^{i\act_1}} &=     \hbar\pi^2R^4 \dint_0^{c^2} \diff{u_2} \left(1-\frac{u_2}{c^2}\right) \nonumber \\
    &= \hbar\frac{\pi^2}{2}R^4 \cos^2\eta_1,
\end{align}
as stated in the main text. As the calculation for the \(e^{i\grt_2}\) term follows identical logic, we simply state the result as
\begin{align}
    \dint_{B^4(R)} \diff^4 r \frac{\hat{L}_{xy} e^{i\grt_2}}{e^{i\grt_2}} &= \hbar\frac{\pi^2}{2}R^4 \sin^2\eta_2.  
\end{align}

    \begin{figure*}[t!]
    \centering
    \begin{minipage}{0.45\textwidth}
        \centering
        \includegraphics[width=\textwidth]{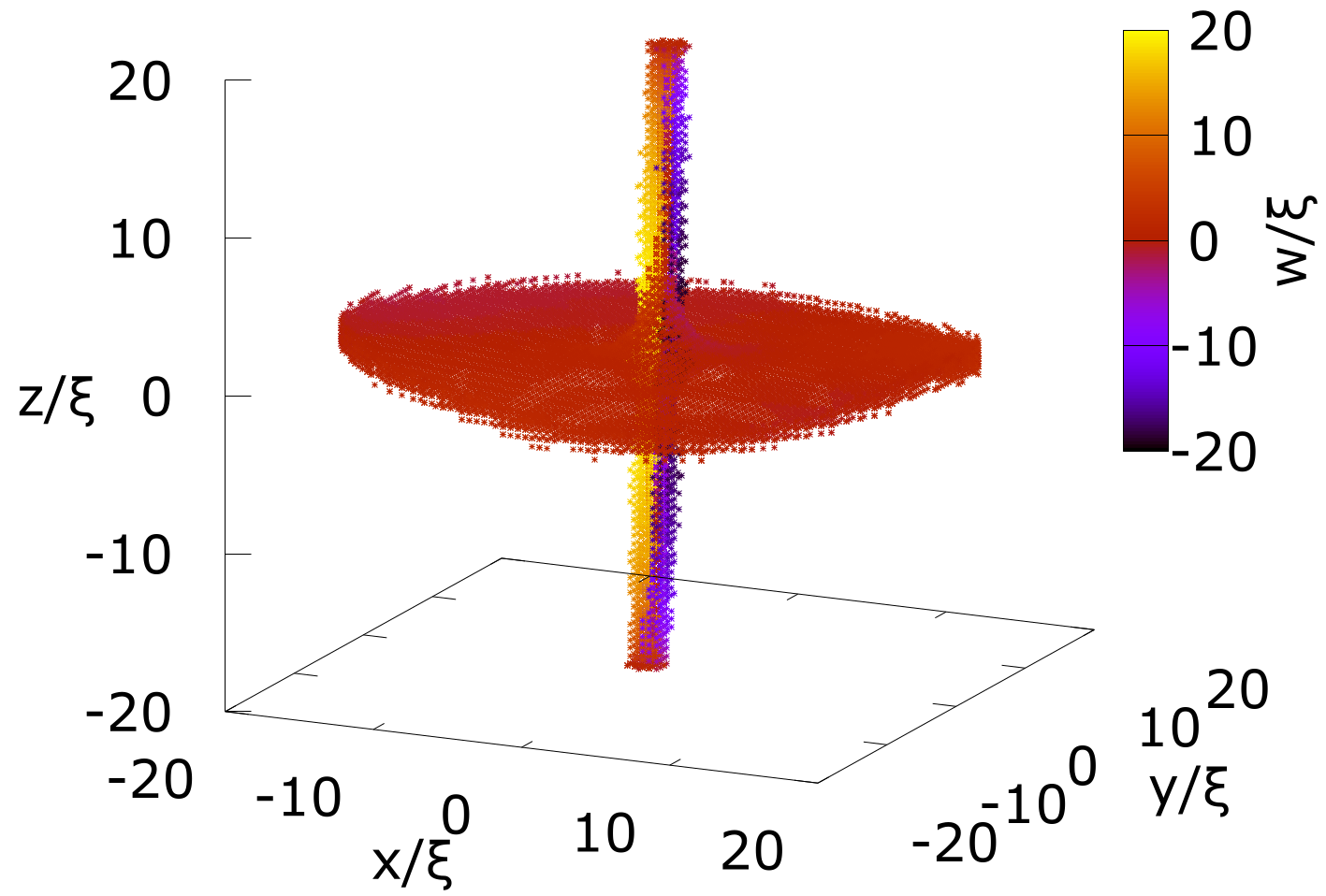}
    \end{minipage}
    \begin{minipage}{0.45\textwidth}
        \centering
        \includegraphics[width=\textwidth]{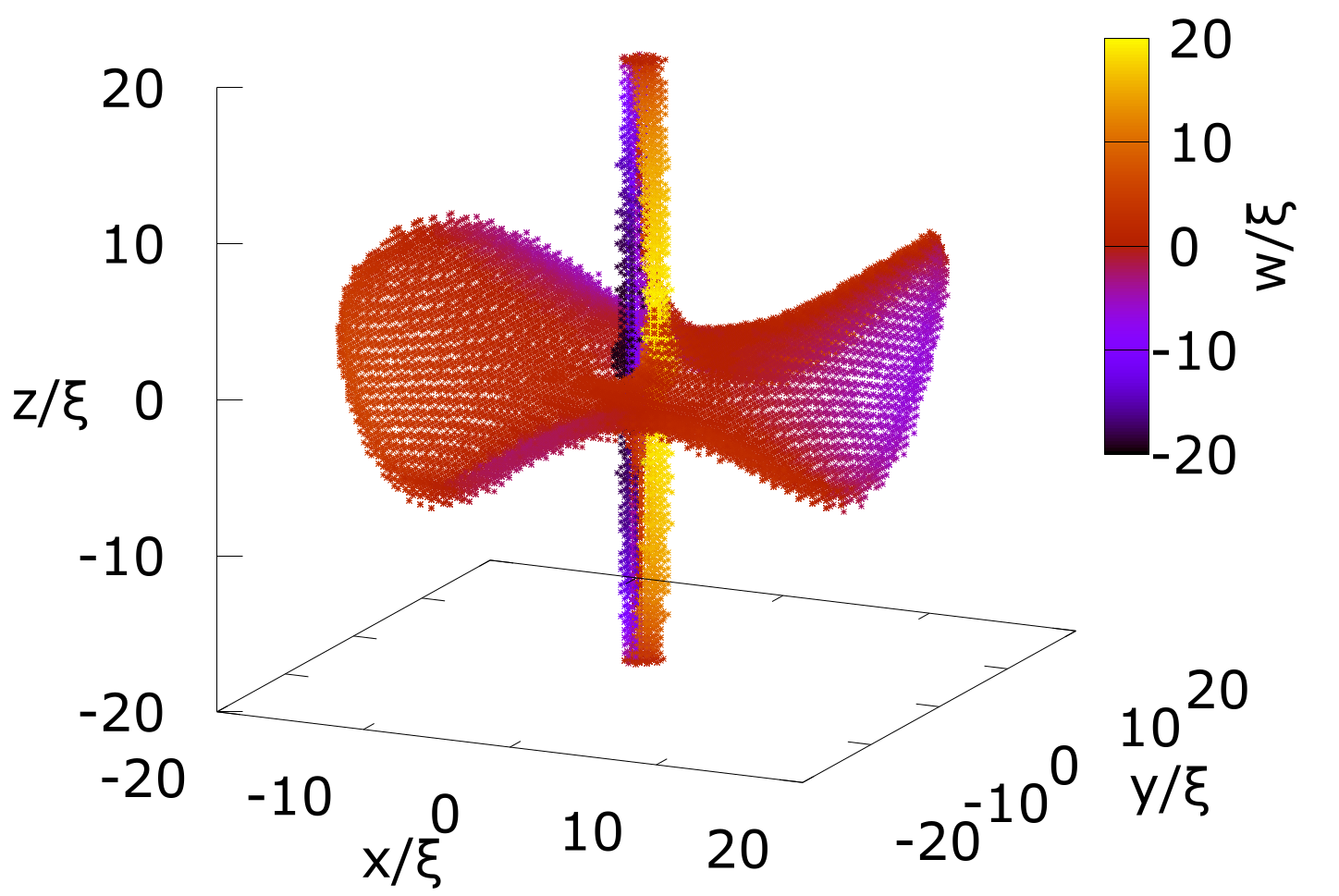}
    \end{minipage}
    \begin{minipage}{0.45\textwidth}
        \centering
        \includegraphics[width=\textwidth]{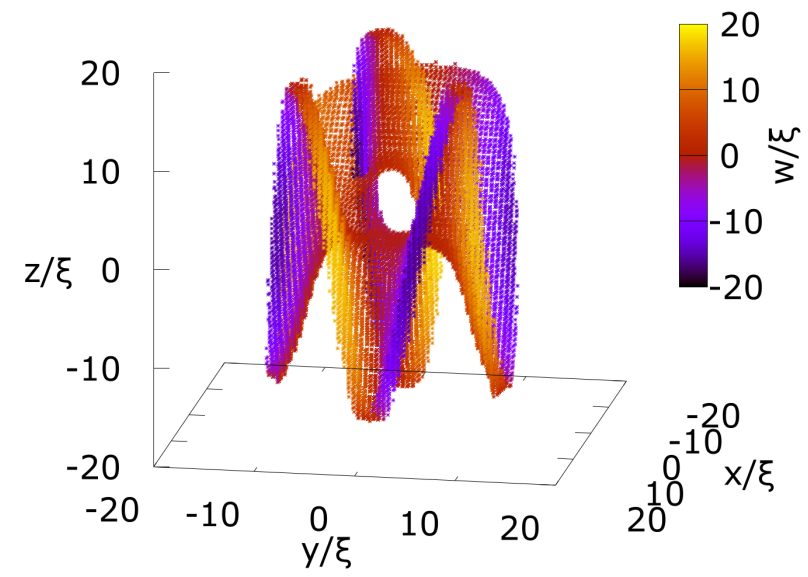}
    \end{minipage}
    \begin{minipage}{0.45\textwidth}
        \centering
        \includegraphics[width=\textwidth]{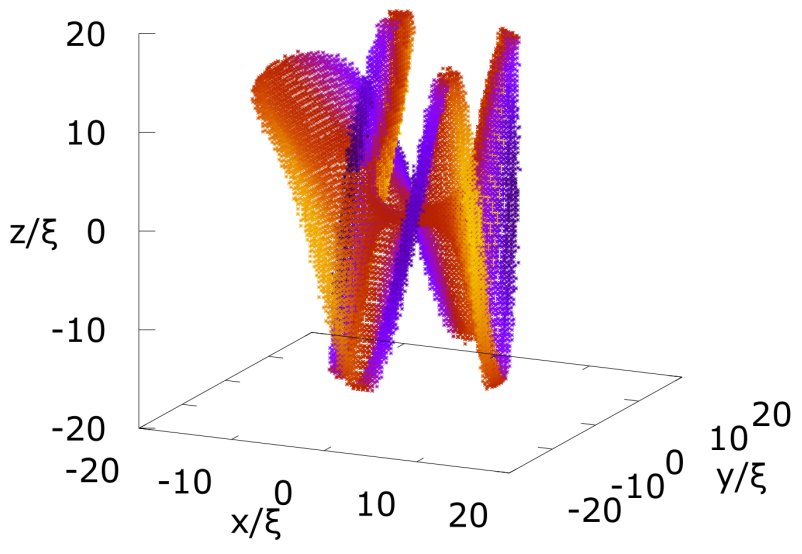}
    \end{minipage}
    \caption{
    Numerical vortex cores in final states of the ITEM-evolved 4D GPE under double rotation [Eq.~\eqref{eq:GPE4DR}]. The spatial step size was \(\Delta x=0.5\xi\), corresponding to \(R\approx20.6\xi\). Rotation frequencies were \(\Omega_{zw}=1.25\Omega_c\), and \(\Omega_{xy}\approx1.35\Omega_c\) (Top-Left), \(\Omega_{xy}\approx1.46\Omega_c\) (Top-Right), and
    \(\Omega_{xy}\approx1.83\Omega_c\) (Bottom), corresponding to predicted skewness values of \(\eta=10\degree\), \(20\degree\), and \(40\degree\), respectively. However, we used an initial phase profile corresponding to asymmetrically tilted planes --- rather than the predicted symmetrically tilted ones --- with added noise, as described in the main text. (Top-Left) Instead of tilting, we see that the vortex core is starting to become slightly curved even at long distances. The energy at these frequencies is \(E\approx0.66 \mu N\), while the predicted skew plane energy from the fit in Fig~\ref{fig:TiltEnergyNumerics} is \(\approx0.65\mu N\). (Top-Right) The surface centred around \(z=w=0\) is now buckled, effectively tilting towards the other plane in different directions depending on the angle \(\theta_1\). The energy is \(E\approx0.66\mu N\) while the skew plane energy is \(\approx0.65\mu N\). (Bottom) The two panels show two different views in \((x,y,z)\) space. The former \(x\en y\) plane is now so curved that it has forced the other surface to become tilted and displaced, and there is an avoided crossing between them as visible in the first panel. Note that the second panel shows that parts of this surface appear to be parallel to each other. Here, the energy is approximately the same as the predicted skew plane energy at \(E\approx0.64\mu N\).}
    \label{fig:Curve}
\end{figure*}

\section{Additional Numerical Results}
\label{app:numerical}

In this appendix, we will present extra numerical results to supplement those in the main text. Some of these data are from simulations not mentioned in the text, and others are additional data from simulations described in the text, to aid in explanation and visualisation.

Fig~\ref{fig:10Skew} shows the final state vortex core for a run of the ITEM with frequencies \(\Omega_{zw}=1.25\Omega_c\), and \(\Omega_{xy}\approx1.35\Omega_c\), which gives a predicted skewness angle of \(\eta=10\degree\). Just as in Fig~\ref{fig:40Skew}, the agreement between theory and numerics is still very good apart from the avoided crossing region near the origin. In particular, the third panel shows a side on view in which the vortex cores lies roughly along a pair of lines, on top of which we have plotted the theoretically predicted lines we expect. As can be seen, the data is still very close to the predicted values, although interestingly there is a small degree of asymmetry visible in the third panel of Fig~\ref{fig:10Skew} --- the data points are not symmetric about the vertical axis --- that can't be seen in Fig~\ref{fig:40Skew}. However, without further investigation we cannot tell whether this is due to numerical inaccuracy or some genuine physical phenomena.

Fig~\ref{fig:HiResSkew} shows the final state for \(\eta=45\degree\), with a spatial resolution of \(\Delta x = 0.2\xi\), compared to \(0.5\xi\) for the previous figures. The agreement between these data and our predictions is as good as before, as can be seen in the third panel, and the data does have mirror symmetry about the vertical axis. However, there are also some interesting qualitative differences to the lower resolution results. 
% Firstly the vortex core is thinner, which is due to the way we designate points on the core. As explained in Sec~\ref{sec:Numerics}, we look for points with a value of \(\abs{\psi}\) less than \(\Delta x/\xi\), in order to target roughly a one gridpoint radius around the vortex core. However, since the gradient of the order parameter near the core is different from unity this does lead to some dependence of these core plots on resolution.
For example, the avoided crossing between the planes has a different orientation to that in the previous two figures, as can be seen by comparing the first two panels. 
Finally, the size of the reconnected region is smaller, which is most likely due to the smaller system size, but could be due to the higher resolution.

Fig~\ref{fig:WeirdThings} shows two strange curved vortex surfaces observed at the isoclinic point for identical parameters. The frequencies were \(\Omega_{zw}=\Omega_{xy}=1.3\Omega_c\), and the spatial step size was \(\Delta x=0.5\xi\), which corresponds to a radius of \(R\approx20.6\xi\). Both of these final states were the last iteration in a sequence of ITEM runs, starting from \(\Delta\Omega=\Omega_c\) down to \(\Delta\Omega=0\) in steps of \(0,1\Omega_c\), with \(\Omega_{zw}=1.3\Omega_c\) fixed. The final state of each run was used as the initial state of the next, so that we could follow the evolution of a particular energy branch. We were attempting to track the predicted skew plane states for a fixed \(\Omega_{zw}=1.3\Omega_c\) and find their energies [c.f. Fig~\ref{fig:TiltEnergyNumerics}], and the two final states at \(\Delta\Omega=\Omega_c\) did in fact closely correspond to these predicted planes. However, both of these states began to deviate from the theoretical states as we decreased \(\Delta\Omega\), becoming more and more curved all the way to the isoclinic point. These isoclinic curved states have approximately the same energy as each other, and as the orthogonal state at the same frequencies (\(E\approx0.65\mu N\)). Since any energy differences are below the precision of our numerics we would need more accuracy to investigate this. If these are indeed low lying excitations, this may be due to the degeneracy associated with isoclinic symmetry, in which case we may expect many more such states.

Finally, Fig.~\ref{fig:Curve} shows the vortex cores obtained using the asymmetric phase profile (with added noise), for \(\Delta\Omega\) values corresponding to \(\eta=10\degree\), \(
20\degree\) and \(40\degree\), respectively. As mentioned in the main text, these are slightly higher in energy than the theoretical states studied in the main text. Already, in the \(10\degree\) case we can see that our assumption of flat vortex planes is broken as there is some long-range curvature of the core. This is then exacerbated as the frequency \(\Delta\Omega\) increases, with the \(20\degree\) state clearly showing that there is almost no overall tilt of the former planes, but instead the plane at \(z=w=0\) has begun to buckle in an approximately threefold symmetric pattern, curving towards the other surface in different directions for different values of the angle \(\theta_1\). Finally, in the \(40\degree\) figure, this curvature has become so extreme that the former plane at \(x=y=0\) appears to have become tilted and displaced as well as curved, leading to a sizeable avoided crossing where these two surfaces come together, as can be seen in the first panel. Interestingly, the second panel appears to show three parts of the vortex core surface that are parallel to each other. This suggests that this bizarre state may be limiting towards a state with multiple vortex planes parallel to the \(z\en w\) plane but separated  in the \(x\en y\) plane. This is the expected lowest energy state for the case of high frequency simple rotation in the \(x\en y\) plane (i.e. with \(\Omega_{zw}=0\)) so it seems reasonable that it should also be the ground state when \(\Omega_{xy}\gg\Omega_{zw}\). However, we are not quite reaching this limit in the \(40\degree\) case, as there we have \(\Omega_{xy}\approx1.46\Omega_{zw}\). We therefore tentatively describe this strange set of states as an instance where the frequencies are large enough that the system wants more than two vortex planes but not enough for three. The planes can curve in order to become larger, thereby fitting a larger area of vortex surface in the system. Whether this is a correct description or not, it is clear that the behaviour of vortex surfaces in 4D is incredibly rich, and there is much more to be explored.

\bibliographystyle{apsrev4-2}

\end{document}